\documentclass[twocolumn]{aastex631}
\hypersetup{linkcolor=red,citecolor=blue,filecolor=cyan,urlcolor=magenta}
\received{xxxx}
\revised{yyyy}
\accepted{zzzz}
\submitjournal{ApJ}
\usepackage[T1]{fontenc}
\usepackage{graphicx}
\usepackage{amsmath}
\usepackage{float}
\usepackage{multirow}

\shorttitle{Galactic impact on multi-transonic accretion}
\shortauthors{Sk et al.}
\graphicspath{{./}{Graphics/}}

\begin{document}

\title{Model-Dependent Galactic Environment Effects on Multi-Transonic Accretion and Emergent Acoustic Gravity around Kerr Black Holes}

\correspondingauthor{Ripon Sk}
\email{riponphysics@gmail.com}

\author[0000-0001-6140-5093]{Ripon Sk}
\affiliation{Department of Physics, West Bengal State University, Berunanpukhuria, West Bengal 700126, India.}

\author[0000-0003-3979-113X]{Sangita Chatterjee}
\affiliation{School of Astrophysics, Presidency University, 86/1 College Street, Kolkata 700073, India.}

\author[0000-0001-6373-7040]{Sankhasubhra Nag}
\affiliation{Department of Physics, Ramakrishna Mission Vivekananda Centenary College, West Bengal 700118, India.}

\begin{abstract}
The hydrodynamics of low angular momentum, multi-transonic, axisymmetric, inviscid accretion flow onto a rotating black hole in presence of a galactic environment has been systematically investigated in detail using three standard disc geometries and two thermodynamic equations of state, within the post-Newtonian framework with a pseudo-potential. In addition to the influence of a centrally located black hole, this study incorporates those due to a multi-component galactic environment -- stellar matter, dark matter, and hot gas. Our analysis reveals that the extent of additional influence of the galactic environment on accretion flow depends severely on the disc model chosen. This effect is particularly pronounced in the vertical equilibrium disc model, whereas it is extremely mild for other two geometries. The same is observed for location of sonic points, parameters related to multi-transonicity and shock formation, and also for the emergent analogue surface gravity calculated. Another observation is that, among different components of galactic environment, contribution of dark matter is most dominant, followed by a mild one due to hot gas; while that due to stellar matter is even milder in determining the flow. The stationary flow characteristics are analysed using semi-analytical numerical methods including the method analogous to critical point analysis used in dynamical systems. Additionally, a time-dependent linear perturbation analysis ensures the stability of stationary accretion flow in all cases, and hence the corresponding acoustic metric and acoustic surface gravity are derived.
\end{abstract}


\section{Introduction}
\label{sec:intro}

Accretion of X-ray-emitting hot gas powers the central supermassive black holes (SMBHs) in massive elliptical galaxies \citep{2003ARA&A..41..191M,2019MNRAS.488.1917G}. The angular momentum of the inflowing gas determines the geometry and dynamics of the accretion flow \citep{2002apa..book.....F}. As gas in realistic galactic environments possesses finite, albeit low, angular momentum \citep{2000MNRAS.311..507D}, the Bondi solution \citep{1952MNRAS.112..195B} is an idealized limit. Consequently, SMBH accretion in elliptical galaxies is more naturally described by low-angular-momentum axisymmetric flows. These flows are inherently transonic and, for suitable combinations of the conserved flow parameters, may admit multiple sonic points. Such solutions can undergo standing or oscillating shock transitions, profoundly modifying the accretion dynamics and providing a natural framework for explaining several AGN phenomena, including hard X-ray emission, relativistic jets, and quasi-periodic oscillations \citep{1980ApJ...240..271L,1981ApJ...246..314A,1982AcA....32....1M,1983AcA....33...79M,1985A&A...148..176L,1986AcA....36....1M,1987PASJ...39..309F,1989PASJ...41..271N,1989MNRAS.240....7C,1990ttaf.book.....C,1992MNRAS.259..259N,1993PASJ...45..167N,1993ApJ...412..254C,1994MNRAS.270..871N,1996PhR...266..229C,1997A&A...321..665L,1998ApJ...504..419P,2004ApJ...611..964F,2008ApJ...689..391N,2009ApJ...696.2026N,2019ApJ...877...65N,Dihingia:2019zos,2021MNRAS.506.3111N,2021Galax...9...21M,2022A&A...662A..77M,2022JApA...43...90M,Das:2022uwy,2023MNRAS.526.5612C}. Consequently, these flows has remained an active area of research, with renewed interest following its success in reproducing the highly variable emission from Sgr A* ~\citep{2018MNRAS.478.3544R,2023ApJ...954...56O,2024ApJ...967....4D}. Early analytical studies established the existence of multi-transonic solutions and shock formation in inviscid accretion flows, which were subsequently generalized to relativistic equations of state, different flow geometries, and extensive parameter-space analyses ~\citep{2016MNRAS.459.3792C,2018PhRvD..98h3004D,2019MNRAS.488.2412D,2020MNRAS.496.3043D,2022PDU....3701120P,2024JCAP...01..060P,2022JCAP...08..048S,2024PDU....4401429U}. More recently, general relativistic hydrodynamic (GRHD) and magnetohydrodynamic (GRMHD) simulations have confirmed the emergence of complex transonic structures, including standing and oscillating shocks, and predicted distinctive timing, spectral, and polarization signatures ~\citet{2017bhns.work..163S,2017MNRAS.472.4327S,2017MNRAS.472..542K,2018CQGra..35i5005M,2019MNRAS.482.3636K,2019MNRAS.487..755P,2023A&A...678A.141O,2024ApJ...971...28M,2025ApJ...990...35D}.

However, almost all existing investigations—whether performed in Newtonian, post-Newtonian, or general relativistic frameworks—model AGN accretion solely under the gravitational influence of the central SMBH \citep{2021PhRvD.103b3023T,2019PhRvD.100d3024T,2016NewA...43...10S,2015NewA...37...81D,1964ApJ...140..796S,1969Natur.223..690L,1972A&A....21....1P,1973A&A....24..337S,2022ApJ...941..131C}. Such an approximation neglects the gravitational field of the host galaxy, which can substantially modify the dynamics of the accretion flow. Indeed, motivated by Hubble Space Telescope observations of elliptical galaxies, \citet{2000ApJ...528..236Q} showed that the host galaxy potential can significantly modify Bondi accretion by incorporating a realistic model for the galactic mass distribution. Extending this idea, \citet{2005MNRAS.363..705M} modeled the gravitational field of a typical massive elliptical galaxy using four principal components—the central SMBH, stellar distribution, diffuse hot gas, and dark matter halo—and demonstrated that the galactic potential significantly influences transonic adiabatic accretion. Subsequently, \citet{2018MNRAS.479.3011R,2021JCAP...05..025R} incorporated dark energy as a fifth component to examine spherical transonic accretion in a five-component galactic potential. While the additional galactic components modify the flow dynamics, these studies, consistent with earlier investigations \citep{2015IJMPD..2450084G,2016PhRvD..94j3513S,2013PhRvD..88h4056M,2013PhRvD..87d4007K}, showed that the influence of dark energy is negligible in the central regions and becomes appreciable only at large galactocentric distances.

In this work, we investigate the influence of the multi-component gravitational potential of the host galaxy, following \citet{2005MNRAS.363..705M}, on the nature and location of transonic low-angular-momentum polytropic accretion onto a Kerr black hole. This work bridges the gap between observationally constrained galactic mass distributions and the transonic structure of axisymmetric black-hole accretion flows. While isolated black-hole accretion models neglect the host-galaxy environment and studies of galactic mass distributions do not address accretion dynamics, combining both within a single framework enables us to quantify how realistic galactic potentials influence flow dynamics with potential implications for observable accretion signatures. Since a fully relativistic treatment that simultaneously incorporates both the Kerr spacetime and the self-consistent gravitational field of the surrounding galaxy is not yet available, we adopt a pseudo-Newtonian framework that provides a physically motivated and computationally tractable description of the system. To describe the black hole, we employ a pseudo-Kerr framework. Existing pseudo-Kerr prescriptions can be broadly classified into phenomenological, metric-derived, and generalized Newtonian formulations, all aiming to reproduce the essential relativistic properties of Kerr spacetime while retaining the simplicity of Newtonian gravity. An early phenomenological model was proposed by \citet{1992MNRAS.256..300C} for transonic accretion studies, although its applicability is limited for rapidly rotating black holes. Subsequently, \citet{1996ApJ...461..565A} introduced a force-based pseudo-Newtonian potential that accurately reproduces the event horizon and marginally stable orbit over a broad range of black-hole spins. Alternative formulations by  \citet{1999A&A...343..325S} were developed to improve the description of relativistic particle dynamics and circular orbits in Kerr geometry. Another advance was made by \citet{2002ApJ...581..427M}, who derived an effective pseudo-Kerr force directly from the Kerr metric, a result later extended by \citet{2003ApJ...582..347M} to recover the angular and epicyclic frequencies relevant to disc oscillations. Further developments include the prescription of \citet{1996ApJ...464..664C} for advective and shocked accretion flows, the vector-potential formulation of \citet{2007ApJ...667..367G} for off-equatorial motion, the generalized Newtonian velocity-dependent potential of \citet{2013MNRAS.433.1930T}, and the pseudo-Newtonian framework of \citet{2014MNRAS.445.4463G}, which provides improved agreement with Kerr spacetime over a wider spin range. In our work, the choice of pseudo-Kerr prescription is guided by the physical problem under consideration. Since we focus on transonic accretion in the combined gravitational field of the host galaxy, a force-based description of the black-hole gravity is particularly advantageous as it allows the relativistic gravitational acceleration to be incorporated naturally alongside the stellar, gaseous, and dark-matter components of the galactic potential. We therefore adopt the pseudo-Kerr formulation of \citet{1996ApJ...461..565A}, which retains the analytical simplicity required for critical point and shock analyses.

To describe the three-dimensional stellar mass density distribution of the chosen galaxy, we follow specific density models: the Sérsic profile ~\citep{1968adga.book.....S}, which characterizes how surface brightness declines smoothly from the center toward the outer regions, with its shape governed by the Sérsic index $n$, allowing it to represent a wide range of galaxy types. Among the various proposed dark matter (DM) density distribution models, the double power-law profile introduced by ~\citet{1995MNRAS.275..720N,1996ApJ...462..563N} (NFW) is most widely used. Subsequent simulations, however, indicated that the inner slope predicted by NFW  could be significantly steeper ~\citep{1999MNRAS.310.1147M,2000ApJ...529L..69J}. In this study, we adopt this steeper model of   ~\citet{2000ApJ...529L..69J} profile to represent the DM component within our multi-component galaxy model. The hot gas found in groups and clusters of galaxies, as well as in large, isolated elliptical galaxies-is in hydrostatic equilibrium ~\citep{2010MNRAS.407.1148C,2012MNRAS.422..686C} and a widely used model for the radial density distribution of this hot gas is the isothermal $\beta$-model, which provides a good empirical fit to X-ray surface brightness profiles of early-type systems ~\citep{2001ApJ...547..154B,2005MNRAS.363..705M}. This model assumes a constant temperature and a spherically symmetric gas distribution, making it particularly suitable for modeling the extended hot gas halo. In our work, we follow ~\citet{2005MNRAS.363..705M,2018MNRAS.479.3011R} and use this $\beta$-model to represent the gas. The resulting gravitational force derived from this density profile is incorporated into the total potential governing the system's dynamical structure.

Within this multi-component framework, we investigate the transonic properties of low-angular-momentum polytropic accretion. In particular, we examine the global flow topology, the existence of multiple sonic points, and the conditions for shock formation in both adiabatic and isothermal flows \citep{1959flme.book.....L,1980ApJ...240..271L,1981ApJ...246..314A,1982AcA....32....1M,1989ApJ...336..304A,2002ApJ...577..880D}. To assess the role of disc geometry, the analysis is carried out for the three canonical axisymmetric configurations—constant-height (CH), conical (CO), and vertical-equilibrium (VE) discs \citep{2001MNRAS.327..808C,2016NewA...43...10S,Bilić2014}.

We further investigate the linear stability of the stationary solutions by perturbing the time-dependent Euler and continuity equations \citep{2003MNRAS.344...83R,2005astro.ph.11018R,2005ApJ...627..368R,2006MNRAS.373..146C,2007MNRAS.378.1407G,2018IJMPD..2750023T}, and explore the resulting analogue acoustic geometry by evaluating the acoustic horizons and their associated surface gravity \citep{1980ApJ...235.1038M,1981PhRvL..46.1351U,2005LRR.....8...12B,2018IJMPD..2750023T}. Finally, we examine how the galactic potential, black-hole spin, and flow parameters all together influence the transonic structure, shock properties, and acoustic characteristics of the accretion flow.

The rest of the paper is planned accordingly: in the \S\ref{govequ} we presents the governing equations ; in \S\ref{galpotsec}, we describe the combined gravitational contribution to the potential of the multi-component galaxy. In \S\ref{galpotd}, we describe the galactic parameters. In \S\ref{flowD}, we describe the hydrodynamics of the flow. The nature of the fixed points are discussed in \S\ref{fixednature}. The time depenedent stability of the solutions are described in \S\ref{stable-F}. We expressed the acoustic surface garvity in\S\ref{Surgrav}. Finally the result and discussion are given in \S\ref{Result}. The concluding remarks are given in \S\ref{conclu}.

\section{Governing Equations}
\label{govequ}

\subsection{Gravitational Potential of the elliptical galaxy}
\label{galpotsec}

Elliptical galaxies, typically approximated as nearly spherical, enable simplified modeling of gravitational dynamics. Observationally, their mass distributions exhibit minimal deviation from spherical symmetry, particularly in massive systems. The gravitational force arising from a spherically symmetric mass distribution-in the case of an elliptical galaxy-results from the combined contribution of mainly four mass components: the central black hole (BH), the stellar content, dark matter (DM), and the surrounding diffuse hot gas. The total gravitational potential can thus be represented as a straightforward linear sum of the individual gravitational potential contributions from each of its distinct components, and can be written as,
\begin{equation}
\Phi_{\mathrm{Gal}}(r) = \Phi_{\mathrm{BH}}(r) + \Phi_{\mathrm{star}}(r) + \Phi_{\mathrm{gas}}(r) + \Phi_{\mathrm{DM}}(r)
\label{eqgalphi},
 \end{equation}

where $\Phi_{\mathrm{Gal}}(r)$ provides the galactic potential and the respective force function can be obtained as $\mathbb{F}_{\mathrm{Gal}}(r)=\frac{d\Phi_{\mathrm{Gal}}(r)}{dr}$. The $\Phi_{\mathrm{BH}}(r)$ defines the Kerr black hole potential. In the present analysis, we employ a pseudo-Newtonian potential as an effective approximation to the Kerr spacetime geometry. Among the various formulations available in the literature, we adopt the potential introduced by  ~\cite{1996ApJ...461..565A}, which offers a suitable representation of the relativistic effects. The explicit expression for the same ~\citep{2012ChPhB..21e0504W} is given by,
\begin{equation}
\Phi_{\mathrm{BH}}(r) = -\frac{1}{(\eta - 1) r_{1}} \left[ \left( \frac{r}{r - r_{1}} \right)^{\eta - 1} - 1 \right] , \label{eqbh}
\end{equation}
where, $r$ is the radial distance measured in units of $r_{g} = {G M_{BH}}/{c^2}$ in which $M_{BH}$ is the central SMBH mass, $G$ is the gravitational constant, and $c$ is the light speed. Note that in this present study we consider $c=G=M_{BH}=1$. The term $\eta=\frac{r_{in}}{r_1}-1$ where $r_1=1 + \sqrt{1 - \rm{a}^2}$ defines the location of the event horizon for Kerr spin $\rm{a}$. Where $r_{in}= 3 +Z_2 - \left[(3-Z_1)(3+Z_1+2Z_2)\right]^{1/2}$, $
Z_1 = 1+ (1-\rm{a}^2)^{1/3}\left[(1+\rm{a})^{1/3}+(1-\rm{a})^{1/3}\right]$, $Z_2 = \left(3\rm{a}^2+Z_1^2\right)^{1/2}$. The rest of the symbols can be found in ~\cite{2018MNRAS.480.3017M}. Although the present analysis employs pseudo-Newtonian framework, the gravitational field is modeled using the above potential, which accurately mimics the general relativistic spacetime of a  Kerr black hole outside the event horizon. The solutions are confined to regions well outside the horizon (\( r \gtrsim 2\text{--}3\,r_{\mathrm{S}} \)), where such approximations remain valid for reproducing the essential transonic behaviour of accretion flows ~\citep{1996ApJ...461..565A,2002ApJ...581..427M,2004PASJ...56..569F,2007ApJ...667..367G,2012NewA...17..254D}.  However, noticeable deviations are exhibited by the Artemova potential for rapidly spinning black holes. In such cases, alternative effective potentials, such as those proposed by \citet{1978A&A....63..209K,2018PhRvD..98h3004D}, may provide a more accurate description. To quantify these deviations, the relative difference in the gravitational-force gradient, $ \delta(r)=100\times
\frac{\left(\frac{d\phi}{dr}\right)_{\rm Artemova}
-\left(\frac{d\phi}{dr}\right)_{\rm Dihingia}}
{\left(\frac{d\phi}{dr}\right)_{\rm Dihingia}},
$
is evaluated using the effective potential of \citet{2018PhRvD..98h3004D} for a black hole spin of $a=0.92$. It is found that $\delta(r)$ increases from $2.79\%$ at $r=100.5\,r_g$ to $13.9\%$ at $r=20.5\,r_g$, reaching $45.05\%$ at $r=6.5\,r_g$. Therefore, although the Artemova potential remains a reasonable approximation in the weak-gravity regime, its deviation becomes substantial in the vicinity of the black hole.

Recent advances in the photometric analysis of elliptical galaxies have shown that traditional models-such as the Hubble-Reynolds law ~\citep{1913MNRAS..74..132R}, King models ~\citep{1962AJ.....67..471K}, and the de Vaucouleurs $ R^{1/4}$ law ~\citep{1948JO.....31..113D} —generally fail to accurately fit their surface brightness profiles. However, a broad consensus has emerged supporting the Sérsic law, a generalization of the $R^{1/4}$ profile, as a superior and widely applicable model for describing the surface photometry of nearly all elliptical galaxies. The $R^{1/4}$ model proposed by Sérsic ~\citep{1968adga.book.....S} is most commonly expressed in the form of an intensity profile : $I(r)=I_0 \exp{\left[-(r/r_s)^{1/n}\right]}$ where $I_0$ is the intensity at the effective radius $r_s$ also known as Sérsic scale radius, $n$ is the Sérsic index. Following ~\cite{2005MNRAS.362...95M,2005MNRAS.363..705M,2005MNRAS.362..197T,2005PASA...22..118G,2018MNRAS.479.3011R}, we get the gravitational potential for the stellar mass distribution in the elliptical galaxy as\citep{2005MNRAS.362...95M},
\begin{multline}
\Phi_{\mathrm{star}}= -\int \mathbb{F}_{\mathrm{star}}(r)\,dr = \\
-\int \frac{G f_{\mathrm{star}} M_{v}}{r^2}
   \cdot \frac{ \xi\left[(3{-}\mu)n,\left({\scriptstyle \frac{r}{r_s}}\right)^{1/n}\right] }
              { \xi\left[(3{-}\mu)n,\left({\scriptstyle \frac{r_v}{r_s}}\right)^{1/n}\right] } \ dr
\label{eqstar},
\end{multline}

where $r_{v}\simeq 206.262 h_{70}^{-1}\left(\frac{h_{70}M_{v}}{10^{12}M_{\odot}}\right)^{\frac{1}{3}}$ kpc is the virial radius and mass enclosed within the virial radius is denoted by $M_v$. $f_{star}$ is stellar mass fraction. The relation between the Sérsic index and $\mu$ is $\mu(n)\simeq 1.0-\frac{0.06097}{n}+\frac{0.05463}{n^2}$. The standard incomplete gamma function is denoted here by, $\xi(t,x)= \int_{0}^{r} t^{x-1} e^{-t} \, dt$
\citep{NIST:DLMF:8.2}. The rest of the notations and their details are defined in ~\cite{1997A&A...321..111P,2005MNRAS.363..705M,2018MNRAS.479.3011R}.

In elliptical galaxies, hot diffuse gas contributes significantly to both X-ray emission and the total baryonic mass, especially at the virial radius. Assuming the cosmic baryon fraction applies, the mass in hot gas may exceed that in stars by a factor of $\sim 4$ ~\citep{2005MNRAS.362...95M,2005MNRAS.363..705M}. This dominance depends on the galaxy's mass-to-light ratio and suggests that hot gas is a crucial component of the baryon budget in massive ellipticals. Assuming hydrostatic equilibrium with the galaxy’s gravitational potential, the spatial distribution of this hot gas is effectively described by the isothermal $\beta$-model ~\citep{1976A&A....49..137C,1986RvMP...58....1S,2001ApJ...547..154B,2005MNRAS.362...95M,2005MNRAS.363..705M}, which assumes constant temperature and radially varying density, reproducing observed X-ray surface brightness profiles. We employ this isothermal $\beta$-model to represent the hot gas component in our analysis. The associated potential profile is obtained by evaluating the integral of the gravitational force arising from this gas distribution within the virial radius, which is expressed as:

\begin{equation}
\Phi_{\mathrm{gas}}= -\int \mathbb{F}_{\mathrm{gas}}(r)\,dr = -\int \frac{G f_{\mathrm{gas}} M_v}{r^2} \cdot \frac{\mathbb{F}_{N}}{\mathbb{F}_D}.dr , \label{eqgas}
\end{equation}
where, \small {$\mathbb{F}_{N}=\left[\left(\frac{1}{3}\left(\frac{r}{r_b}\right)^3\right)^{-\delta} + \left(\frac{2}{3}\left(\frac{r}{r_b}\right)^{3/2}\right)^{-\delta}\right]^{-1/\delta}$ and $\mathbb{F}_D=\left[\left(\frac{1}{3}\left(\frac{r_v}{r_b}\right)^3\right)^{-\delta} + \left(\frac{2}{3}\left(\frac{r_v}{r_b}\right)^{3/2}\right)^{-\delta}\right]^{-1/\delta}$.} Within the virial radius the hot gas mass fraction is given by $f_{gas}$ and scale radius $r_b\simeq \mathcal{R}_{eff}/q$, here $\mathcal{R}_{eff}$ is the effective radius, $q\simeq10$ and $\delta=2^{\frac{1}{8}}$.

Constraints on dark matter in elliptical galaxies are significantly improved by combining internal stellar kinematics with either X-ray observations or strong gravitational lensing. On the theoretical front, high-resolution dissipationless cosmological N-body simulations have converged on halo structures characterized by an outer density slope of $-3$ and inner slopes ranging from $-1$ (NFW) to $-1.5$ (Moore et al. 1999)~\citep{1999MNRAS.310.1147M}. Jing \& Suto (2000) (JS-1.5) ~\citep{2000ApJ...529L..69J} proposed a generalized double power-law profile that accurately reproduces these simulated halo properties, which is given by,

\begin{equation*}
\rho_{\mathrm{DM}}(r) \propto \frac{1}{\left(r/r_d\right)^\gamma \left[1+\left(r/r_d\right)\right]^{3-\gamma}}, \quad \gamma = \frac{3}{2}~(\text{JS-1.5}),
\end{equation*}
with scale radius $r_d = r_{v}/c_0$, where $c_0$ is the concentration parameter ~\citep{2006AJ....132.2685M}.
The gravitational potential associated with this JS-1.5 model's dark matter mass distribution is given by,

\begin{multline}
 \Phi_{\mathrm{DM}} = -\int \mathbb{F}_{\mathrm{DM}}(r)\,dr = \\
   -\int \frac{G f_{DM}M_{v}}{r^2} \frac{1}{\mathcal{G}(c_0)}2[sinh^{-1}(\sqrt{r/r_d})-\sqrt{\frac{r/r_d}{(1+r/r_d)}}].dr,
 \label{eqdm}
\end{multline}
where
\begin{equation}
 \mathcal{G}(c_0)=2 \left[\sinh^{-1}\left(\sqrt{c_0}\right)-\sqrt{\frac{c_0}{(1+c_0)}}\right].
\end{equation}

Now using equations (\ref{eqbh})-(\ref{eqdm}) in equation (\ref{eqgalphi}) we get the host galaxy's gravitational potential $\Phi_{\mathrm{Gal}}$.
The details of the various galactic scale parameters are provided in the next subsection \S \ref{galpotd}.

\subsection{Galactic Potential Parameters and the Role of $\Upsilon_B$}
\label{galpotd}

The global mass--to--light ratio $\Upsilon_B$ is defined as the ratio of the total virial mass of a galaxy to its $B$--band luminosity,
\begin{equation}
\Upsilon_B \equiv \frac{M_{\rm v}}{L_B},
\end{equation}
and therefore depends on two global quantities: the virial mass $M_{\rm v}$, determined by the dark matter halo and the total baryonic content of the system, and the luminosity $L_B$, which is fixed by the stellar population properties such as the initial mass function, age, metallicity, and star formation history. $\Upsilon_B$ is not a radially dependent quantity but a global halo parameter evaluated at the virial radius. Different values of $\Upsilon_B$ correspond to different assumptions regarding the amount of dark matter present per unit stellar light, and thus $\Upsilon_B$ effectively acts as a proxy for the halo mass at fixed luminosity. Variations in $\Upsilon_B$ do not alter the functional forms of the adopted density profiles for the stellar, dark matter, and gaseous components; instead, they determine the total mass normalization assigned to each component. This ratio $\Upsilon_B$ therefore provides a natural measure of whether a galaxy is baryon-dominated or dark-matter-dominated. The virial mass is related to the luminosity through ~\citep{2015IJMPD..2450084G,2010MNRAS.407.1148C}

\begin{equation}
M_v = b_{\Upsilon}\,\bar{\Upsilon}_B\,L_B ,
\end{equation}
where $\bar{\Upsilon}_B \simeq 390 M_\odot/L_\odot$ is the cosmic mean B-band mass-to-light ratio, $M_\odot$ and $L_\odot$ are the solar mass and solar luminosity. $b_\Upsilon=\Upsilon_B/\bar{\Upsilon}_B$ is the mass-to-light bias factor.

The stellar contribution is described by the stellar mass-to-light ratio in the B-band, taken as $\Upsilon_{\ast B} \simeq 6.5$, giving the stellar mass fraction ~\citep{2006MNRAS.370.1581M,2005MNRAS.363..705M}
\begin{equation}
f_{star} = \frac{\Upsilon_{\ast B}}{b_{\Upsilon}\bar{\Upsilon}_B}.
\end{equation}

For a fixed stellar mass--to--light ratio $\Upsilon_{*B}$, the stellar mass is given by
\begin{equation}
M_{star} = \Upsilon_{*B} L_B,
\end{equation}
and is therefore independent of $\Upsilon_B$, while the total virial mass satisfies $M_{\rm v} = \Upsilon_B L_B$. Consequently, an increase in $\Upsilon_B$ leads to a corresponding increase in the dark matter mass
with the gas mass adjusted to satisfy the assumed baryon fraction within virial radius,
\begin{equation}
f_b = \frac{M_{star} + M_{\rm gas}}{M_{\rm v}}.
\end{equation}
Although $\Upsilon_B$ is defined globally at the virial radius, the relative mass contribution of stars, gas, and dark matter varies with radius because each component follows a different density profile (e.g., the dark matter profile modeled by  Jing \& Suto (2000) (JS-1.5) profile ).

The deviation in baryon fraction within the virial radius from the universal mean $\bar{f}_b=\Omega_b/\Omega_m\simeq0.14$ is quantified by the baryon fraction bias, $b_b=f_b/\bar f_b$. We consider here the ratio of baryon density to critical density $\Omega_b=0.041$  and the ratio of matter density to critical density $\Omega_m=0.3$, following the big-bang nucleosynthesis measurement by ~\cite{2001ApJ...552..718O} and from the WMAP cosmic microwave background (CMB) experiment by ~\cite{2003ApJS..148..175S}.

The dark matter and gas mass fractions then follow as ~\citep{2005MNRAS.363..705M,2015IJMPD..2450084G,2018MNRAS.479.3011R}
\begin{equation}
f_{\mathrm{DM}} = 1 - b_b\bar f_b , \qquad
f_{\mathrm{gas}} = 1 - (f_{star} + f_{\mathrm{DM}}),
\end{equation}
where it has been assumed that the fraction of BH mass to the stellar mass is much less than unity.

The central black hole mass is estimated from the bulge mass $M_{\text{bulge}}=f_{star}M_v$ using the empirical correlation  ~\citep{2000ApJ...539L..13G,2002ApJ...574..740T,2013ApJ...764..151G,2004ApJ...604L..89H}
\begin{equation}
\log \!\left(\frac{M_{\mathrm{BH}}}{M_\odot}\right) = (8.20 \pm 0.10) + (1.12 \pm 0.06)\,\log \!\left(\frac{M_{\text{bulge}}}{10^{11} M_\odot}\right).
\end{equation}

In this study, we adopt a fiducial luminosity $L_B \simeq 2 \times 10^{10}L_\odot$ and examine four representative values of the global M/L ratio, $\Upsilon_B=14, 33, 100,$ and $390$, which span the full range from baryon-dominated to dark-matter-dominated systems. The lowest values ($\sim 14{-}17$) correspond to galaxies with little or no dark matter, such as NGC~821, while the highest value ($\Upsilon_B\simeq 390$)~\citep{2003Sci...301.1696R} reflects the cosmic mean, consistent with virialized halos. Intermediate cases ($\Upsilon_B=33$ and $100$) describe systems with progressively larger dark matter fractions, such as NGC~3379. For the case $\Upsilon_B=100$, we assume $f_b=\bar f_b\simeq0.14$, while for $\Upsilon_B=33$ a similar gas-to-star ratio is adopted.

The resulting fractions are:
\begin{itemize}
     \item \textbf{Set 1 ($\Upsilon_B=14$)}: $b_\Upsilon=0.0359$, $f_{star}=0.4643$, $b_b=7.1429$, $f_{\mathrm{DM}}=0.0$, $f_{\mathrm{gas}}=0.5357$.\\
    Galaxy Type : Baryon-rich, gas-dominated, negligible DM (e.g.,NGC 821 ~\citep{2003Sci...301.1696R}).
     \item \textbf{Set 2 ($\Upsilon_B=33$)}: $b_\Upsilon=0.08462$, $f_{star}=0.1969$, $b_b=3.0307$, $f_{\mathrm{DM}}=0.5757$, $f_{\mathrm{gas}}=0.2272$.\\
    Galaxy Type : Mixed, significant stars and gas (e.g.,NGC 3379~\citep{2003Sci...301.1696R}).
     \item \textbf{Set 3 ($\Upsilon_B=100$)}: $b_\Upsilon=0.25641$, $f_{star}=0.065$, $b_b=1.0$, $f_{\mathrm{DM}}=0.860$, $f_{\mathrm{gas}}=0.075$.\\
    Galaxy Type : Intermediate, still DM-dominated (e.g.,NGC 4494~\citep{2003Sci...301.1696R}).
    \item \textbf{Set 4 ($\Upsilon_B=390$)}: $b_\Upsilon=1.0$, $f_{star}=0.0167$, $b_b=1.0$, $f_{\mathrm{DM}}=0.860$, $f_{\mathrm{gas}}=0.123$.\\
    Galaxy Type : Dark-matter-dominated, faint stellar content (e.g.,cluster-central ellipticals such as M87 (NGC 4486) in Virgo ~\citep{1990AJ.....99.1823M}).
\end{itemize}

These sets clearly demonstrate the role of $\Upsilon_B$: low values correspond to luminous, baryon-rich galaxies with strong stellar and gaseous components, while high values describe dark-matter-dominated systems with faint stellar populations. For more details regarding the galactic parameters and other significant scaling parameters, please see ~\cite{2005MNRAS.362...95M,2005MNRAS.363..705M,2018MNRAS.479.3011R}.

\begin{figure}
\centering

\begin{minipage}{0.48\linewidth}
\centering
\includegraphics[width=1.1\linewidth,height=4cm]{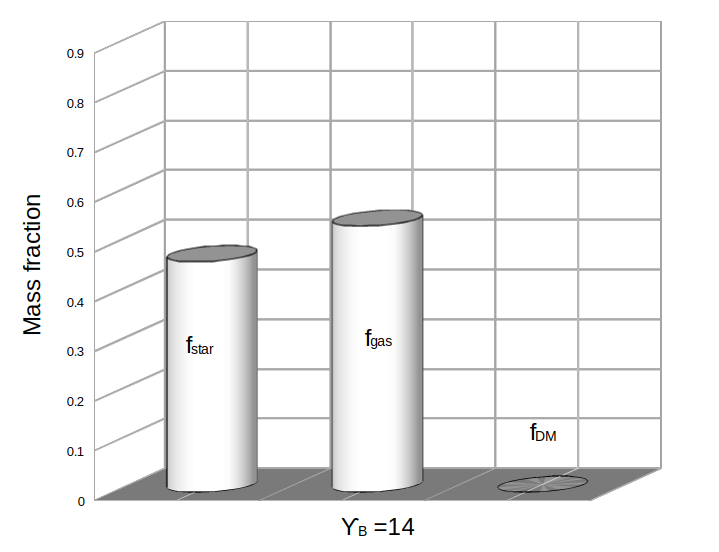}
\small (a)
\label{para141}
\end{minipage}
\hfill
\begin{minipage}{0.48\linewidth}
\centering
\includegraphics[width=1.1\linewidth,height=4cm]{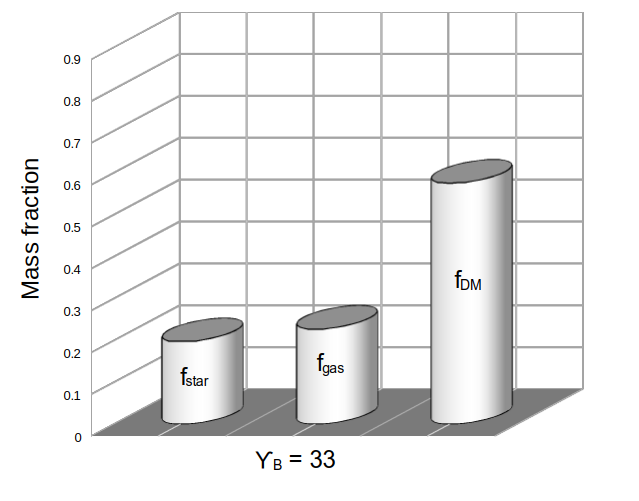}
\small (b)
\label{para331}
\end{minipage}
\vspace{2ex}
\begin{minipage}{0.48\linewidth}
\centering
\includegraphics[width=1.1\linewidth,height=4cm]{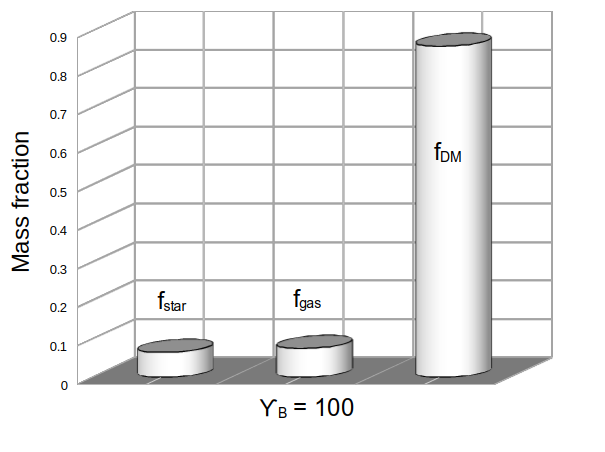}
\small (c)
\label{para1001}
\end{minipage}
\hfill
\begin{minipage}{0.48\linewidth}
\centering
\includegraphics[width=1.1\linewidth,height=4cm]{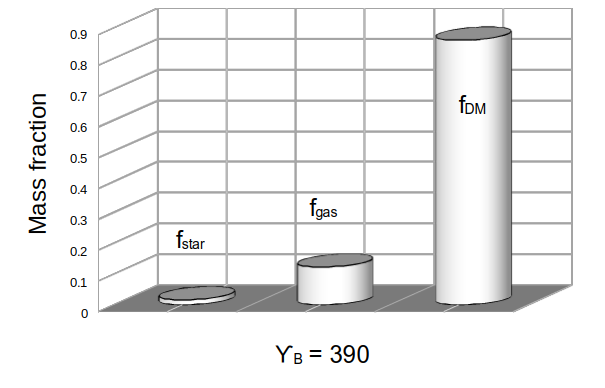}
\small (d)
\label{para3901}
\end{minipage}
\caption{In these figures, we represent four different galactic configurations with the relative contribution of different parameters, as given in subsection \S\ref{galpotd}. In Figure (a), we consider Set 1 where $\Upsilon_{B}=14$. As can be seen here, the maximum mass fraction was contributed by the hot gas mass fraction  ($f_{gas}$), followed by the stellar mass fraction within the virial radius($f_{star}$). The dark matter contribution, denoted by the term $f_{DM}$, is zero in this scenario. In figure (b), Set 2, where $\Upsilon_{B}=33$, the contribution to the mass fraction from the dark matter distribution $f_{DM}$ leads here and followed by  $f_{gas}$ and $f_{star}$. The third and fourth figures shown in figures (c) \& (d) respectively represent the cases of Set 3 and 4 when $\Upsilon_{B}=100$ and $\Upsilon_{B}=390$. We can see that in both these cases, $f_{DM}$ dominates the total mass fraction compared to the other terms present here.}
\label{ellpgal}
\end{figure}

For the convenience of the reader, we also present in Fig.~\ref{ellpgal} a graphical illustration of the mass distributions corresponding to the four selected models: in Fig.~\ref{ellpgal}(a) the hot gas dominates, in Fig.~\ref{ellpgal}(b) stars and gas are comparable, while in Fig.~\ref{ellpgal}(c) and Fig.~\ref{ellpgal}(d) the dark matter contribution dominates. In the following analysis, we employ these four cases to explore how variations in $\Upsilon_B$ and the associated mass decomposition impact the galactic potential and the dynamics of accretion flows.

In the next section, we describe the axisymmetric flow under the impact of the galactic potential  $\Phi_{\mathrm{Gal}}$ for adiabatic and isothermal thermodynamic equations of state.

\subsection{Hydrodynamic Equations : Stationary solutions and fixed points  }
\label{flowD}
\subsubsection{Adiabatic flows}
We investigate stationary solutions due to the analytical challenges posed by the time-dependent Euler equation and continuity equation. For a steady axisymmetric inviscid flow, in the framework of a  Kerr black hole, we can write the Euler equation and the mass conservation (continuity) equation as ~\citep{2002apa..book.....F},
\begin{equation}
v \frac{\mathrm{d} v}{\mathrm{d} r} + \frac{1}{\rho} \frac{\mathrm{d} P}{\mathrm{d} r} + \Phi'_{\mathrm{Gal}}(r) - \frac{\lambda^{2}}{r^{3}} = 0,
\label{euler2}
\end{equation}

\begin{equation}
\frac{\mathrm{d}}{\mathrm{d} r} \left( \rho v r H \right) = 0,
\label{cont2}
 \end{equation}

where \(\rho\) is the mass density, \(v\) is the radial velocity. \(\lambda\) is the conserved specific angular momentum, and \(\Phi'_{\mathrm{Gal}}(r)=\frac{d\Phi_{\mathrm{Gal}}(r)}{dr}\) is the spatial derivative with respect to radial distance $r$ and obtained from equation \eqref{eqgalphi}. $H$ denotes the height of the accretion disc. The pressure \(P(\rho)\) is related to the mass density by the equation of state:
\begin{equation}
P = K \rho^\gamma,
\label{adia1}
\end{equation}
where  \( \gamma \) is the adiabatic index, and \( K \) is a constant related to the entropy of the flow ~\citep{1939isss.book.....C}. Using this equation of state, the specific energy equation, also called the Bernoulli equation, can be obtained from the first integral of equation \eqref{euler2},
\begin{equation}
\mathcal{E} = \frac{v^{2}}{2} + \frac{c_{s}^{2}}{\gamma - 1} + \frac{\lambda^{2}}{2r^{2}} + \Phi_{\mathrm{Gal}},
\label{ber}
\end{equation}
where \(c_s = \sqrt{\frac{\partial P}{\partial \rho}}\) is the local adiabatic sound speed. Another constant quantity obtained from the first integral of the continuity equation and using the relation of $P$ and $\rho$ is known as the entropy accretion rate \(\dot{\mathcal{M}}\) ~\citep{1990ttaf.book.....C} that measures the inward flux of entropy per unit time, given by,
\begin{equation}
\dot{\mathcal{M}}^{2} = 4 \pi^{2} c_{s}^{4n} v^{2} r^{2} H^{2} \label{ent},
\end{equation}
where $\dot{\mathcal{M}} = (\gamma K)^{n} \dot{m}$, $n = (\gamma - 1)^{-1}$.
\(\dot{m}\) is the mass accretion rate.
To locate the critical points of the flow, we differentiate  Eqs. \eqref{ber} and \eqref{ent} and then add them to obtain,
\begin{equation}
 \frac{\mathrm{d}v^{2}}{\mathrm{d}r}
= \frac{2v^{2}}{r\left(v^{2} - c_{\mathrm{s}}^{2}\right)} \left[
\frac{\lambda^{2}}{r^{2}} - r \Phi'_{\mathrm{Gal}} + c_{\mathrm{s}}^{2} \left(1 + r \frac{H'}{H} \right)
\right]
\label{cri1},
\end{equation}
where \(H'=\frac{dH}{dr}\). The critical points of the flow are determined by using the condition that \(\dfrac{\mathrm{d}v^{2}}{\mathrm{d}r} = \dfrac{0}{0}\). Explicitly expressing and rearranging the terms, we obtain the two critical point conditions as:
\begin{equation}
v_{\mathrm{c}}^{2}=c_{\mathrm{sc}}^{2}=\frac{\left[r_{\mathrm{c}} \Phi'_{\mathrm{Gal}}\left(r_{\mathrm{c}}\right)-\frac{\lambda^{2}}{r_{\mathrm{c}}^{2}}\right]}{\left[1+r_{\mathrm{c}} \frac{H^{\prime}\left(r_{\mathrm{c}}\right)}{H\left(r_{\mathrm{c}}\right)}\right]}.
\label{vc}
\end{equation}
For a given set of fixed values of \([\mathcal{E}, \lambda, \gamma]\), the location of the critical point for a specific flow model can be determined by substituting the relevant critical point condition (as described in Eqs. (\ref{vc}) into the energy first integral  (\ref{ber}) for the galactic potential,
\begin{equation}
\mathcal{E} = \frac{1}{2}\left(\frac{\gamma+1}{\gamma-1}\right)\frac{\left[r_{\mathrm{c}} \Phi'_{\mathrm{Gal}}\left(r_{\mathrm{c}}\right)-\frac{\lambda^{2}}{r_{\mathrm{c}}^{2}}\right]}{\left[1+r_{\mathrm{c}} \frac{H^{\prime}\left(r_{\mathrm{c}}\right)}{H\left(r_{\mathrm{c}}\right)}\right]}+\Phi_{\mathrm{Gal}}\left(r_{\mathrm{c}}\right)+\frac{\lambda^{2}}{2 r_{\mathrm{c}}^{2}}.  \label{E-space}
\end{equation}
From this expression, it is evident that the solutions for \(r_c\) can also be obtained as a function of \(\lambda\) and \(\mathcal{E}\), i.e., \(r_c = f_1(\lambda, \mathcal{E})\). After substitution, the energy first integral transforms into an algebraic equation in terms of \(r_c\). Alternatively, \(r_c\) can be expressed in terms of \(\lambda\) and \(\dot{\mathcal{M}}\). By making use of the critical point conditions in Eq.\eqref{ent}, one can write:
\begin{equation}
\dot{\mathcal{M}}^{2} = 4 \pi^{2} r_{\mathrm{c}}^{2} H^{2}\left(r_{\mathrm{c}}\right)\left\{\frac{\left[r_{\mathrm{c}} \Phi'_{\mathrm{Gal}}\left(r_{\mathrm{c}}\right)-\frac{\lambda^{2}}{r_{\mathrm{c}}^{2}}\right]}{\left[1+r_{\mathrm{c}} \frac{H^{\prime}\left(r_{\mathrm{c}}\right)}{H\left(r_{\mathrm{c}}\right)}\right]}\right\}^{2 n+1} \label{mmen},
\end{equation}
with the critical radius as a function \(r_c = f_2(\lambda, \dot{\mathcal{M}})\), with the obvious implication that the location of the critical point depends explicitly on both angular momentum and entropy accretion rate.

In astrophysics, especially when studying accretion discs around compact objects or active galactic nuclei, the disc height denotes the vertical scale or geometrical thickness of the disc as a function of radial distance from the central object. This height can be described in different ways, depending on the physical assumptions and the dynamical regime of the accretion flow. The three flow geometries considered in this work and their corresponding equations and critical point conditions are summarized in Appendix \ref{Vcric}.

\subsubsection{ Isothermal flows}
\label{isoequations}

An isothermal flow refers to a fluid flow in which the temperature remains constant, and the pressure $P$ and density $\rho$ are related by :
\begin{equation}
 P=\frac{\rho k_B T}{\mu m_{H}}
\end{equation}
where, $k_B$, $\mu$, and $m_H$ denotes the Boltzmann’s constant, reduced mass \& the mass of a hydrogen atom respectively. $T$ is the constant temperature. For isothermal flow, the integral solution of the time-independent Euler equation provides the following first integral of motion
\begin{equation}
\frac{v^2}{2} + c_{\mathrm s}^2 \ln \rho
+ \frac{\lambda^2}{2 r^2} + \Phi_{\mathrm{Gal}} (r) = {\rm C}
\label{iso1}
\end{equation}
Obviously, this $C=$ constant of motion can not be identified with the
specific energy of the flow. The mass accretion rate, another first integral of motion of the accreting system of the aforementioned kind, may be obtained for three different flow geometries as
Mass accretion rates for three models

\begin{eqnarray}
\dot{\mathcal{M}}_{\rm CH} &= \rho v r \\
\dot{\mathcal{M}}_{\rm CO} &= D \rho v r^2 \\
\dot{\mathcal{M}}_{\rm VE} &= c_s \rho v r^{3/2} \left( \Phi'_{\mathrm{Gal}} \right)^{-1/2}
\end{eqnarray}
The respective derivatives of the flow variables at the sonic point for isothermal cases are given in Appendix \ref{Vcriciso}.
More detailed Discussions and derivations of all the above-mentioned equations can be found in ~\cite{2006MNRAS.373..146C,Bilić2014,2016NewA...43...10S}.

To study flow trajectories or phase portraits, we performed numerical integration starting from the critical values of flow variables and its derivative at saddles (where transonic flow trajectories intersect one another). This procedure yields a quantitative (numerical) picture of possible flow structures, either continuous or shock-forming depending on boundary conditions and flow parameters (see Figures \ref{PP1} and \ref{Piso}). Still, drawing these topologies alone cannot capture the underlying qualitative (descriptive) nature of the flow. Numerical integration will always generate a trajectory, but it does not determine whether the corresponding critical point is physically realizable. For this reason, a dynamical systems analysis is given in \S \ref{fixednature}  under the influence of a multi-component galactic potential to uncover the true qualitative nature of the flow.

\subsection{Nature and Classification of Fixed Points: A Dynamical Systems Analysis}
\label{fixednature}
The governing fluid equations are first-order nonlinear differential equations. Due to their complexity, an exact analytical solution is generally not feasible. In most practical scenarios, one must rely on numerical integration techniques to explore the characteristics of such flow solutions. An alternative strategy involves reformulating these equations into a standard first-order autonomous dynamical system. This method is widely adopted in fluid mechanics and has proven particularly useful in studies of accretion processes ~\citep{jordan1999nonlinear,strogatz1994nonlinear,1993JFM...254..635B,2005astro.ph.11018R,2003MNRAS.344...83R,2003ApJ...592..354A,2006MNRAS.373..146C,2012NewA...17..285N,2003MNRAS.344...83R,2002PhRvE..66f6303R}. To implement this, one begins by expressing the stationary polytropic flow equations (see equation \eqref{cri1}) in a parametrised form suitable for constructing a coupled dynamical system of first-order differential equations,

\begin{eqnarray}
\frac{\mathrm{d}v^{2}}{\mathrm{d} \tau} = 2v^{2} \left[ \frac{\lambda^{2}}{r^{2}} - r \Phi'_{\mathrm{Gal}} + c_{\mathrm{s}}^{2} \left(1 + r \frac{H'}{H} \right) \right] \nonumber\\
\frac{\mathrm{d}r}{\mathrm{d} \tau} = r \left(v^{2} - c_{\mathrm{s}}^{2} \right),
\end{eqnarray}
where $\tau$ is an arbitrary mathematical parameter. It is important to note that in both cases, the parameter \( \tau \) does not explicitly appear on the right-hand side. Expanding about the fixed-point values, a perturbation scheme of the kind,
\begin{equation}
v^{2} = v_{\mathrm{c}}^{2} + \delta v^{2}, \quad
c_{\mathrm{s}}^{2} = c_{\mathrm{sc}}^{2} + \delta c_{\mathrm{s}}^{2}, \quad
r = r_{\mathrm{c}} + \delta r.
\
\end{equation}
one could derive a set of two autonomous first-order linear differential equations
in the $\delta r$-$\delta v^2$  plane, with $\delta c_{\mathrm{s}}^{2} $
 having to be first expressed in terms of $\delta r$ and $\delta v^2$,
\begin{equation}
\delta c_{\mathrm{s}}^{2} = -c_{\mathrm{sc}}^{2} (\gamma - 1) \left\{
\frac{\delta v^{2}}{2 v_{\mathrm{c}}^{2}} +
\left[1 + r_{\mathrm{c}} \frac{H^{\prime}\left(r_{\mathrm{c}}\right)}{H\left(r_{\mathrm{c}}\right)}\right]
\frac{\delta r}{r_{\mathrm{c}}}
\right\}.
\end{equation}
The resulting coupled set of linear equations in $\delta r$ and $\delta v^2$ will follow simply as,

\begin{multline}
 \frac{\mathrm{d}}{\mathrm{d} \tau}\left(\delta v^{2}\right) =
 -(\gamma - 1) \left[ 1 + r_{\mathrm{c}} \frac{H^{\prime}(r_{\mathrm{c}})}{H(r_{\mathrm{c}})} \right] c_{\mathrm{sc}}^{2} \delta v^{2}  - \\ 2 c_{\mathrm{sc}}^{2} \, \Big[
 \frac{2 \lambda^{2}}{r_{\mathrm{c}}^{3}}
 + \Phi'_{\mathrm{Gal}}(r_{\mathrm{c}})
 + r_{\mathrm{c}} \Phi''_{\mathrm{Gal}}(r_{\mathrm{c}}) \nonumber \\
 + (\gamma - 1) \left\{ 1 + r_{\mathrm{c}} \frac{H^{\prime}(r_{\mathrm{c}})}{H(r_{\mathrm{c}})} \right\}^{2} \frac{c_{\mathrm{sc}}^{2}}{r_{\mathrm{c}}} \\
 - c_{\mathrm{sc}}^{2} \left\{ \left( \ln H(r_{\mathrm{c}}) \right)^{\prime} + r_{\mathrm{c}} \left( \ln H(r_{\mathrm{c}}) \right)^{\prime \prime} \right\}
 \Big] \delta r,
\end{multline}

\begin{equation}
\frac{\mathrm{d}}{\mathrm{d} \tau}(\delta r) = \left(\frac{\gamma+1}{2}\right) r_{\mathrm{c}} \delta v^{2} + (\gamma-1) \left[1 + r_{\mathrm{c}} \frac{H^{\prime}\left(r_{\mathrm{c}}\right)}{H\left(r_{\mathrm{c}}\right)}\right] \delta r.
\end{equation}
We consider type of solutions : $\delta v^2 \sim \exp(\Omega \tau)$ and $\delta r \sim \exp(\Omega \tau)$ in the above equations ~\citep{2012NewA...17..285N}, and calculate the eigenvalues $\Omega$ that provides the growth rates of $\delta v^2$ and $\delta r$ — as
\begin{multline}
\Omega^2_{\rm CH} = (\gamma-1)^{2} c_{\mathrm{sc}}^{4} - \\ (\gamma+1) r_{\mathrm{c}} c_{\mathrm{sc}}^{2} \left[ \frac{2 \lambda^{2}}{r_{\mathrm{c}}^{3}} + \Phi'_{\mathrm{Gal}}\left(r_{\mathrm{c}}\right) \right.  + \left. r_{\mathrm{c}} \Phi''_{\mathrm{Gal}}\left(r_{\mathrm{c}}\right) + (\gamma-1) \frac{c_{\mathrm{sc}}^{2}}{r_{\mathrm{c}}} \right] \label{OCH},
\end{multline}

\begin{multline}
\Omega^2_{\rm CO} = 4 (\gamma-1)^{2} c_{\mathrm{sc}}^{4} -\\ (\gamma+1) r_{\mathrm{c}} c_{\mathrm{sc}}^{2} \left[ \frac{2 \lambda^{2}}{r_{\mathrm{c}}^{3}} + \Phi'_{\mathrm{Gal}}\left(r_{\mathrm{c}}\right)\right.  \left. + r_{\mathrm{c}} \Phi''_{\mathrm{Gal}}\left(r_{\mathrm{c}}\right) + 4 (\gamma-1) \frac{c_{\mathrm{sc}}^{2}}{r_{\mathrm{c}}} \right] \label{OCO},
\end{multline}

\begin{multline}
\Omega^2_{\rm VE} = \frac{4 r_{\mathrm{c}} \Phi'_{\mathrm{Gal}}\left(r_{\mathrm{c}}\right) c_{\mathrm{sc}}^{2}}{(\gamma+1)^{2}} \times \\
\Bigg\{
\left[(\gamma-1) \mathcal{A} - 2 \gamma (4 + \mathcal{A})
+ 2 \gamma \mathcal{B} \left(1 + \frac{3}{\mathcal{A}}\right) \right] \\
- \frac{\lambda^{2}}{\lambda_{\mathrm{K}}^{2}\left(r_{\mathrm{c}}\right)}
\left[ 4 \gamma + (\gamma-1) \mathcal{A}
+ 2 \gamma \mathcal{B} \left(1 + \frac{3}{\mathcal{A}}\right) \right]
\Bigg\}
\label{OVE1},
\end{multline}
where \( \lambda_{\mathrm{K}}^{2}(r) = r^{3} \Phi'_{\mathrm{Gal}}(r) \), and with:
\[
\mathcal{A} = r_{\mathrm{c}} \frac{\Phi''_{\mathrm{Gal}}\left(r_{\mathrm{c}}\right)}{\Phi'_{\mathrm{Gal}}\left(r_{\mathrm{c}}\right)} - 3, \quad \mathcal{B} = 1 + r_{\mathrm{c}} \frac{\Phi'''_{\mathrm{Gal}}\left(r_{\mathrm{c}}\right)}{\Phi''_{\mathrm{Gal}}\left(r_{\mathrm{c}}\right)} - r_{\mathrm{c}} \frac{\Phi''_{\mathrm{Gal}}\left(r_{\mathrm{c}}\right)}{\Phi'_{\mathrm{Gal}}\left(r_{\mathrm{c}}\right)}.
\]
The nature of the possible critical points can also be predicted from the form of $\Omega^2$. If $\Omega^2 > 0$, then the critical point will be a saddle point. On the other hand, if $\Omega^2 < 0$, then it will be a centre-type point, with $\Omega^2$ having real values in both cases. For isothermal flows, we obtain the eigenvalues $\Omega^2$ just by putting $\gamma = 1$. Saddles are chosen as initial points for numerical integration to find flow profiles.

After obtaining the stationary solutions and identifying the sonic points (see \S \ref{flowD}), it becomes natural to investigate the possibility of Rankine-Hugoniot type shock formation. While stationary solutions describe the overall structure of the flow, the location of sonic points reveals whether smooth transonic transitions are feasible. In cases where multiple sonic points are present, the flow may undergo a discontinuous yet physically admissible transition—namely, a shock—that connects distinct transonic branches. In the Appendix \ref{shok-F} we explore this possibility in detail.

It is necessary to assess the stability of stationary solutions (\S \ref{flowD}) once they have been obtained. The flow may be disturbed by even small disturbances. We investigate whether the transonic solutions and shocks are still dynamically sustainable by varying the time-dependent continuity and Euler equations. This ensures that the global solutions are physically achievable in astrophysical accretion processes.

In the next section, we focus on the stability analysis of these stationary solutions.

\subsection{Time-Dependent Stability Analysis of Stationary Solutions}
\label{stable-F}

We consider a galactic potential and want to obtain the stationary flow behaviour in an AGN. To ensure the stability of these stationary solutions, one must check the validity by considering the time dependent flow dynamics ~\citep{2009MNRAS.398..841B,2006MNRAS.373..146C,2007MNRAS.378.1407G,1980MNRAS.191..571P,1992ApJ...384..587T}. In this regard the governing equations i.e. the time dependent Euler equation and continuity equation can be written as,

\begin{equation}
\frac{\partial v}{\partial t} + v \frac{\mathrm{d} v}{\mathrm{d} r} + \frac{1}{\rho} \frac{\mathrm{d} P}{\mathrm{d} r} + \Phi'_{\mathrm{Gal}}(r) - \frac{\lambda^{2}}{r^{3}} = 0,
\label{euler11}
\end{equation}

\begin{equation}
\frac{\partial \rho}{\partial t} + \frac{1}{r H} \frac{\partial}{\partial r}
\left( \rho v r H \right) = 0 \,.
\label{cont111}
\end{equation}
We introduced here a time dependent perturbative analysis to check the time viability of the solutions and expressed the perturbed quantities as, $v(r,t) = v_0(r) + \tilde{v}(r,t)$ and
$\rho (r,t) = \rho_0 (r) + \tilde{\rho}(r,t)$. From equ \eqref{cont111}, we introduce the variable \( f = \rho v r H \) whose stationary value is a constant \( f_0 \) also known as matter flux rate, yields us a relation that connects all the three
fluctuating quantities, $\tilde{v}$,  $\tilde{\rho}$
and $\tilde{f}$, with one another and given by,
\begin{equation}
\label{effprimeh}
\frac{\tilde{f}}{f_0} = \frac{\tilde{\rho}}{\rho_0}
+ \frac{\tilde{v}}{v_0},
\end{equation}
where $\sim$ symbol denotes the small time-dependent perturbations and quantities with subscript 0 denotes stationary background flow variables. Using equation \eqref{cont111} \& \eqref{effprimeh} we get the relation for CH and CO model,
\begin{equation}
\label{flucdenh}
\frac{\partial \tilde{\rho}}{\partial t} +
\frac{v_0 \rho_0}{f_0}
\left(\frac{\partial \tilde{f}}{\partial r}\right)=0 \,.
\end{equation}

For VE model, since the disc height $H$ depends on both the density $\rho$ and the radial coordinate $r$, the above equation get modified to ~\citep{2006MNRAS.373..146C,2012NewA...17..285N},
\begin{equation}
\label{volden}
\frac{\partial}{\partial t} \left[\rho^{(\gamma +1)/2}\right]
+ \frac{\sqrt{\Phi^{\prime}_{\text{Gal}}}}{r^{3/2}}
\frac{\partial}{\partial r} \left[ \rho^{(\gamma +1)/2} \, v \,
\frac{r^{3/2}}{\sqrt{\Phi^{\prime}_{\text{Gal}}}} \right] = 0 \,,
\end{equation}
The $f$ get modified therefore for VE model as,
\begin{equation}
f=\frac{\rho^{(\gamma +1)/2} \, v \, r^{3/2}}{\sqrt{\Phi^{\prime}_{\text{Gal}}}} \,,
\end{equation}
and we can write,
\begin{equation}
\label{effprime}
\frac{\tilde{f}}{f_0} = \left( \frac{\gamma +1}{2} \right)
\frac{\tilde{\rho}}{\rho_0} + \frac{\tilde{v}}{v_0} \,.
\end{equation}
Therefore, equ \eqref{flucdenh}, now beocmes
\begin{equation}
\label{flucden}
\frac{\partial \tilde{\rho}}{\partial t} + \beta^2
\frac{v_0 \rho_0}{f_0} \left(\frac{\partial \tilde{f}}{\partial r}\right)=0 \,,
\end{equation}
where $\beta^2 = \frac{2}{\gamma +1}$. The velocity fluctuations can be written as
\begin{equation}
\label{flucvel}
\frac{\partial \tilde{v}}{\partial t}= \frac{v_0}{f_0}
\left(\frac{\partial \tilde{f}}{\partial t}+ v_0
\frac{\partial \tilde{f}}{\partial r}\right), \,
\end{equation}
which, upon a further partial differentiation with respect to time,
will give
\begin{equation}
\label{flucvelder2}
\frac{{\partial}^2 \tilde{v}}{\partial t^2}=\frac{\partial}{\partial t} \left[
\frac{v_0}{f_0} \left(\frac{\partial \tilde{f}}{\partial t}\right) \right]
+ \frac{\partial}{\partial t} \left[ \frac{v_0^2}{f_0} \left(
\frac{\partial \tilde{f}}{\partial r}\right) \right] .\,
\end{equation}
The linearised fluctuating part from the time-dependent Euler equation (\ref{euler11})
can be written as,
\begin{equation}
\label{fluceuler}
\frac{\partial \tilde{v}}{\partial t}+ \frac{\partial}{\partial r}
\left( v_0 \tilde{v} + c_{\mathrm{s0}}^2
\frac{\tilde{\rho}}{\rho_0}\right) =0, \,
\end{equation}
with $c_{\mathrm{s0}}$ being the speed of sound in the steady state. Differentiating Eq.~(\ref{fluceuler}) with respect to $t$, and depending on the chosen flow geometry using the respective equations (either Eq.\eqref{flucdenh} or Eq.\eqref{flucden}) with the velocity fluctuations and its derivative  we get,
\begin{multline}
\label{interm}
\frac{\partial}{\partial t} \left[\frac{v_0}{f_0}
\left( \frac{\partial \tilde{f}}{\partial t} \right) \right]
+ \frac{\partial}{\partial t} \left[\frac{v_0^2}{f_0}
\left( \frac{\partial \tilde{f}}{\partial r} \right) \right] \\
+ \frac{\partial}{\partial r} \left[\frac{v_0^2}{f_0}
\left( \frac{\partial \tilde{f}}{\partial t} \right) \right]
+ \frac{\partial}{\partial r} \left[\frac{v_0}{f_0}
\left(v_0^2 - \zeta c_{\mathrm{s0}}^2 \right)
\frac{\partial \tilde{f}}{\partial r} \right] = 0,
\end{multline}
where $\zeta =1$ for CH or CO and $\zeta = \beta^2$ for VE model. The Eq.~(\ref{interm}) can be written in a compact format as,
\begin{equation}
\label{compact}
\partial_\mu \left( {\mathrm{f}}^{\mu \nu} \partial_\nu
\tilde{f}\right) = 0, \,
\end{equation}
where $\mu,\nu =t,r$,
We can write the above equation using a symmetric matrix as,
\begin{equation}
\label{matrix}
\mathrm{f}^{\mu\nu} = \frac{v_0}{f_0}
\left(
  \begin{array}{cc}
    1 & v_0 \\[4pt]
    v_0 & v_0^2 - \zeta\, c_{\mathrm{s}0}^2
  \end{array}
\right).\,.
\end{equation}

In curved spacetime, the d'Alembertian (wave operator) acting on a scalar field \( \Phi \) is given by ~\citep{1998CQGra..15.1767V}
\begin{equation}
\square \Phi = \nabla^\mu \nabla_\mu \Phi
= \frac{1}{\sqrt{-g}} \, \partial_\mu \left( \sqrt{-g} \, g^{\mu\nu} \, \partial_\nu \Phi \right),
\label{Alembert}
\end{equation}
where \( g^{\mu\nu} \) is the inverse of the metric tensor \( g_{\mu\nu} \), and \( g = \det(g_{\mu\nu}) \).
By defining a tensor density
\[
f^{\mu\nu} = \sqrt{-g} \, g^{\mu\nu},
\]
this operator simplifies to
\begin{equation}
\square \Phi = \frac{1}{\sqrt{-g}} \, \partial_\mu \left( f^{\mu\nu} \, \partial_\nu \Phi \right).
\end{equation}

So the metric determinant can be expressed as \( g = \det(f^{\mu\nu}) \), directly linking the density form of the wave operator to the spacetime geometry. It is immediately possible to set down an effective metric for the propagation of an acoustic disturbance as

\begin{equation}
\label{metric}
\mathrm{g}^{\mu \nu}_{\mathrm{eff}} =
\left(
  \begin{array}{cc}
    1 & v_0 \\[4pt]
    v_0 & v_0^2 - \zeta\, c_{\mathrm{s}0}^2
  \end{array}
\right)\,.
\end{equation}

This give $v_0^2 = \zeta c_{\mathrm{s0}}^2$ as the horizon condition of an acoustic black hole for inflow solutions. Finally, a readjustment of terms in Eq.~(\ref{interm}) will
give the following perturbation equation
\begin{equation}
\label{eq:wave_eq}
\frac{\partial^2 \tilde{f}}{\partial t^2}
+ 2 \frac{\partial}{\partial r} \left( v_0 \frac{\partial \tilde{f}}{\partial t} \right)
+ \frac{1}{v_0} \frac{\partial}{\partial r} \left[ v_0 \left(v_0^2 - \zeta c_{\mathrm{s0}}^2 \right)
\frac{\partial \tilde{f}}{\partial r} \right] = 0 \,,
\end{equation}
whose detailed solution has been given in earlier works ~\citep{2003MNRAS.344.1085R,2006MNRAS.373..146C,2007MNRAS.378.1407G}.

Remarkably, the resulting wave equation closely resembles the d’Alembert equation \eqref{Alembert} for a massless scalar field, which in turn mirrors the Schwarzschild metric near a black hole’s event horizon.

In a classical, inviscid, inhomogeneous, transonic fluid, the propagation of acoustic perturbations naturally gives rise to an analogue event horizon at the transonic point. A collection of such points forms a sonic surface, which acts as a trapping boundary for outgoing phonons. Beyond this region, the flow becomes supersonic, and acoustic perturbations carried by the fluid can no longer propagate upstream or escape through the sonic surface, which resembles the black hole event horizon. Here, the gradient of the flow velocity at the sonic surface plays a role analogous to the surface gravity of a black hole, governing the strength of the acoustic horizon.

 ~\cite{1981PhRvL..46.1351U} introduced the concept of acoustic geometry in supersonic fluids and showed that an analogue surface gravity can be associated with such horizons. A remarkable implication of this analogy is that the acoustic horizon is expected to emit Hawking-like radiation in the form of thermal phonons, directly paralleling Hawking radiation.

In the next section, we focus on the stability analysis of acoustic surface gravity.

\subsection{ACOUSTIC SURFACE GRAVITY}
\label{Surgrav}
The acoustic surface gravity $\kappa$ for the stationary background fluid accreting under the influence of a post-Newtonian black hole potential can be obtained as ~\citep{1998CQGra..15.1767V,1999PThPS.136....1J,1981PhRvL..46.1351U,2014CQGra..31c5002B,2016NewA...43...10S}
\begin{multline}
 \kappa = \left| \sqrt{(1 + 2\Phi_{\mathrm{Gal}}(r))
\left( 1 - \frac{\lambda^{2}}{r^{2}}
- 2\Phi_{\mathrm{Gal}}(r) \frac{\lambda^{2}}{r^{2}} \right) } \right. \\ \left. \times  \left( \frac{1}{1 - c^{2}_{sc}}
\left[ \left. \frac{dv}{dr} \right|_{r_{c}}
- \left. \frac{dc_{s}}{dr} \right|_{r_{c}} \right] \right) \right|
\label{eq.501}.
\end{multline}
The surface gravity $\kappa \equiv \kappa[\gamma, \mathcal{E}, \lambda, \rm{a}]$. The bracketed terms indicate its functional dependence. We aim to investigate how the acoustic surface gravity depends on the black hole spin $\rm{a}$ and the galactic parameter. For this purpose, we calculate $\kappa$ for different values of $\rm{a}$ while varying the galactic parameter, keeping the other parameters $[\gamma, \mathcal{E}, \lambda]$ fixed.

This work explores the transonic nature and dynamical stability of accretion flows subjected to a composite galactic potential. We examine how the flow behaviour varies across three disc geometries and under two thermodynamic frameworks—adiabatic and isothermal. The conditions for shock formation are evaluated in both regimes, revealing several intriguing features. We also compute the corresponding surface gravity linked to the emergent acoustic structure. Notably, the inclusion of galactic-scale gravitational components leads to shifts in the positions of critical points and shock fronts when compared to models considering only a central black hole. Although these shifts are relatively modest in magnitude, they carry important physical consequences. All results and their implications are illustrated and discussed in detail through a series of plots in the following section.

\section{Results \& Discussions}
\label{Result}

In this section, we investigate the effects of the galactic potential on transonic accretion flows for different accretion disc geometries. In \S\ref{cpa}, we analyze the critical-point structure and the corresponding $\mathcal{E}$--$\lambda$ parameter space. We then compare the resulting phase portraits with those of the isolated black hole flow, while the nature of the critical points is discussed separately in \S\ref{cpb}.  In \S\ref{cpc} and \S\ref{cpd}, we investigate the shock-permitting parameter space and analyze the properties of the resulting shocks, including their locations, strengths, and associated flow variables. Finally, in \S\ref{cpe}, we study the behaviour of the acoustic surface gravity, $\kappa$, in the presence of the galactic potential.

\subsection{Critical point analysis and multi-critical parameter space}
\label{cpa}

\begin{figure*}
 \centerline{\includegraphics[width=1.0\linewidth,clip]{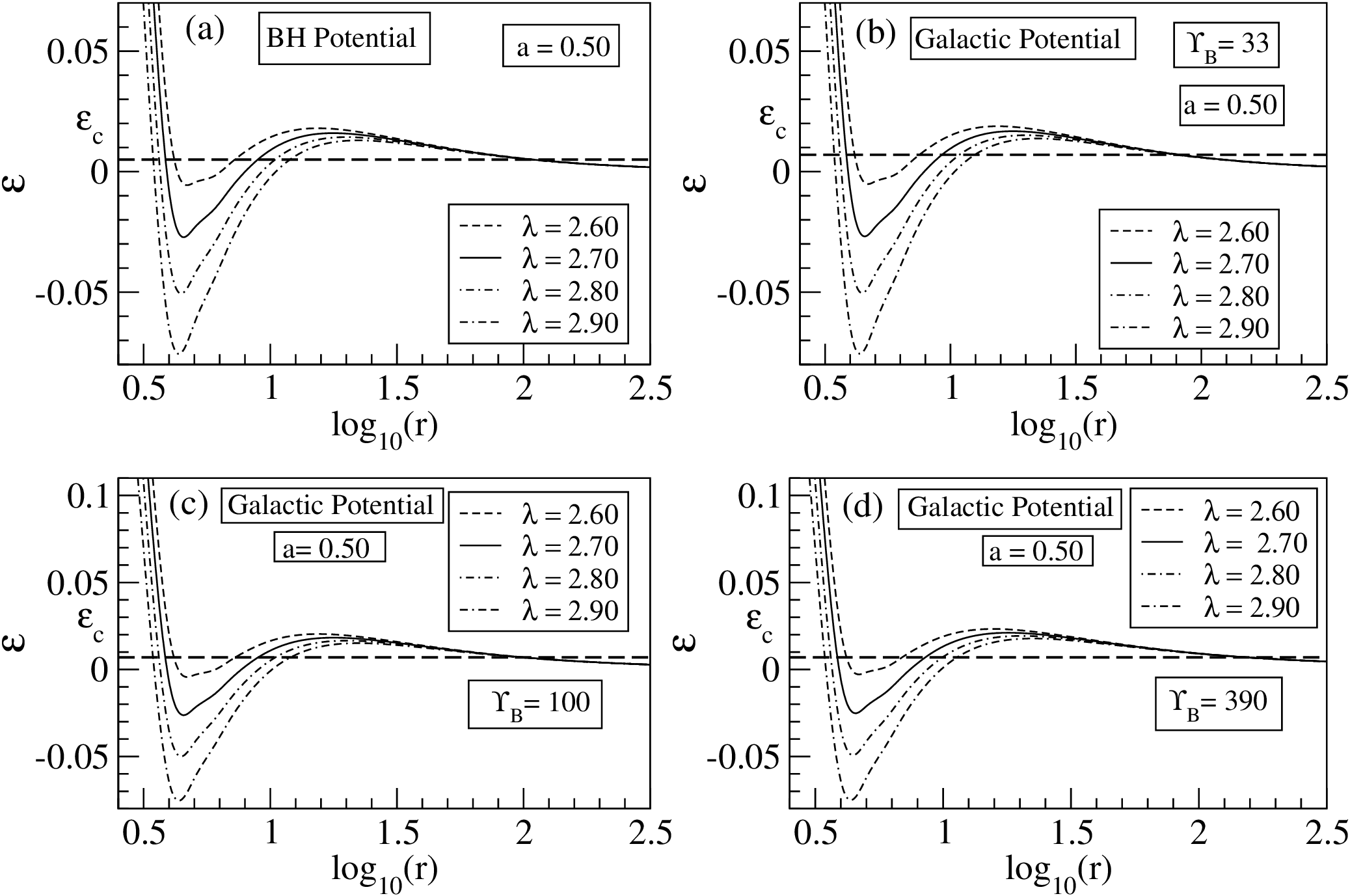}}
 \caption{ In this figure, we plot the variation of the specific energy ($\mathcal{E}$) as a function of radial distance ($r$) for the vertical equilibrium (VE) disc model. All panels are computed for a fixed black hole spin parameter $\rm{a} = 0.50$.  Panel (a) depicts the black hole (BH) potential, whereas panels (b), (c), and (d) correspond to galactic potentials with $\Upsilon_{B}=33$, $100$, and $390$, respectively. In each panel, the specific energy $\mathcal{E}$ is plotted as a function of radial distance $r$ for different values of the specific angular momentum $\lambda$, represented by distinct line styles (e.g., solid for $\lambda=2.70$, dashed for $\lambda=2.60$, and so on). A horizontal line at a fixed value of $\mathcal{E}$ denotes the sonic energy $\mathcal{E}_{c}$. The intersections of this line with the energy curves define the critical points $r_{c}$, and the total number of such intersections indicates the number of critical points, namely $r_{c1}$, $r_{c2}$, and $r_{c3}$, for the chosen energy and angular momentum. The comparative analysis among the black hole and the three galactic configurations is given in Table \ref{tab:critical-points-VE}.
}
\label{E-r}
\end{figure*}
We first investigate the transonic property of the accretion flow under the galactic potential and compare it with the BH potential in Fig ~\ref{E-r}.
For this study, we consider the VE disc model and an adiabatic equation of state. All panels (a--d) are computed for a fixed black hole spin parameter $\rm{a} = 0.50$.   Panel (a) corresponds to the black hole (BH) potential, while panels (b), (c), and (d) represent galactic potentials with $\Upsilon_B = 33$, $100$, and $390$, respectively. In each panel, we plot specific energy $\mathcal{E}$ \eqref{EVE} as a function of the radius $r$ for varying specific angular momentum $\lambda$, shown by different shades of curved lines, e.g., solid line for $\lambda=2.70$, dashed line for $\lambda= 2.60$ and so on. A horizontal line drawn at a fixed value of $\mathcal{E}$, representing the sonic energy:$\mathcal{E}_{c}$ intersects these curves at multiple locations which are known as critical points $r_c$, and the number of such intersections corresponds to the number of admissible critical points $r_{c1}$, $r_{c2}$ \& $r_{c3}$ respectively for the chosen energy and angular momentum. The comparative study from these four panels of Fig ~\ref{E-r} is given in
Table~\ref{tab:critical-points-VE}, which summarizes the numerical values of the critical point locations from the plot corresponding to the representative energy $\mathcal{E}_{c}=0.003$ and specific angular momentum $\lambda=2.60$. For each potential configuration, three critical points are obtained: an inner ($r_{c1}$), a middle ($r_{c2}$), and an outer ($r_{c3}$) critical point. We can see that the inner and middle critical point locations exhibit only marginal shifts; however, the outer critical point shows a pronounced outward displacement for different  $\Upsilon_B$. We can see from the figure as well as from the table that the inner critical point $r_{c1}$ increases only marginally from $4.36r_g$ for the black hole–only potential to $4.43r_g$ for $\Upsilon_B = 390$, whereas the middle critical point $r_{c2}$ shifts inward from $6.86r_g$ to $6.45r_g$. In contrast, the outer critical point $r_{c3}$ exhibits a substantial outward shift, increasing from $190.28r_g$ to $807.03r_g$. This behaviour indicates that the galactic environment has a greater impact at larger radii, while the near-horizon flow structure remains predominantly governed by the central black hole potential. An important outcome of this analysis is that the accretion flow continues to exhibit multitransonic behaviour even in the presence of a multi-component galactic potential. Although the deviations in the inner and middle critical point locations are relatively small, they are significant for the outer critical point, which indicates that galactic components can influence the positions of the critical points.

\begin{table}
  \centering
  \begin{tabular}{|l|c|c|c|c|c|}
    \hline

    \textbf{Potential} &
    $\boldsymbol{\mathcal{E}_c}$ &
    $\boldsymbol{\lambda}$ &
    $\boldsymbol{r_{c1}(r_g)}$ &
    $\boldsymbol{r_{c2}(r_g)}$ &
    $\boldsymbol{r_{c3}(r_g)}$ \\
    \hline
      BH                    & 0.003 & 2.60 & 4.36 & 6.86 & 190.28 \\
    \hline
      $\Upsilon_{B=33}$     & 0.003 & 2.60 & 4.38 & 6.75 & 217.85 \\
    \hline
      $\Upsilon_{B=100}$    & 0.003 & 2.60 & 4.41 & 6.57 & 290.68 \\
    \hline
      $\Upsilon_{B=390}$    & 0.003 & 2.60 & 4.43 & 6.45 & 807.03 \\
    \hline
  \end{tabular}
  \caption{ In this table we present a comparison of critical points location of the accretion flow for the BH potential and different galactic potential indicated by $\Upsilon_B$ for a selected sonic energy $\mathcal{E}_c=0.003$ and specific angular momentum $\lambda=2.60$, among the cases presented in the Fig ~\ref{E-r}. The quantities $r_{c1}$, $r_{c2}$, and $r_{c3}$ denote the inner, middle, and outer critical points, respectively.
  }
  \label{tab:critical-points-VE}
\end{table}

\begin{figure*}
 \centerline{\includegraphics[width=0.85\linewidth,clip]{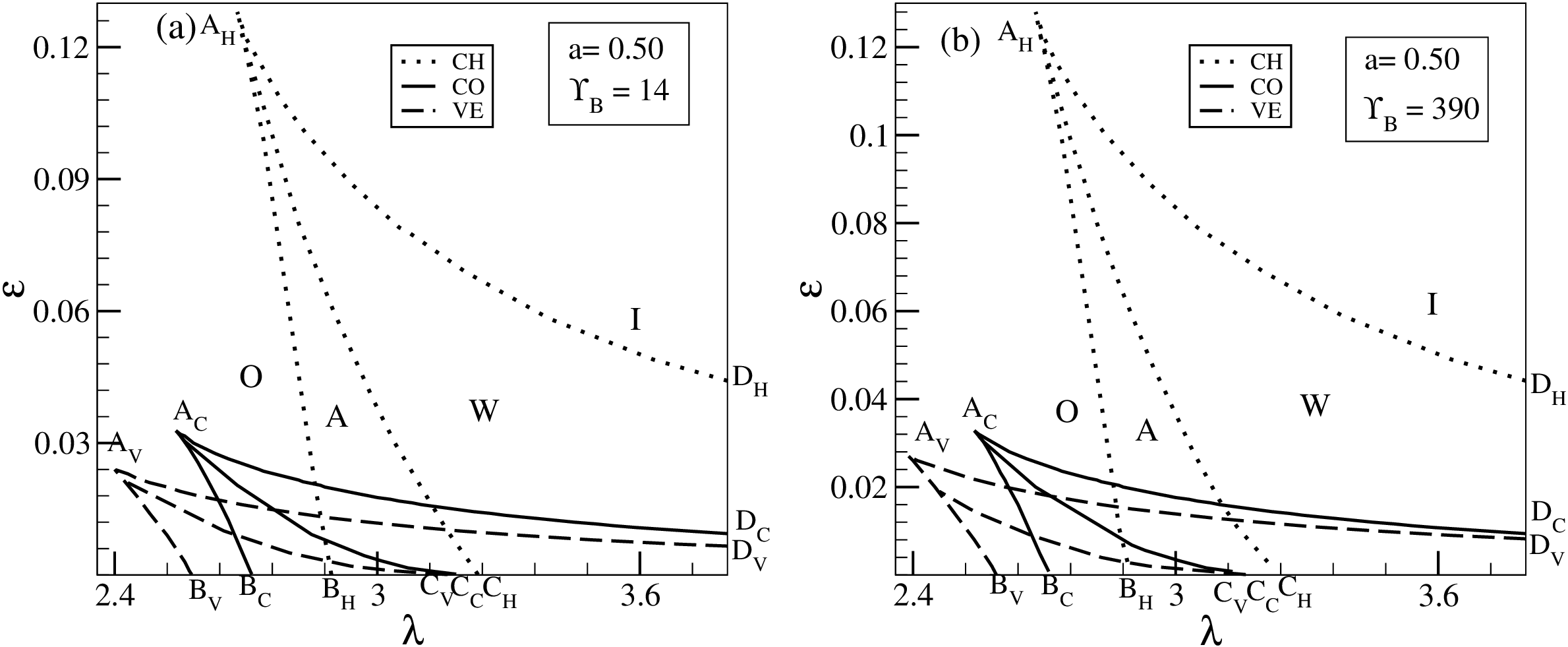}}
  \caption{ In these figures we investigated different regions in the parameter space of specific energy ($\mathcal{E}$) and specific angular momentum ($\lambda$), corresponding to the number and nature of critical points in adiabatic flows. Each panel is computed for a fixed black hole spin parameter $\rm{a} = 0.5$, and the corresponding parameter space is classified into four distinct regions, labeled as $\boldsymbol{\textbf{O}}$, $\boldsymbol{\textbf{I}}$, $\boldsymbol{\textbf{A}}$, and $\boldsymbol{\textbf{W}}$, based on the number and nature of the critical points. These diagram shows the classification of these regions based on three accretion disc geometries: dotted lines represent the CH model, solid lines correspond to CO model, and dashed lines indicate  VE model. For each geometry, the parameter space is shown for two chosen values of the galactic parameter: $\Upsilon_{B} = 14$, and $\Upsilon_{B} = 390$. The structure and topology of the critical point regions vary with both the geometry and the galactic background, illustrating how the transonic properties of the flow are influenced by the combined effects of disc geometry and galactic environment.
}
\label{PMSCHCOVE}
\end{figure*}
In Fig.~\ref{E-r}, we fix the parameters $f(\mathcal{E},\lambda)$ for the VE disc model and determine the corresponding critical points of the flow. To identify the complete parameter space (PMS) in the $\mathcal{E}$–$\lambda$ plane where the accretion flow exhibits multi-transonic behaviour, we analyse Eq.~\eqref{E-space} and map the admissible region, as shown in Fig.~\ref{PMSCHCOVE}. We further present a comparative study of the PMS for the CH model by dotted lines, CO model by solid lines, and VE model by dashed lines, assuming an adiabatic equation of state. The influence of the galactic background is modeled by choosing the potential scaling parameter $\Upsilon_{B} = 14$ \& $390$. Each panel corresponds to a black hole spin parameter $\rm{a} = 0.5$, and the parameter space is divided into four regions labeled as $\boldsymbol{\textbf{O}}$, $\boldsymbol{\textbf{I}}$, $\boldsymbol{\textbf{A}}$, and $\boldsymbol{\textbf{W}}$, depending on the number and type of critical points. The $\boldsymbol{\textbf{O}}$ region corresponds to flows possessing a single outer saddle-type critical point ($r_\text{out}$), located far from the black hole horizon. The $\boldsymbol{\textbf{I}}$ region denotes solutions with a single inner saddle-type critical point ($r_\text{in}$), closer to the horizon. In contrast, the $\boldsymbol{\textbf{A}}$ and $\boldsymbol{\textbf{W}}$ regions correspond to flows with three critical points: an inner and an outer saddle-type point ($r_\text{in}$ and $r_\text{out}$), separated by a center-type point ($r_\text{mid}$). While $r_\text{mid}$ exists mathematically, it does not support transonic flow and thus has limited physical relevance in global accretion solutions. The difference between $\boldsymbol{\textbf{A}}$ and $\boldsymbol{\textbf{W}}$ lies in the entropy distribution at the critical points. In the $\boldsymbol{\textbf{A}}$ region, the entropy at the inner critical point is greater than that at the outer one, i.e., $\mathcal{\dot{M}}(r_\text{in}) > \mathcal{\dot{M}}(r_\text{out})$, making shock formation in accretion-type flows possible. Conversely, in the $\boldsymbol{\textbf{W}}$ region, the entropy satisfies $\mathcal{\dot{M}}(r_\text{in}) < \mathcal{\dot{M}}(r_\text{out})$, favoring shock formation in wind-type flows. The dividing line between these two regions, at which entropy values at the two saddle points are equal, is marked by the central curve within the wedge-shaped multi-critical region in each panel. For the CH disc model, the multi-critical region is bounded by points $A_H$, $B_H$, and $D_H$, and contains two subregions. The subregion enclosed by $A_H$, $B_H$, and $C_H$ satisfies $\dot{\mathcal{M}}_\text{in} > \dot{\mathcal{M}}_\text{out}$, while the region enclosed by $A_H$, $C_H$, and $D_H$ shows the opposite entropy trend. Along the curve $A_H C_H$, the entropy at both critical points is identical. Similar wedge-shaped multi-critical zones appear for the CO and VE models, labeled by $A_C$–$D_C$ and $A_V$–$D_V$, respectively. Thus, for low $\mathcal{E}$, there exists a finite range in $\lambda$ that allows three critical points to form. This makes multi-transonic behavior possible under certain conditions, particularly when $\dot{\mathcal{M}}_\text{in} > \dot{\mathcal{M}}_\text{out}$ and the Rankine–Hugoniot shock conditions are satisfied. Outside the multi-critical wedges, only single critical points are found, limiting the flow to mono-transonic solutions. For the $\Upsilon_{B}=390$ value, the multi-critical regions in the $\mathcal{E}$–$\lambda$ space change; though very small changes are observed, this indicates the sensitivity of critical point structure to both flow geometry and galactic potential. We present the corresponding changes in the PMS for the two $\Upsilon_B$ cases in terms of variations in the wedge-shaped ABC region, which delineates the accretion domain for each flow geometry. These results are summarised in tabular form in Table \ref{tab:ABC_geometry1}. We can see from the table that, for CH and CO disc models, the shifts are either absent or negligible. In contrast, for the VE disc model, all three points exhibit shifts, which, although small, are noticeable.

\begin{table}
\centering
\renewcommand{\arraystretch}{1.4}
\setlength{\tabcolsep}{3pt}
\begin{tabular}{|l|c|c|c|c|c|c|}
\hline
Points & \multicolumn{6}{|c|}{Flow Geometry} \\
\cline{2-7}
& \multicolumn{2}{|c|}{CH}
& \multicolumn{2}{|c|}{CO}
& \multicolumn{2}{c|}{VE} \\
\cline{2-7}
 & $\Upsilon_{B=14}$ & $\Upsilon_{B=390}$
 & $\Upsilon_{B=14}$ & $\Upsilon_{B=390}$
 & $\Upsilon_{B=14}$ & $\Upsilon_{B=390}$ \\
\hline
A & 0.1279 & 0.1279 & 0.0320 & 0.0320 & 0.0235 & 0.0320 \\
B & 2.892  & 2.849  & 2.712  & 2.650  & 2.595  & 2.510  \\
C & 3.69   & 3.05   & 3.69   & 3.05   & 3.72   & 3.20  \\
\hline
\end{tabular}
\caption{This table quantifies the change in the size of the parameter space by numerically representing the shifts of the wedge-shaped regions connecting points A, B, and C. While the corresponding shifts in the PMS are illustrated graphically in Fig.~\ref{PMSCHCOVE}, here we present their numerical values for the accretion case. The variations of points A, B, and C in the PMS for different flow geometries—constant height (CH: $A_H, B_H, C_H$), conical height (CO: $A_C, B_C, C_C$), and vertical equilibrium (VE: $A_V, B_V, C_V$)—are listed for $\Upsilon_B = 14$ and $390$. }
\label{tab:ABC_geometry1}
\end{table}

\begin{figure*}
 \centerline{\includegraphics[width=0.8\linewidth,clip]{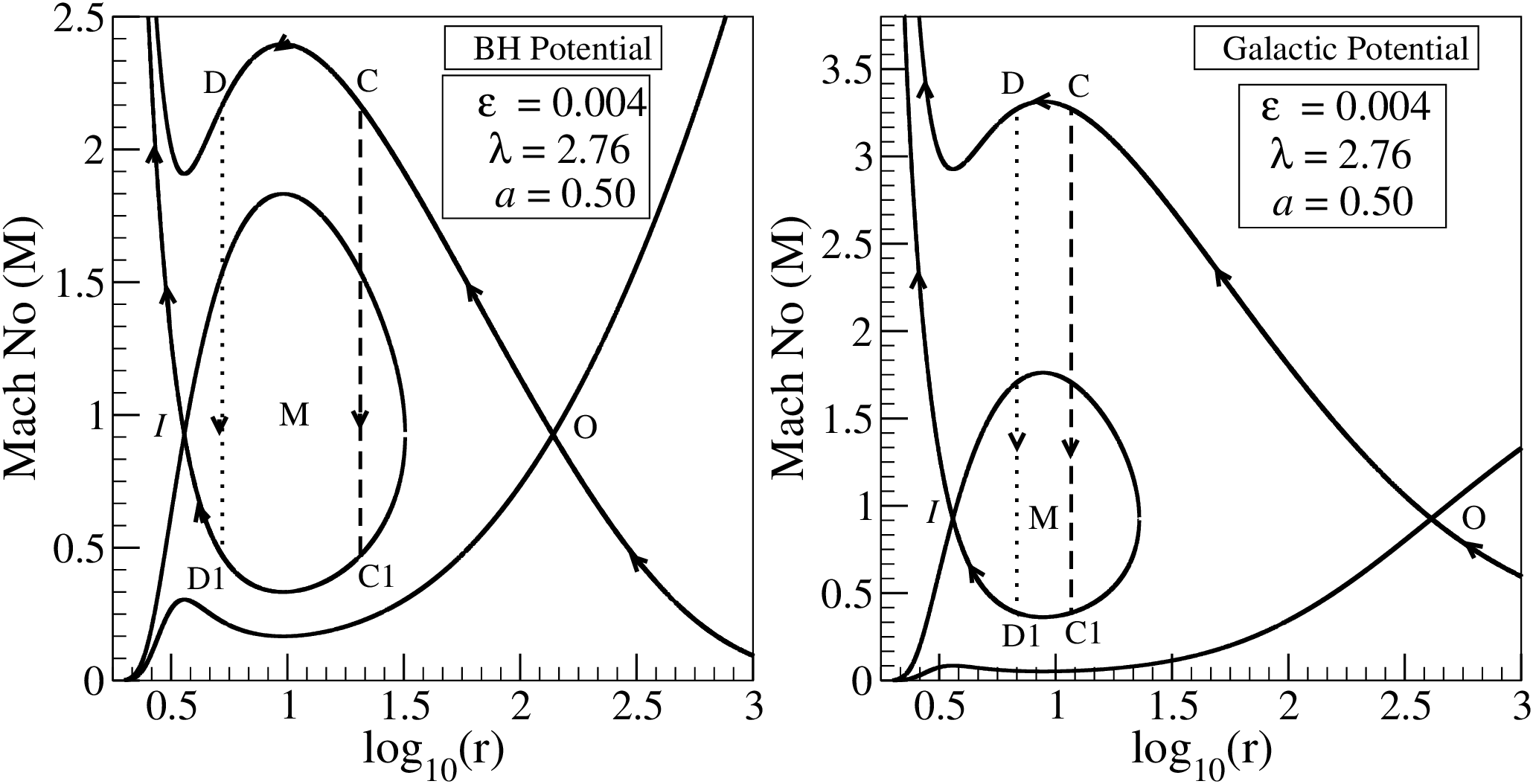}}
 \caption{ This figure presents a comparison of the phase portrait of adiabatic transonic accretion in the vertical equilibrium (VE) disc geometry for the Kerr BH potential and the framework of a galactic potential. The galactic parameter is set to $\Upsilon_{B} = 390$. In the figure, both inner and outer sonic points, as well as the corresponding inner and outer shock locations, are clearly marked. $CC1$ and $DD1$ represent the discontinuous shock transition occurring in the accretion flow. The phase space diagram is plotted for the parameter set: [$\mathcal{E}$, $\lambda$, $\rm{a}$] = [0.004, 2.76, 0.50]. The comparison of the critical point location and shock location is given in Table \ref{tab:comparison_quantities}.
}
\label{PP1}{}
\end{figure*}

\begin{figure}
 \centerline{\includegraphics[width=1.0\linewidth,clip]{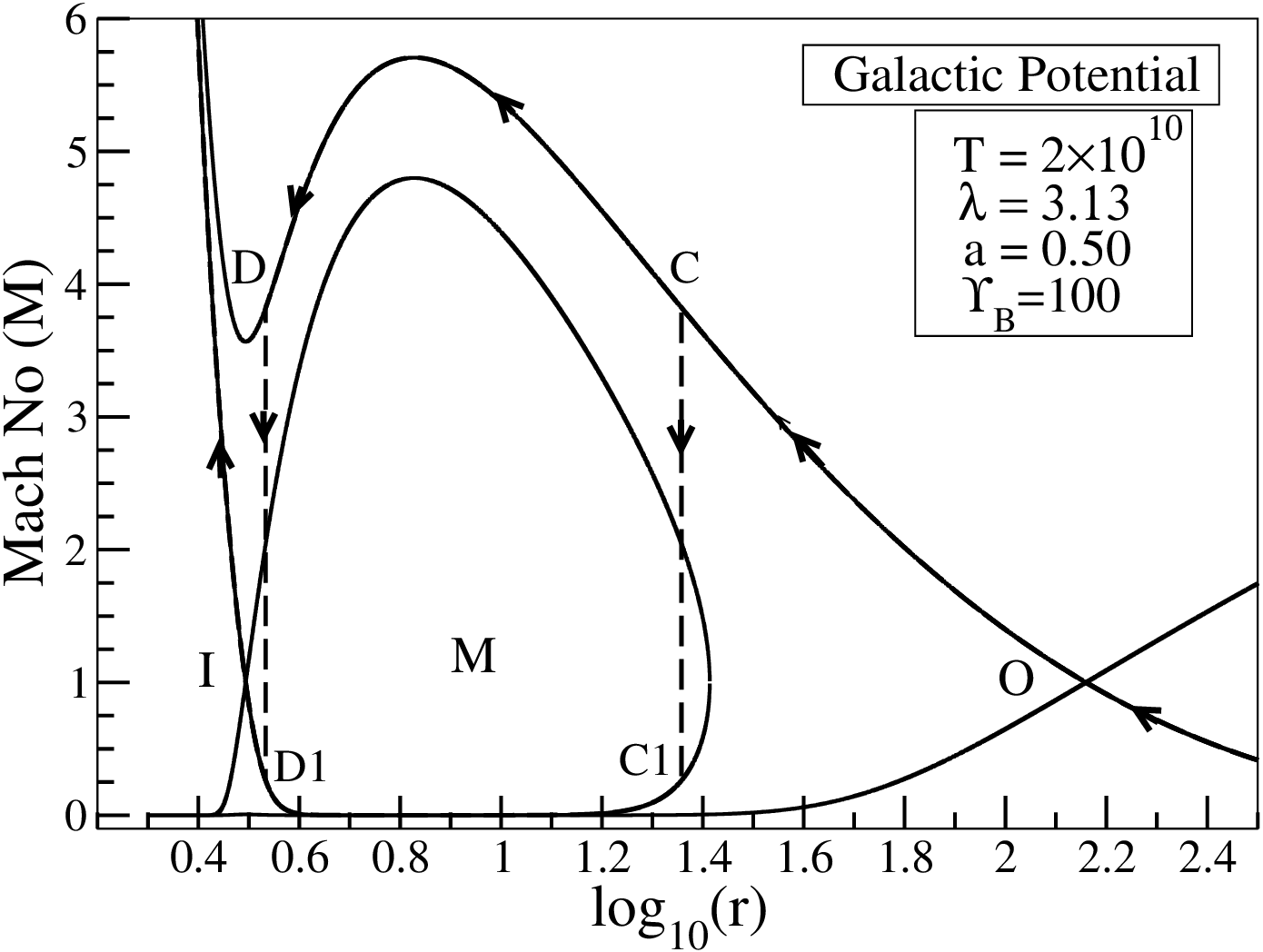}}
\caption{
A phase portrait of transonic accretion in the conical (CO) flow disc geometry, considering an isothermal equation of state under the influence of a galactic potential, is presented here. The figure shows a variation of Mach number ($\rm{M}$) with the radial distance ($r$) on a log scale. The three sonic points, inner, middle, and outer, are shown by the symbols $\mathbf{I}$, $\mathbf{M}$, and  $\mathbf{O}$. The galactic parameter is set to $\Upsilon_{B} = 100$. The other parameters are chosen as: [T, $\lambda$, a] = $[2\times10^{10}, 3.13, 0.50]$ where the symbols are defined in text. $CC1$ and $DD1$ represent the shock transition occurring in the accretion flow. Here $[I,M,O,CC1,DD1]=[3.11,\,6.72,\,148.47,\,24.05,\,3.46]$.
}
\label{Piso}
\end{figure}

We can now readily identify the regions corresponding to types I, O, A, and W, along with their associated $\mathcal{E}$–$\lambda$ domains. Next, to identify the nature of the solutions, we plot the phase portrait (Mach number, $M$, vs $\log_{10}(r)$) for adiabatic flow in the VE model. Figure~\ref{PP1} compares the BH potential (left panel) and galactic potential (right panel). We use $[\mathcal{E}, \lambda, a, \Upsilon_{B}] = [0.004, 2.76, 0.50, 390]$. In both cases, the flow has three sonic points: inner (I), middle (M), \& outer (O). The vertical jumps ($CC1$ \& $DD1$) mark shock (\ref{RH3}). Thus, the flow is a multi-transonic and shocked accretion flow. The corresponding values are listed in Table~\ref{tab:comparison_quantities}. The galactic potential shifts the outer sonic point outward by about $273.09r_g$ compared to the BH. In contrast, the inner sonic point \& the inner shock location change very little. This shows that the inner flow is mainly controlled by the SMBH gravity. However, the outer shock moves closer to the event horizon by about $9.9r_g$ in the galactic potential case. This shows that the galactic potential mainly affects the outer accretion flow. Similarly, we plot the phase diagram for isothermal flow in Figure~\ref{Piso}, which shows the results for CO disc geometry under galactic potential. We use $[T, \lambda, a, \Upsilon_{B}] = [2 \times 10^{10}, 3.13, 0.50, 100]$, where $T$ is the flow temperature. The flow again shows multi-transonic behaviour. The flow also undergoes a dissipative shock transition \eqref{disshok}.

\begin{table}
\centering
\renewcommand{\arraystretch}{1.0}
\setlength{\tabcolsep}{4.5pt}
\begin{tabular}{|l|r|r|r|r|r|}
\hline
 \textbf{Potential}
 &
$\mathbf{I}$ &
$\mathbf{M}$ &
$\mathbf{O}$ &
$\mathbf{CC1}$ &
$\mathbf{DD1}$ \\
\hline
BH        & 3.604 & 9.589 & 138.31 & 21.38 & 5.62 \\
\hline
$\Upsilon_{B=390}$   & 3.619 & 8.810 & 411.40 & 11.48 & 6.91 \\
\hline
\end{tabular}
\caption{Comparison of the locations of the inner (I), middle (M), and outer (O) critical points, along with the corresponding shock locations CC1 and DD1, as shown in Figure~\ref{PP1}.
}
\label{tab:comparison_quantities}
\end{table}

To determine the nature of the critical points located over the entire PMS, we use the dynamical systems approach described in \S\ref{fixednature}. The eigenvalue analysis identifies each critical point as either saddle-type or centre-type. The results are presented in the next subsection.

\subsection{Nature and classification of critical points}
\label{cpb}
\begin{figure*}
\centering
\begin{minipage}{0.32\linewidth}
    \centering
    \includegraphics[width=\linewidth,height=6cm]{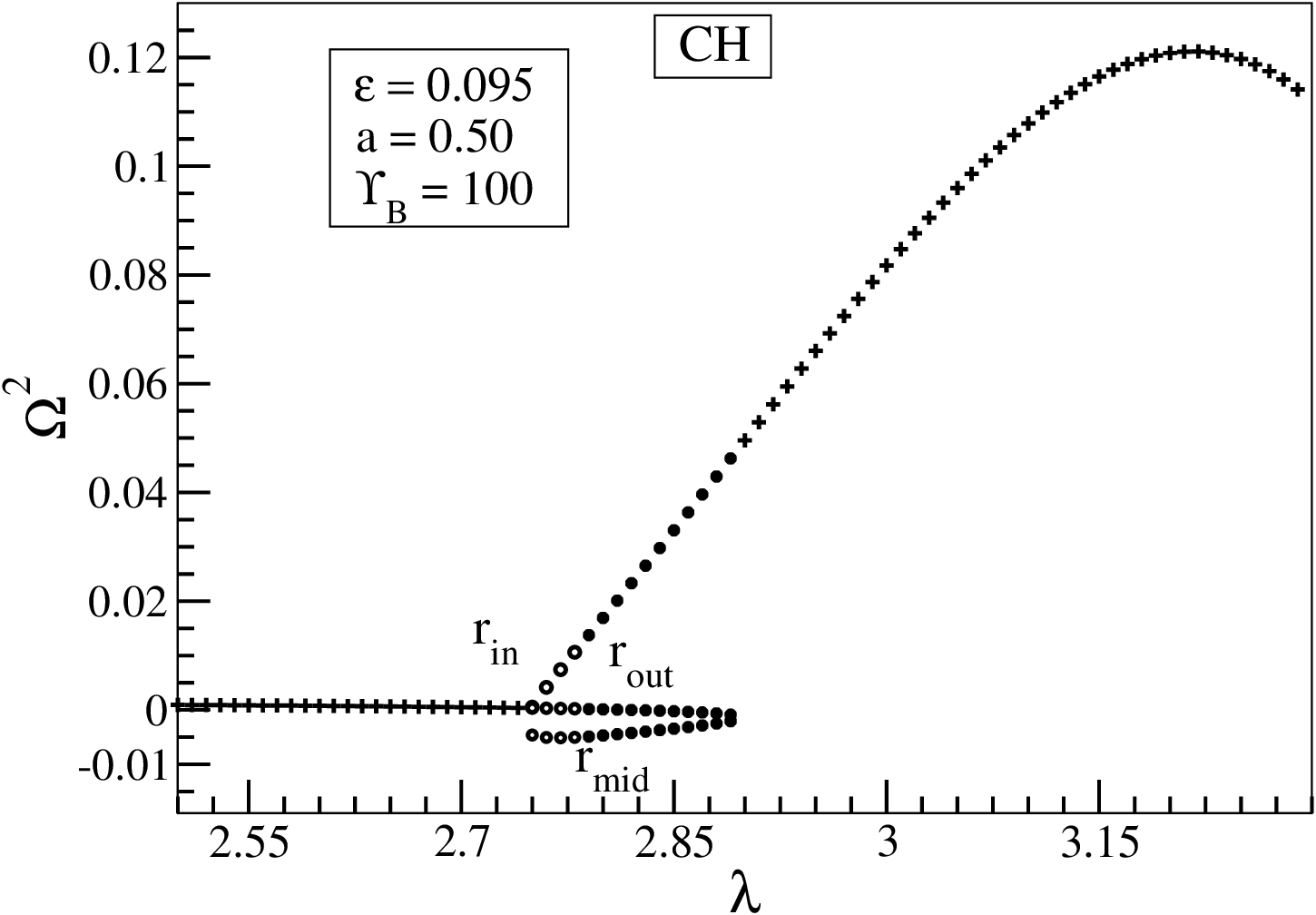}\\[1ex]
    {\small (a)}
    \label{OmegaCH}
\end{minipage}
\hfill
\begin{minipage}{0.32\linewidth}
    \centering
    \includegraphics[width=\linewidth,height=6cm]{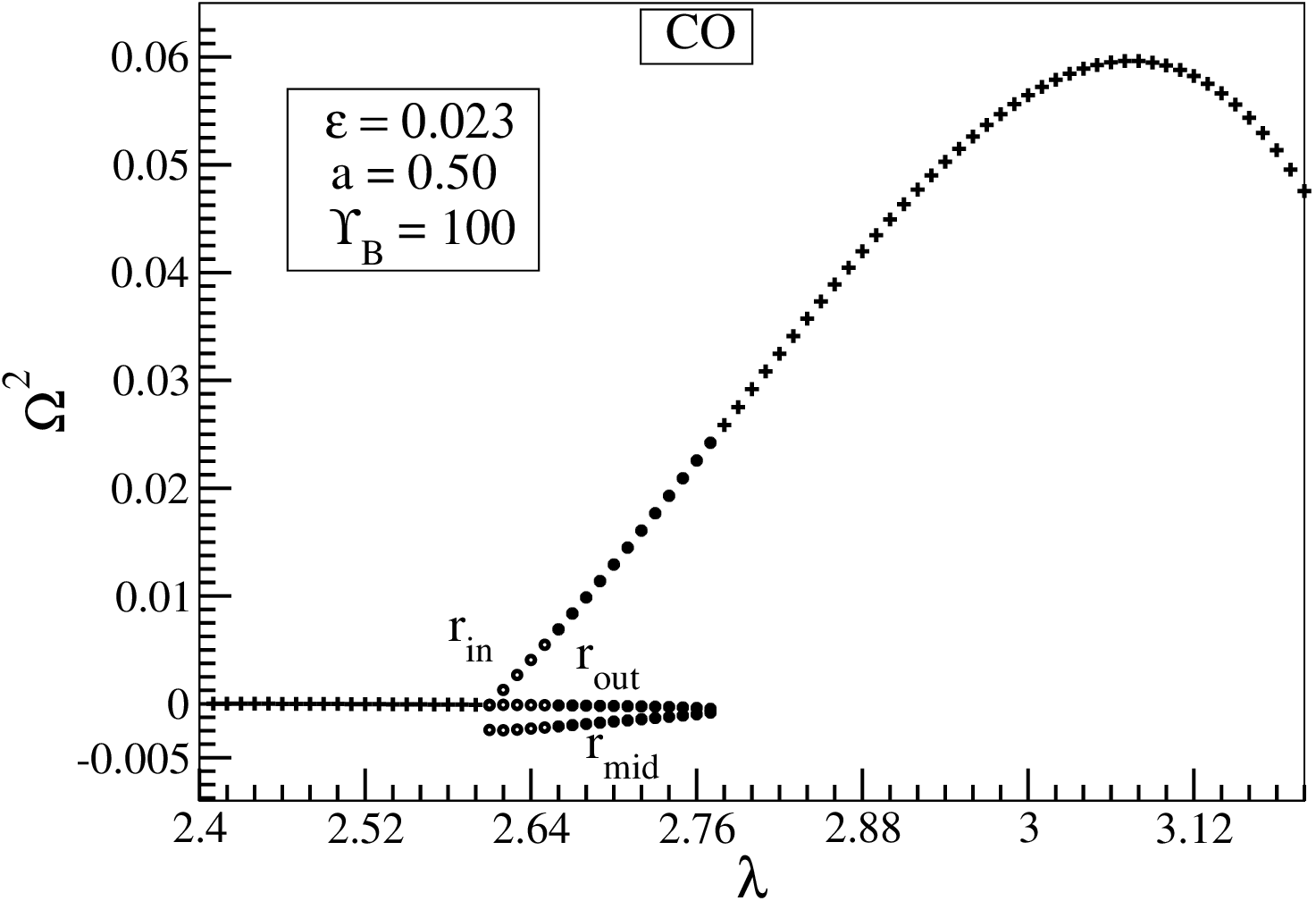}\\[1ex]
    {\small (b)}
    \label{OmegaCO}
\end{minipage}
\hfill
\begin{minipage}{0.32\linewidth}
    \centering
    \includegraphics[width=\linewidth,height=6cm]{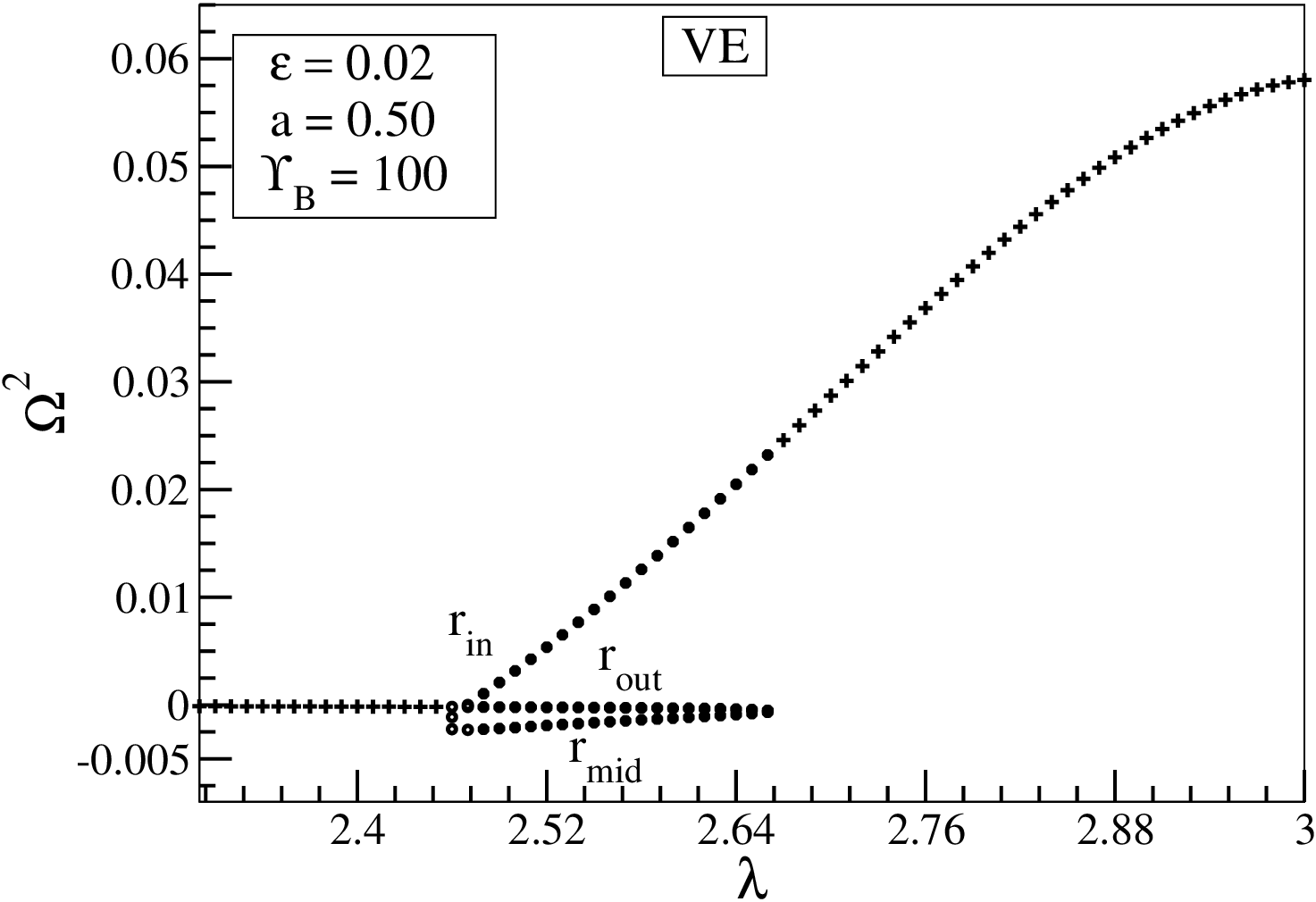}\\[1ex]
    {\small (c)}
    \label{OmegaVE}
\end{minipage}

\caption{This figure represents the nature of the critical points obtained for three disc flow models, e.g., (a) constant height, (b) conical height, and (c) vertical height. In all the subplots, when $\Omega^2$ is positive, it suggests a saddle point; when it is negative, it indicates a center.  All of these plots share the characteristic that, at first, there is a region with a single saddle point, after which a center-type point and another saddle point are born (saddle-centre type bifurcation); at a higher value of $\lambda$, the center-type point merges with the other saddle point (the outer one), and the two of them eradicate each other (a different saddle-centre kind bifurcation, instead this one occurring in the reverse orientation), allowing the last saddle point (outer) to persist beyond the critical value of $\lambda$. In all three plots, a single saddle point is shown by a $+$ sign line curve; the behavior of $\Omega^2$ for the three critical points is represented by $\bullet$ sign line curves in the plot, and as $\lambda$ increases further, a homoclinic loop opens up, and a similar homoclinic connection forms, marked by a $ \ circ$ sign line curve. Figures (a), (b), and  (c) respectively consider the initial parameters as $\mathcal{E}=0.095,0.023,0.02$. The spin parameter and the galactic parameters for all the cases are equal to $\rm{a}=0.50$ and $\Upsilon_{B} = 100$.}
\label{nature-eigen-CHCOVE}
\end{figure*}

\begin{figure}
 \centerline{\includegraphics[width=1.0\linewidth,clip]{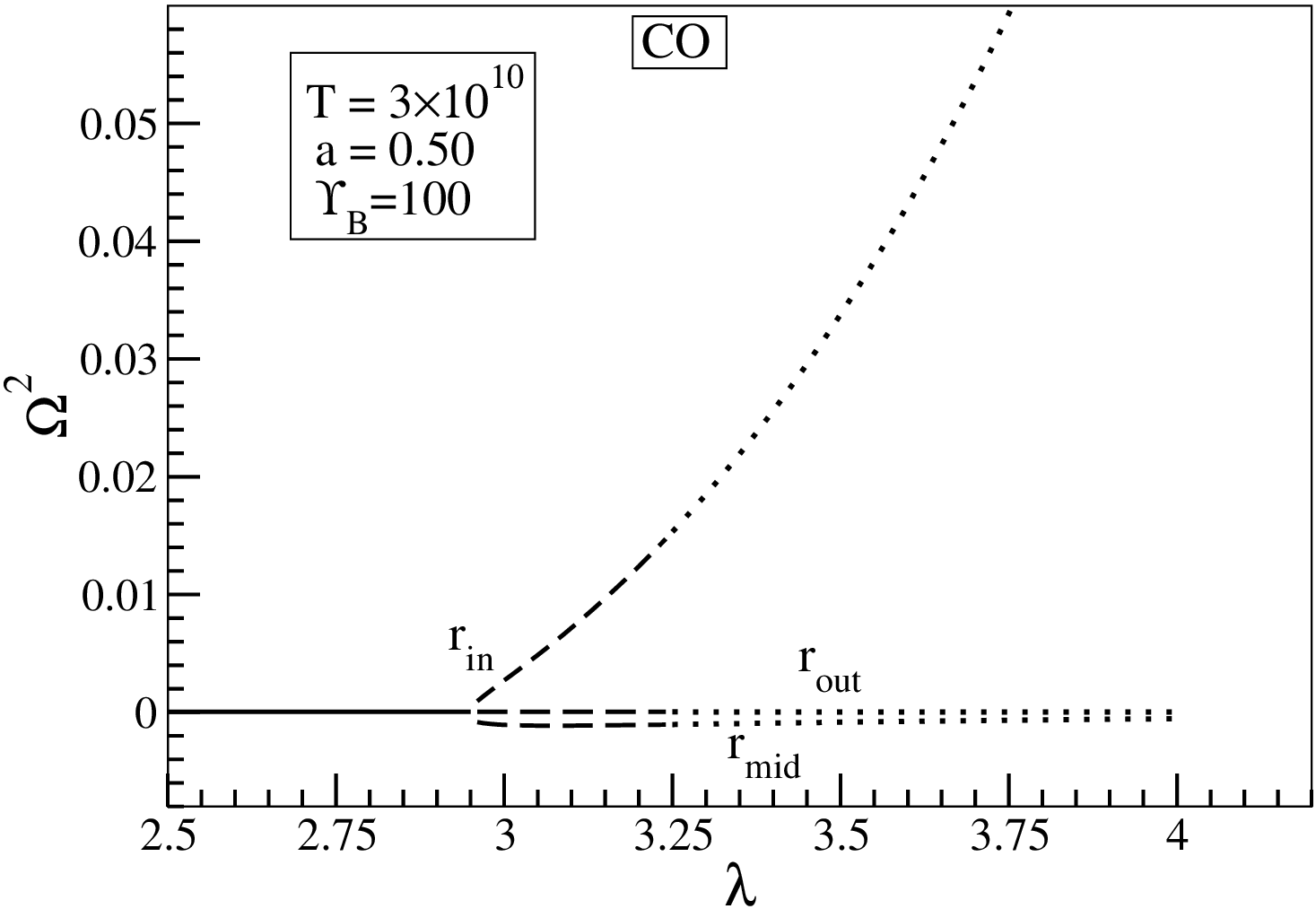}}
  \caption{ For conical isothermal flow, the variation of $\Omega^2$ with angular momentum ($\lambda$) is shown here. Temperature, Kerr parameter, and galactic parameter values are given by $[T,\rm{a},\Upsilon_{B}] = [3\times10^{10},0.50,100]$. In this plot, the solid line shows the single saddle point; the behavior of $\Omega^2$ for the three critical points is represented by dotted line curves, and the homoclinic connection is shown by dash line curve.}
\label{ISO-eigen} 
\end{figure}

To determine the nature of the critical points, we examine the eigenvalues of the stability matrix (\S\ref{fixednature}). Fig. ~\ref{nature-eigen-CHCOVE} \& \ref{ISO-eigen} show variation of $\Omega^2$ for adiabatic \& isothermal flows, respectively. For the adiabatic case, the CH, CO, \& VE disc models are chosen, while the CO model is shown for the isothermal case. We use $\rm{a}=0.50$ \& $\Upsilon_{B}=100$. $\Omega^2$ is calculated from Eqs.~\eqref{OCH}, \eqref{OCO}, \& \eqref{OVE1}. $\Omega^2>0$ identifies a saddle point, whereas $\Omega^2<0$ identifies centre point. At low $\lambda$, flow has single saddle point ($+$ curves in Fig.~\ref{nature-eigen-CHCOVE} and solid curves in Fig.~\ref{ISO-eigen}). As $\lambda$ increases, a saddle--centre bifurcation occurs, producing three critical points: an inner saddle, a centre, and an outer saddle. In this region, $\dot{M}_{\rm in}>\dot{M}_{\rm out}$ for adiabatic flow (Eq.~\ref{ent}) \& $C_{\rm in}<C_{\rm out}$ for isothermal flow (Eq.~\ref{iso1}). These solutions are shown by $\bullet$ curves in Fig.~\ref{nature-eigen-CHCOVE} \& dashed curves in Fig.~\ref{ISO-eigen}. With a further increase in $\lambda$, the solution changes to $\circ$ curves (dotted curves for isothermal). At still higher $\lambda$, the centre point merges with the outer saddle and disappears. The flow then returns to a single inner saddle point, shown again by $+$ (solid) curves. Boundary between two multicritical subregions corresponds to $\dot{M}_{\rm in}=\dot{M}_{\rm out}$ (or $C_{\rm in}=C_{\rm out}$), where a heteroclinic connection forms between two saddle points \citep{2012NewA...17..285N,2002PhRvE..66f6303R}.

Having determined the nature of the critical points, we next examine the possibility of shock formation. In the presence of multiple critical points, the flow can undergo a standing shock that connects two transonic branches. In the following subsection, we present the shock-permitting regions and examine the effect of the galactic potential on the shock properties.

\subsection{Shock permitting region of accretion flow under the influence of galactic potential}
\label{cpc}

\begin{figure*}
\centering
\begin{minipage}{0.48\linewidth}
\centering
\includegraphics[width=\linewidth]{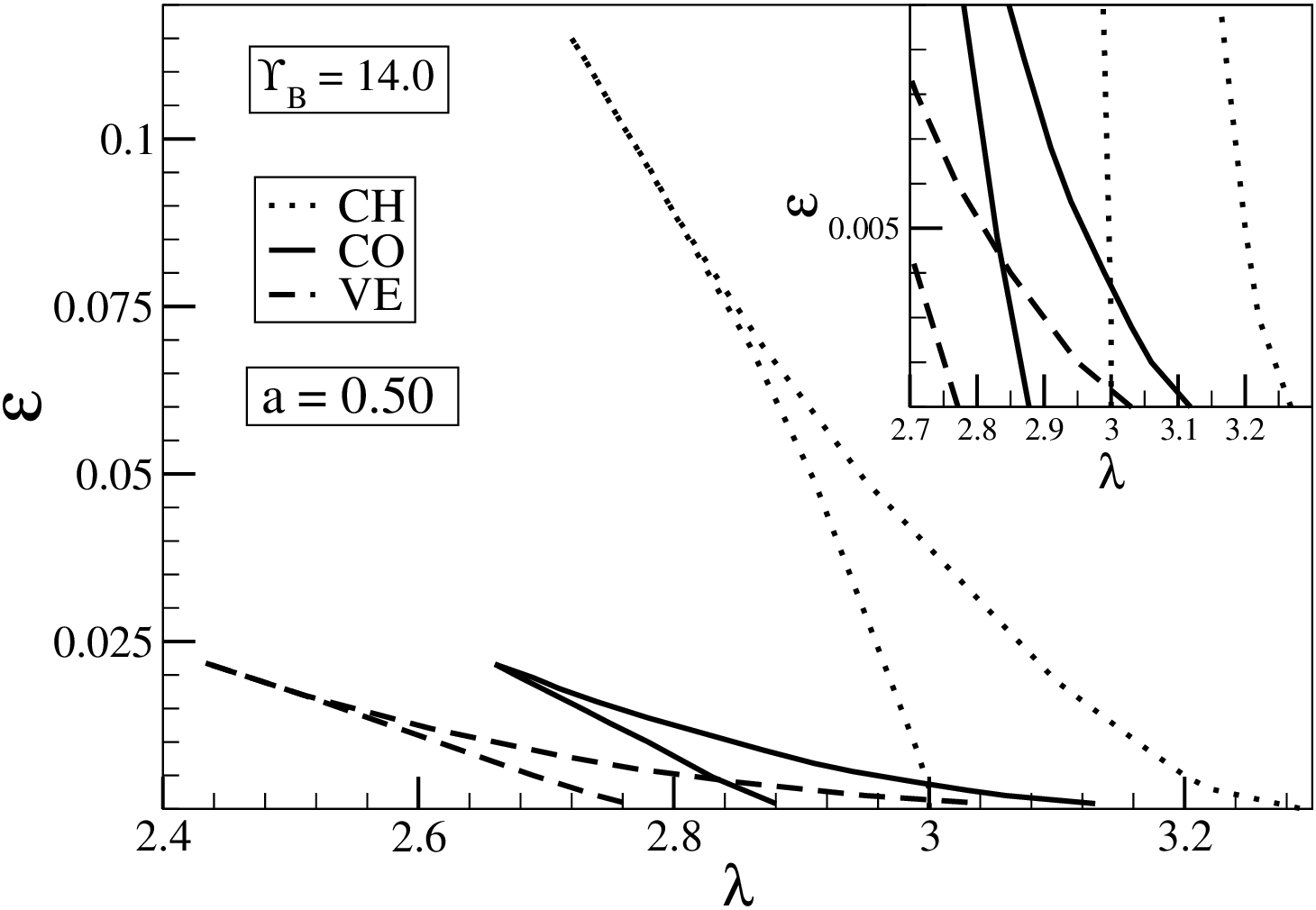}
\small (a)
\label{shok14}
\end{minipage}
\hfill
\begin{minipage}{0.48\linewidth}
\centering
\includegraphics[width=\linewidth]{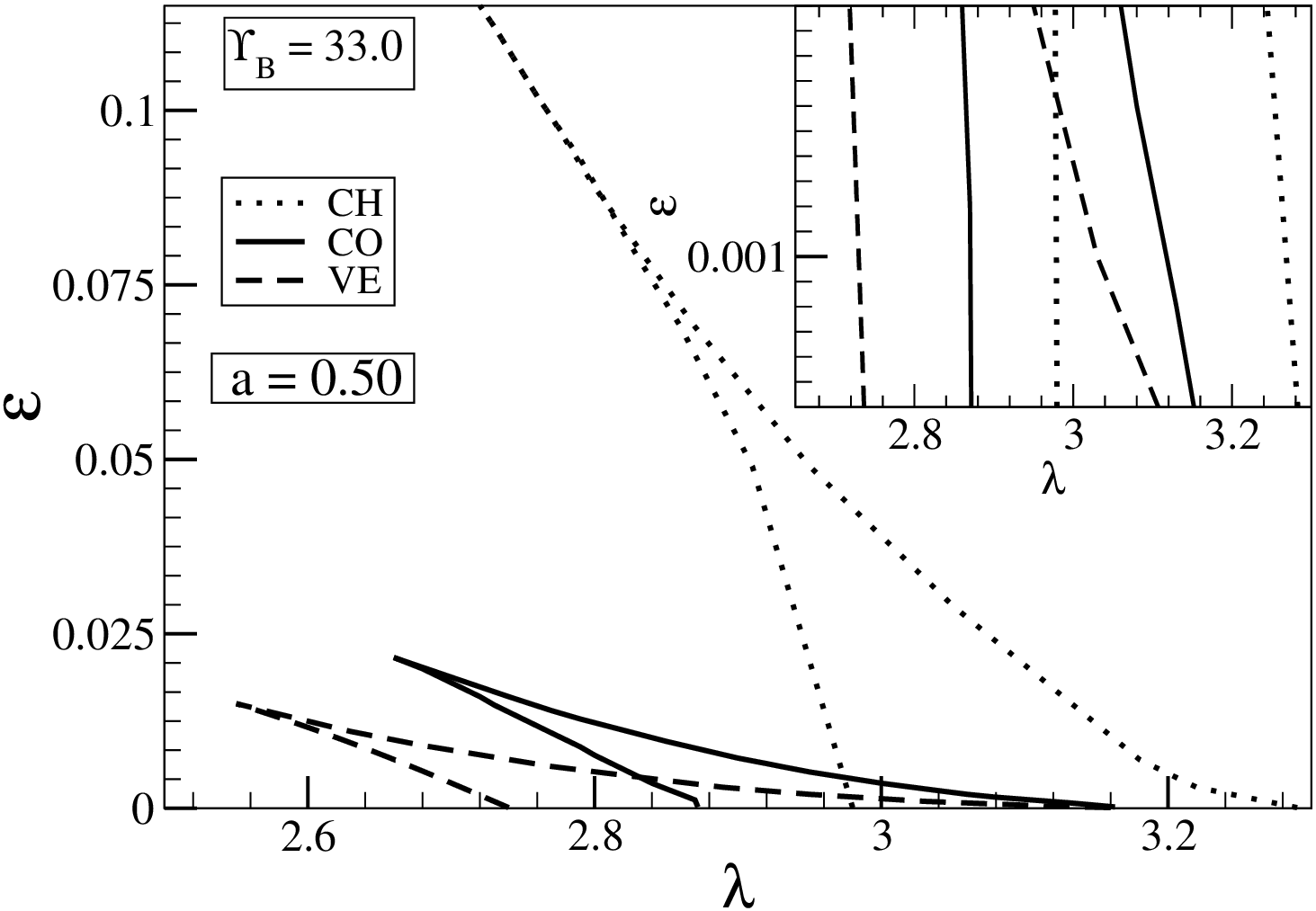}
\small (b)
\label{shok33}
\end{minipage}

\vspace{2ex}

\begin{minipage}{0.48\linewidth}
\centering
\includegraphics[width=\linewidth]{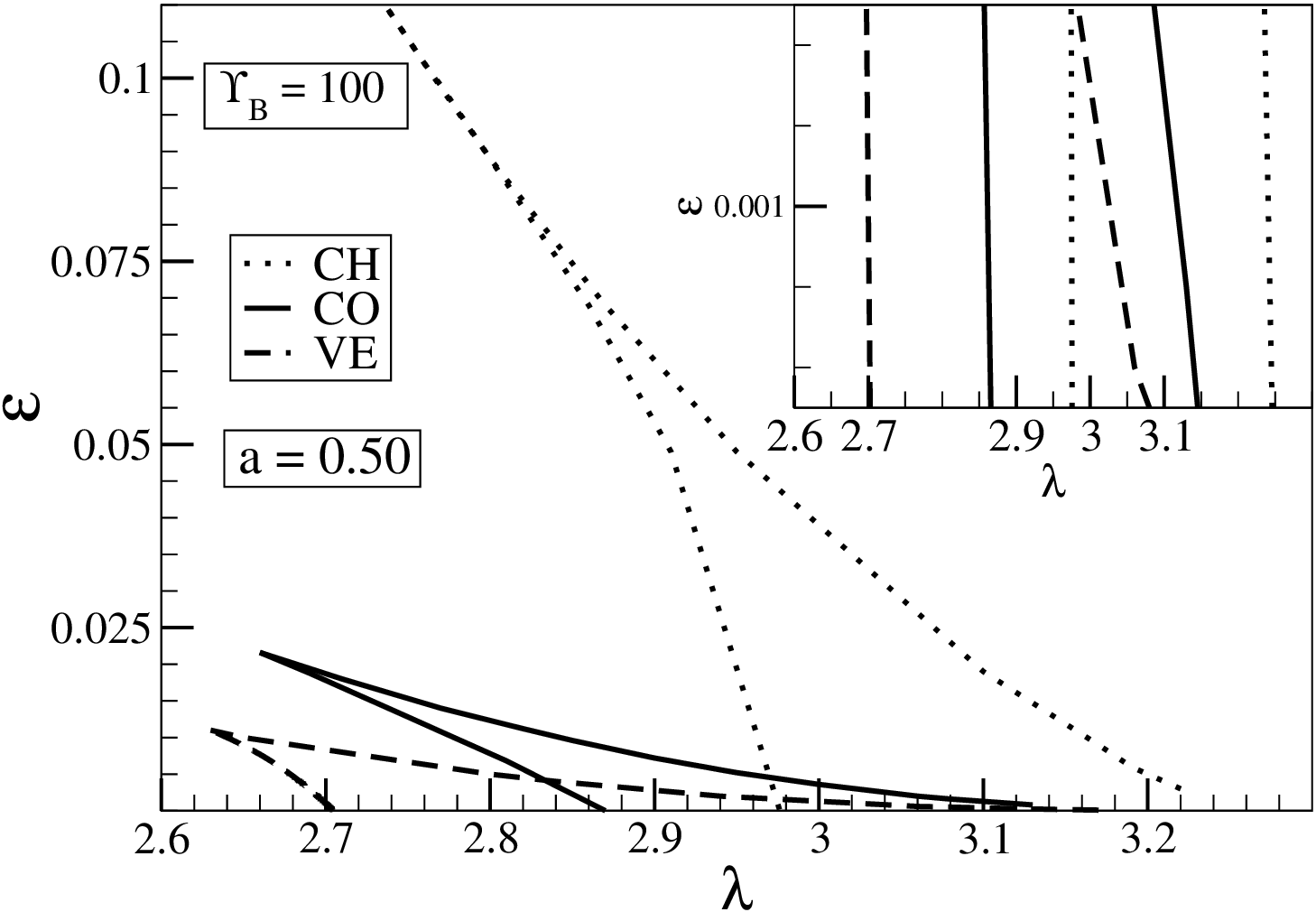}
\small (c)
\label{shok100}
\end{minipage}
\hfill
\begin{minipage}{0.48\linewidth}
\centering
\includegraphics[width=\linewidth]{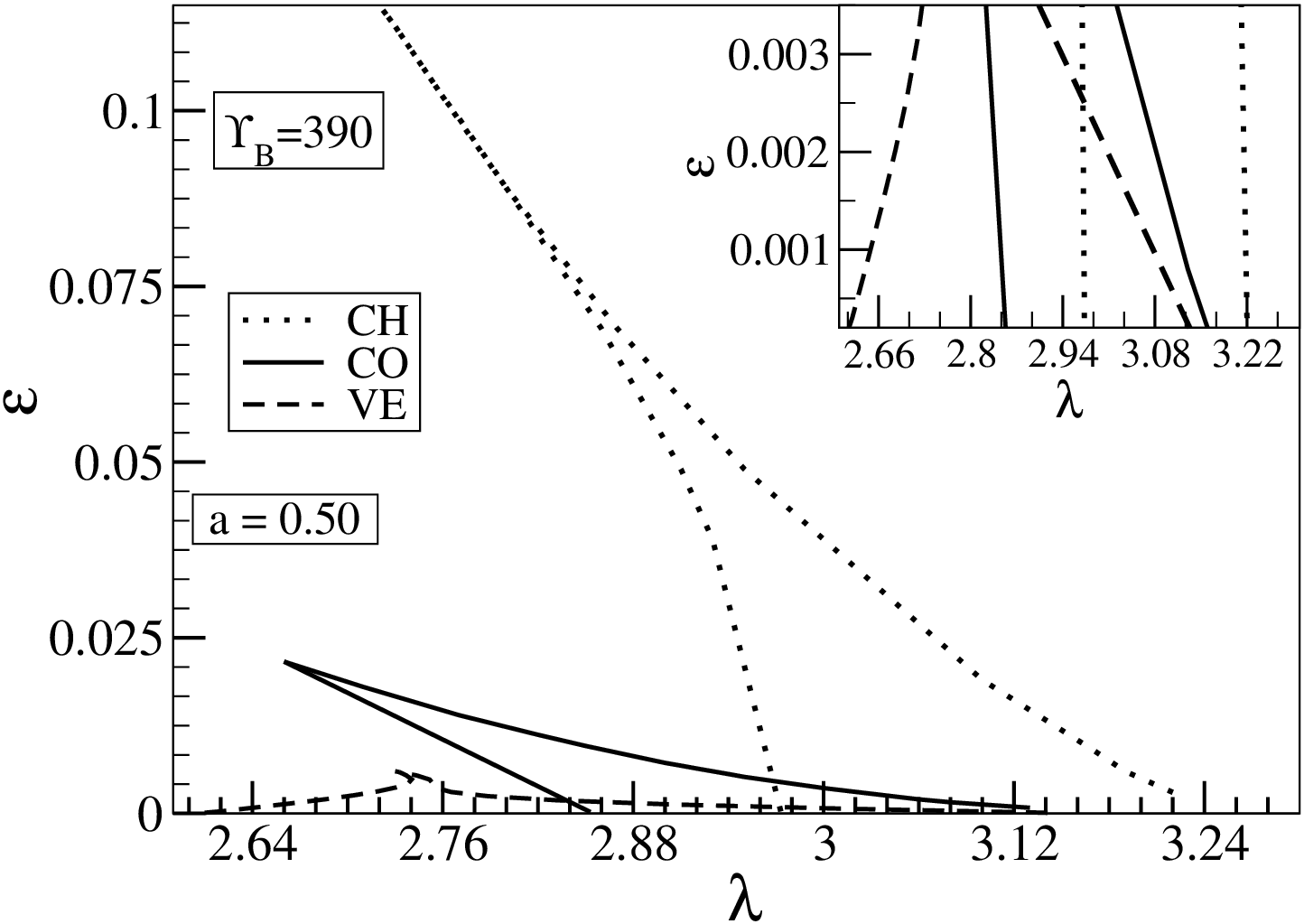}
\small (d)
\label{shok390}
\end{minipage}
\caption{In these figures, we compare the shock-permitting regions in the parameter space ($\mathcal{E}-\lambda$) for the three different accretion disc geometries: CH, CO, and VE, respectively. The dotted lines represent the CH model, the solid lines correspond to the CO model, and the dashed lines indicate the VE model. Each panel corresponds to different values of the galactic parameter $\Upsilon_{B} = 14$, $33$, $100$, and $390$ in panels (a), (b), (c), and (d), respectively. The subplots of each panel provide a zoomed-in view of the region where all three models exhibit overlapping shock-allowed domains, indicating their comparative behavior and the influence of $\Upsilon_{B}$ on shock formation in the context of a multi-component galactic potential. The comparison is also presented with numerical detail in Table \ref{tab:Elambda_geometry_Upsilon}. }
\label{threeshockCHCOVE}
\end{figure*}

As described in \S \ref{shok-F}, shock formation in accretion is required to conserve energy in the case of an adiabatic equation of state and to follow the Rankine-Hugoniot conditions. We investigate and present a comparative analysis of such shock-permitting regions for the adiabatic case for the three different accretion disc geometries—CH, CO, and VE, respectively, in Fig \ref{threeshockCHCOVE}.  The shock-allowed regions are delineated using different line styles: dotted for the CH model, solid for the CO model, and dashed for the VE model. Each panel represents a distinct value of the galactic parameter $\Upsilon_{B}$, chosen as $\Upsilon_{B} = 14$, $33$, $100$, and $390$, to account for varying degrees of galactic background influence. For different sets of galactic configurations indicated by the galactic parameter $\Upsilon_{B}$, the shock-permitting regions in the $\mathcal{E}$–$\lambda$ parameter space progressively shrink and shift toward the central region. This trend is consistently observed across all three disc geometries considered—CH, CO, and VE; however, we see greater impact for the VE model (see the figure and the table \ref{tab:Elambda_geometry_Upsilon}). The subplot of each panel presents a magnified view of the central portion of the parameter space, where the shock-allowed zones corresponding to all three geometries intersect. These zoomed-in views allow for a clearer visualization of the differences and commonalities among the disc models under identical galactic conditions.

\begin{table}
\centering
\renewcommand{\arraystretch}{0.9}
\setlength{\tabcolsep}{4pt}
\begin{tabular}{|c|c|c|c|c|c|}
\hline
Geometry & Quantity & $\Upsilon_{B=14}$ & $\Upsilon_{B=33}$ & $\Upsilon_{B=100}$ & $\Upsilon_{B=390}$ \\
\hline
CH & $\mathcal{E}$
 & 0.1200 & 0.1150 & 0.1070 & 0.1060 \\
\cline{2-6}
   & $\lambda$
 & 3.000 & 2.970 & 2.960 & 2.955 \\
\hline
CO & $\mathcal{E}$
 & 0.0200 & 0.02115 & 0.0220 & 0.0240 \\
\cline{2-6}
   & $\lambda$
 & 2.880 & 2.870 & 2.850 & 2.850 \\
\hline
VE & $\mathcal{E}$
 & 0.0240 & 0.0140 & 0.0100 & 0.0040 \\
\cline{2-6}
   & $\lambda$
 & 2.760 & 2.730 & 2.700 & 2.600 \\
\hline
\end{tabular}

\caption{This table presents the shift in the shock region for the adiabatic equation of state in terms of the variation of specific energy $\mathcal{E}$ and angular momentum $\lambda$ for different flow geometries with the galactic parameter $\Upsilon_B$; see Figure \ref{threeshockCHCOVE}.}
\label{tab:Elambda_geometry_Upsilon}
\end{table}

\begin{figure}
 \centerline{\includegraphics[width=1.0\linewidth,clip]{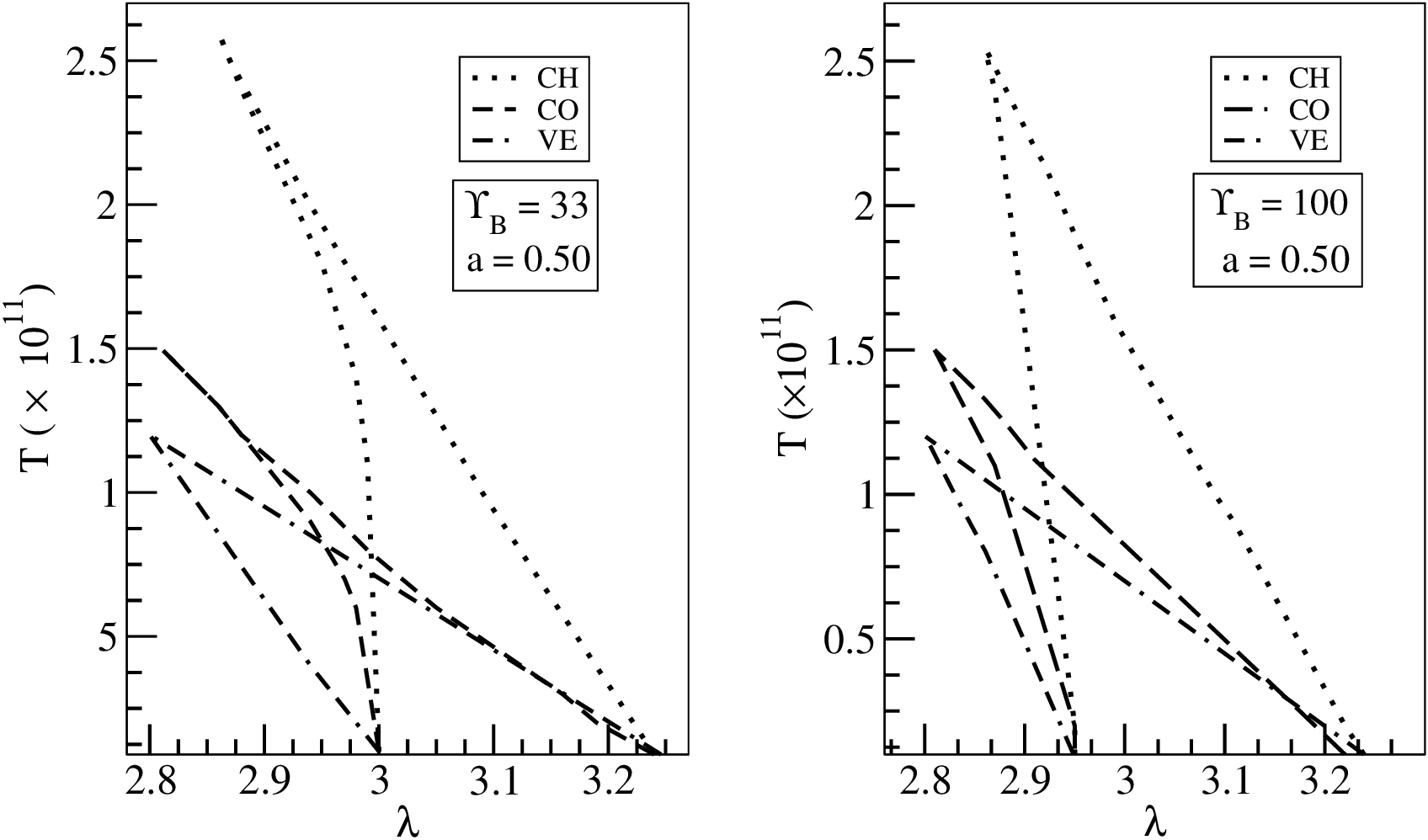}}
 \caption{ In these plot, we compare the shock-allowing regions for isothermal equation of state in the parameter space of temperature ($T$) and specific angular momentum ($\lambda$) for the three disc model e.g., CH, CO, VE model, considering the galactic potential $\Upsilon_{B} = 33$ in the left panel and $\Upsilon_{B} = 100$ in the right panel. We consider the spin parameter $\rm{a} = 0.50$. In both panels, the dotted lines represent results obtained using the CH model, while the dashed line and the dash-dot line correspond to the CO and VE flow models. We see from the figure that the constant height flow under $\Upsilon_B$ also allows a larger region for shock formation than others. Here we also see a substantial overlapping range for the creation of the isothermal shock. The differences that appear due to galactic parameter changes are given in detail in Table \ref{tab:Tlambda_geometry_Upsilon}.
}
\label{shock-ISO}
\end{figure}

Unlike the adiabatic case, for isothermal flows the shocks are dissipative in nature, as energy is not conserved in this case. As described earlier, the shock formation for such flows does not depend explicitly on the chosen disc geometry. We present the shock-permitting regions for such flows in Figure \ref{shock-ISO} in the $T-\lambda$ parameter space for three disc geometries. The CH, CO, and VE  models are presented by dotted lines, dashed lines, and the dashed-dotted line, respectively. The left panel is drawn for $\Upsilon_{B} = 33$, and that in the right panel is $\Upsilon_{B} = 100$. The spin parameter is $\rm{a}=0.50$. We can see from the figure that with different  $\Upsilon_{B}$, the shock region shifts slightly towards the central region, and the impacts are higher for the VE disk model. The two panels present the impact of the galactic potential and demonstrate how it changes the size and location of the shock-allowing regions in the parameter space, which are given in detail in Table \ref{tab:Tlambda_geometry_Upsilon}.

\begin{table}
\centering
\renewcommand{\arraystretch}{1.2}
\setlength{\tabcolsep}{6pt}
\begin{tabular}{|c|c|c|c|}
\hline
Geometry & Parameter & $\Upsilon_{B=33}$ & $\Upsilon_{B=100}$ \\
\hline
CH & $T \times 10^{11}$ & 1.24 & 1.24 \\
\cline{2-4}
   & $\lambda$          & 3.00 & 2.92 \\
\hline
CO & $T \times 10^{11}$ & 1.50 & 1.50 \\
\cline{2-4}
   & $\lambda$          & 3.00 & 2.92 \\
\hline
VE & $T \times 10^{11}$ & 2.75 & 2.58 \\
\cline{2-4}
   & $\lambda$          & 3.00 & 2.92 \\
\hline
\end{tabular}
\caption{This table present the shift in shock region for isothermal flows in terms of the variation of temperature $T$ and angular momentum $\lambda$ for different flow geometries with galactic parameter $\Upsilon_B$ see \ref{shock-ISO}.}
\label{tab:Tlambda_geometry_Upsilon}
\end{table}

To clearly demonstrate the impact of galaxy potential, we compare the shock region for different galactic parameters with the black hole potential in the next figure for the adiabatic case. The figure \ref{shockBHvsVE} illustrates the comparison of shock-permitting regions in the specific energy–specific angular momentum ($\mathcal{E}$–$\lambda$) parameter space for the vertical equilibrium (VE) disc configuration, at a fixed black hole spin of $\rm{a} = 0.50$.  In Figure \ref{shockBHvsVE}, dashed curves correspond to shock-allowing regions under the black hole potential, while solid curves represent those under the galactic potential represented in terms of $\Upsilon_{B} = 14$, $33$, $100$, and $390$. When $\Upsilon_{B} = 14$, both potentials yield nearly overlapping shock regions. However, for the other galactic configuration given by $\Upsilon_{B}$, the shock-allowed domains systematically shrink and shift inward in the $\mathcal{E}$–$\lambda$ space, reflecting the increasing dominance of the galactic potential in altering the conditions necessary for steady shock formation in low angular momentum accretion flows. In Table \ref{tab:Elambda_relative}, we present these results numerically. The shift in $\mathcal{E}$ and $\lambda$  with different $\Upsilon_B$ with respect to the black hole potential is noticeable. We presents these differences in terms of $\Delta \mathcal{E}=\mathcal{E}^{\rm BH}-\mathcal{E}^{\Upsilon_B}$ and $\Delta \lambda =\lambda^{\rm BH}-\lambda^{\Upsilon_B}$. To provide the galactic effect sizes, we also present the fractional change of the shock-permitting area in parameter space in terms of $\Delta \mathcal{E}/\mathcal{E}^{\rm BH}$ and $\Delta \lambda/\lambda^{\rm BH}$. We see that for $\Upsilon_B=390$, the fractional change in energy is almost $70\%$ and in angular momentum is nearly $6\%$, which is the maximum impact of the galactic potential.

We can see that the response of the accretion flow to the galactic potential is strongly dependent on the adopted geometry. While the CH and CO models exhibit only mild variations in the shock-permitting energy and angular momentum with various $\Upsilon_B$, the VE geometry shows a pronounced reduction in the allowed range of both parameters. The pronounced sensitivity of the VE geometry to the galactic potential can be understood from the disc height prescription
$H(r)=c_{\mathrm{s}}\left(r/(\gamma\,\Phi'_{\mathrm{Gal}})\right)^{1/2}$, which directly couples the vertical structure of the flow to the radial gradient of the galactic potential.
For different $\Upsilon_B$, the respective mass distribution from different components of the galaxy changes (as can be seen from \S \ref{galpotd} and Fig \ref{ellpgal}). Depending on the selected fraction of the mass distribution, the overall galactic potential changes and, as a result, $\Phi'_{\mathrm{Gal}}$ also gets affected. This alters the disc thickness and changes the vertical pressure confinement, leading to different effective disc geometries of the flow and hence changing the shock-permitting energy and angular momentum.  In contrast, the CH and CO geometries lack this explicit dependence on $\Phi'_{\mathrm{Gal}}$, resulting in a much weaker response to the galactic environment (see Table \ref{tab:Elambda_geometry_Upsilon},\ref{tab:Tlambda_geometry_Upsilon}).
It can be noted that, in our studies, we see a greater impact of the galactic environment on the flow geometries for sets 3 and 4 of the $\Upsilon_B$ parameter. Among the four sets of $\Upsilon_B$, Sets 1 and 2, i.e., parameters 14 and 33, contain effectively lower dark matter contributions, and Sets 3 and 4, i.e 100 and 390, contain a very high contribution from dark matter. So far, from our plots, we see that the galaxy potential and its impact become higher with increasing fraction of dark matter content, e.g., Set-3 ($\Upsilon_B=100$). Set-4, where $\Upsilon_B=390$, shows even more impact, where the dark matter content is the same as $\Upsilon_B=100$ however contains a slightly higher hot gas contribution than the $100$ case.

\begin{figure*}
 \centerline{\includegraphics[width=0.95\linewidth,clip]{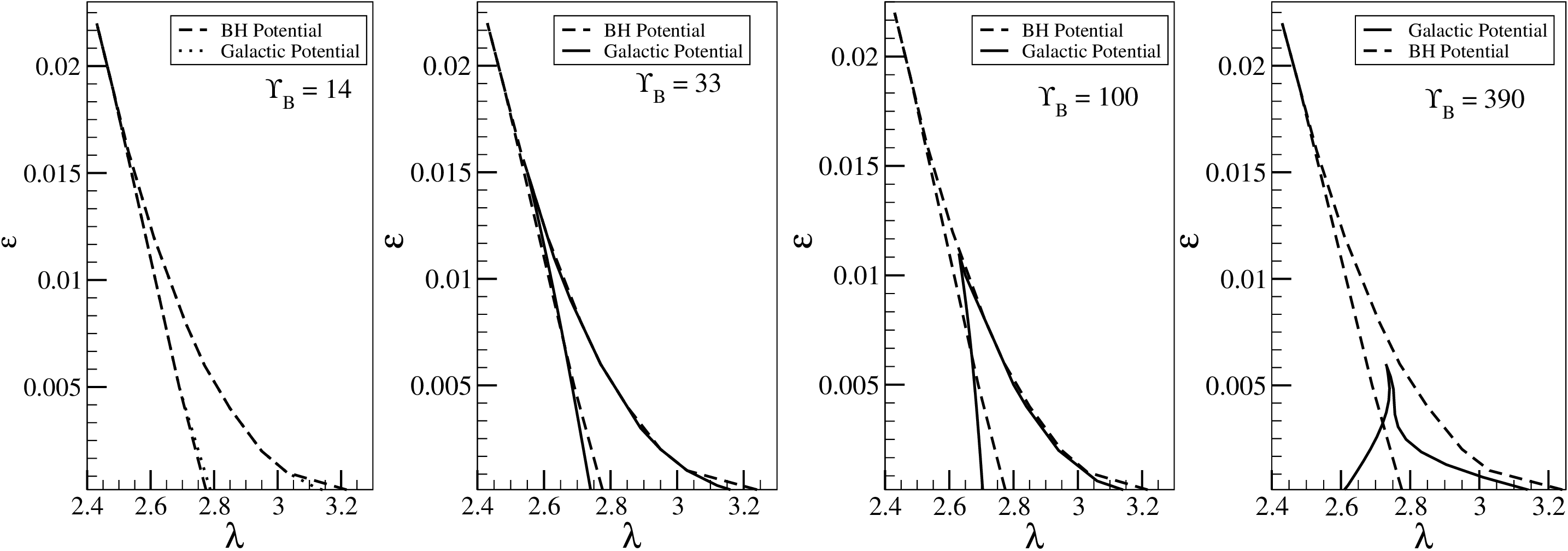}}
 \caption{
These figures compare the shock-allowing regions in the parameter space of specific energy ($\mathcal{E}$) and specific angular momentum ($\lambda$) for the vertical equilibrium  (VE) disc model, considering the black hole (BH) potential and the galactic potential, for the spin parameter $\rm{a} = 0.50$. The dashed lines represent the case for the BH potential, while the solid lines correspond to cases where the galactic potential is included. The galactic contribution is characterized by different values of the parameter: $\Upsilon_{B} = 14$, $\Upsilon_{B} = 33$, $\Upsilon_{B} = 100$, and $\Upsilon_{B} = 390$. The comparison demonstrates how the inclusion of additional galactic components—such as dark matter, stellar mass, and hot gas- modifies the size and location of the shock-allowing regions in the parameter space. The comparison is discussed numerically in detail in the text and in Table \ref{tab:Elambda_relative}.
}
\label{shockBHvsVE}
\end{figure*}

\begin{table}[H]
\centering
\renewcommand{\arraystretch}{1.4}
\setlength{\tabcolsep}{3pt}
\begin{tabular}{|l|c|c|c|c|c|c|}
\hline
  Potential
& $\mathcal{E}$
& $\lambda$
& $\Delta \mathcal{E}$
& $\Delta \mathcal{E}/\mathcal{E}^{\rm BH}$
& $\Delta \lambda$
& $\Delta \lambda/\lambda^{\rm BH}$ \\
\hline
BH
& 0.0225
& 2.78
& --
& --
& --
& -- \\
\hline
$\Upsilon_{B=14}$
& 0.0225
& 2.78
& 0.0000
& 0.0000
& 0.00
& 0.0000 \\
\hline
$\Upsilon_{B=33}$
& 0.0183
& 2.75
& 0.0042
& 0.1867
& 0.03
& 0.0108 \\
\hline
$\Upsilon_{B=100}$
& 0.0116
& 2.70
& 0.0109
& 0.4844
& 0.08
& 0.0288 \\
\hline
$\Upsilon_{B=390}$
& 0.0066
& 2.60
& 0.0159
& 0.7067
& 0.18
& 0.0647 \\
\hline
\end{tabular}
\caption{This table presents the changes in shock-permitting regions for the VE disc model and the adiabatic case in the presence of a galactic potential in terms of $\mathcal{E}$ and $\lambda$  with respect to the  BH case. We consider the spin parameter $\rm{a}=0.50$. Here $\Delta \mathcal{E}=\mathcal{E}^{\rm BH}-\mathcal{E}^{\Upsilon_B}$ and $\Delta \lambda =\lambda^{\rm BH}-\lambda^{\Upsilon_B}$. The fractional changes with respect to the BH case are given by $\Delta \mathcal{E}/\mathcal{E}^{\rm BH}$ and $\Delta \lambda/\lambda^{\rm BH}$. For more details, see Figure \ref{shockBHvsVE}.}
\label{tab:Elambda_relative}
\end{table}

\subsubsection{Impact of black hole spin on shock region}
\begin{figure*}
 \centerline{\includegraphics[width=0.8\linewidth,clip]{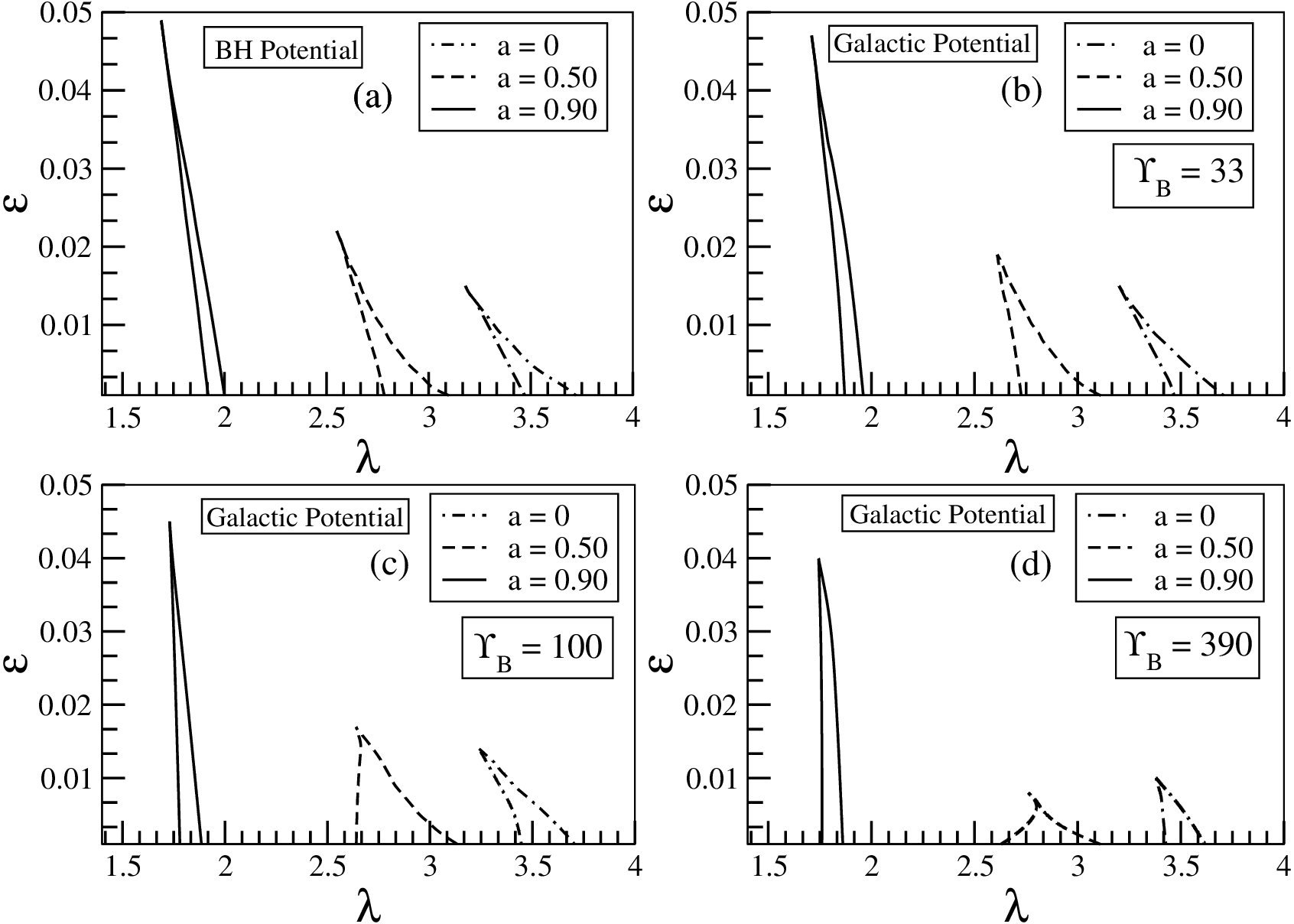}}
 \caption{
These figures compare the shock-allowing regions in the parameter space of specific energy ($\mathcal{E}$) and specific angular momentum ($\lambda$) for the vertical equilibrium  (VE) disc model, considering the black hole (BH) potential and the galactic potential, for the spin parameter $\rm{a} = 0,0.50,0.90$. The solid lines represent the case for $\rm{a} = 0.90$, while the long-dashed lines correspond to the case $\rm{a} = 0.50$ and the dash-dotted line for $\rm{a} = 0$. Panel (a) represents the black hole (BH) potential, and panels (b),(c), and (d) are for different values of the parameter: $\Upsilon_{B} = 33$, $\Upsilon_{B} = 100$, and $\Upsilon_{B} = 390$, respectively. The comparison is discussed numerically in detail in the text and in Table \ref{tab:Elambda_forspin}.
}
\label{spinVE059}
\end{figure*}

To examine the effect of black hole spin, we compare the shock parameter space for the isolated BH potential and the combined galactic potential. Figure~\ref{spinVE059} presents the shock parameter space in the $\mathcal{E}$--$\lambda$ plane for different values of the spin parameter. Panel (a) corresponds to the isolated BH potential, while panels (b)--(d) show the galactic potential cases with $\Upsilon_B=33$, $100$, and $390$, respectively. In each panel, the results are shown for $a=0.0$ (dot-dash), $0.50$ (dashed), and $0.90$ (solid).
We also examine the case of $a=0.92$. In the presence of the galactic potential, the shock-permitting region shrinks to a nearly one-dimensional curve, enclosing essentially very small area. This indicates that the parameter space supporting standing shocks vanishes at sufficiently high black hole spin. All results are obtained for the VE disc geometry. As the black hole spin increases, the shock parameter space shifts towards lower $\lambda$ and higher $\mathcal{E}$. The same trend is found for both the isolated BH and galactic potential cases. However, the inclusion of the galactic potential shifts the entire shock region to lower $\lambda$ and slightly reduces the range of $\mathcal{E}$ and $\lambda$ that permits standing shocks. This effect becomes more pronounced as $\Upsilon_B$ increases. The numerical values for panels (a) and (d) are listed in Table~\ref{tab:Elambda_forspin}. For the high-spin case ($a=0.90$), the specific energy changes by about $18\%$, while the specific angular momentum changes by about $7\%$. The observed trend can be understood from the balance between gravity and centrifugal support. As the black hole spin increases, frame-dragging enhances the effective rotational support of the flow. Therefore, standing shocks can form at lower values of $\lambda$, shifting the shock parameter space towards lower angular momentum. The galactic potential adds an additional gravitational contribution to the effective potential. This reduces the range of flow parameters that admit standing shocks and shifts the entire shock-permitting region relative to the isolated black hole case. It should also be noted that adopting a more accurate effective potential, such as that proposed by \citet{2018PhRvD..98h3004D}, instead of the \citet{1996ApJ...461..565A} potential, has a direct impact on the shock-permitting parameter space for rapidly spinning black holes. For the case of $a=0.92$, the minimum specific angular momentum required for standing-shock formation is $\lambda \simeq 1.71$ when the Artemova potential is used, whereas it decreases to $\lambda \simeq 1.56$ for the potential of \citet{2018PhRvD..98h3004D}, corresponding to a relative deviation of approximately $9.6\%$.
\begin{table}[H]
\centering
\renewcommand{\arraystretch}{1.4}
\setlength{\tabcolsep}{4pt}
\begin{tabular}{|l|c|c|c|c|c|c|}
\hline
Potential
& $\mathcal{E}$
& $\lambda$
& $\Delta \mathcal{E}$
& $\Delta \mathcal{E}/\mathcal{E}^{\rm BH}$
& $\Delta \lambda$
& $\Delta \lambda/\lambda^{\rm BH}$ \\
\hline
BH
& 0.0489
& 1.90
& --
& --
& --
& -- \\
\hline
$\Upsilon_{B=390}$
& 0.0399
& 1.76
& 0.0090
& 0.1840
& 0.14
& 0.0737 \\
\hline
\end{tabular}
\caption{This table presents the changes in shock-permitting regions for the VE disc model and the adiabatic case in the presence of a galactic potential in terms of $\mathcal{E}$ and $\lambda$  with respect to the  BH case. We consider the spin parameter $\rm{a}=0.90$. Here $\Delta \mathcal{E}=\mathcal{E}^{\rm BH}-\mathcal{E}^{\Upsilon_B}$ and $\Delta \lambda =\lambda^{\rm BH}-\lambda^{\Upsilon_B}$. The fractional changes with respect to the BH case are given by $\Delta \mathcal{E}/\mathcal{E}^{\rm BH}$ and $\Delta \lambda/\lambda^{\rm BH}$. For more details see figure \ref{spinVE059}.}
\label{tab:Elambda_forspin}
\end{table}

\subsection{Shock-strength, shock-induced flow variables and shock location}
\label{cpd}

In this section, we study the properties of shocks in the presence of a galactic potential by introducing the concept of shock strength, which provides a quantitative measure of how different the pre-shock and post-shock flow states are. The shock strength determines the magnitude of compression, entropy generation, and post-shock pressure across the shock front. Small compression ratios correspond to weak shocks, whereas strong shocks are associated with large entropy gradients. Physically, the shock strength quantifies how abruptly and irreversibly the accretion flow is restructured at the shock front by measuring the efficiency of compression and the conversion of kinetic energy into thermal energy, thereby fixing the thermodynamic and dynamical character of the post-shock region under the galactic potential.

\begin{figure*}
 \centerline{\includegraphics[width=1.00\linewidth,clip]{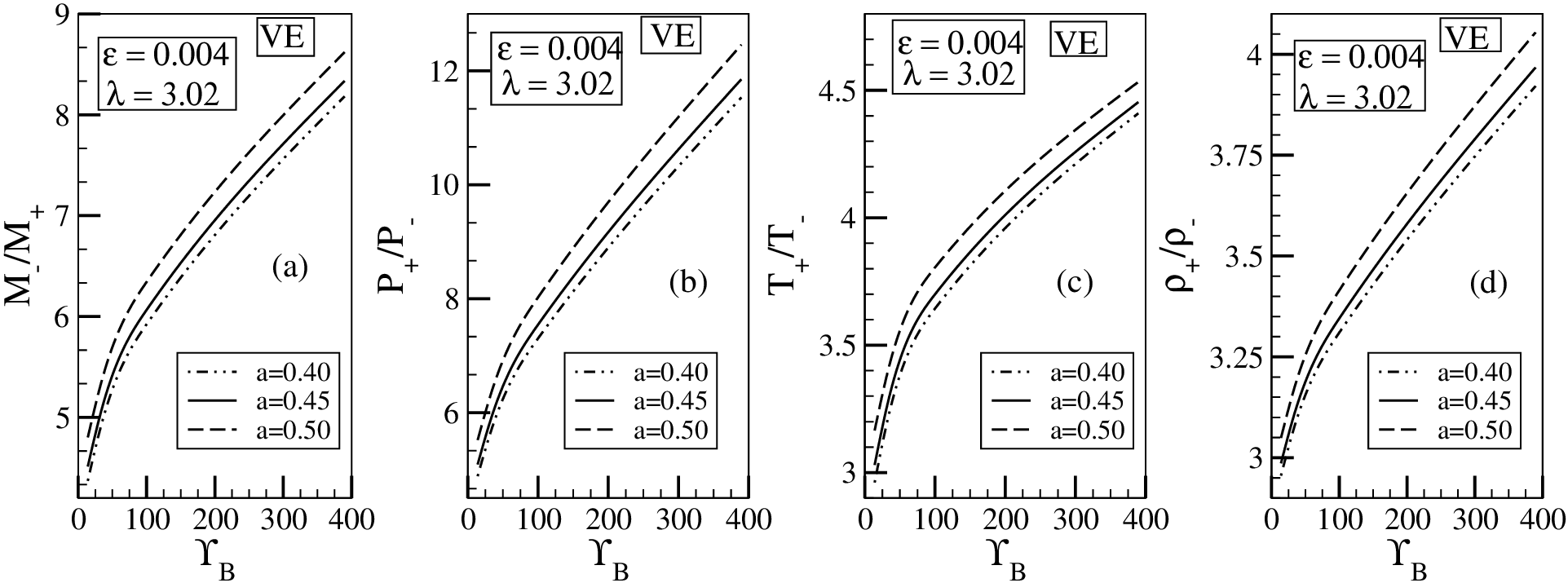}}
 \caption{
Comparison of shock strength in the VE disc model as a function of different values of the galactic parameter: $\Upsilon_{B} = 14$, $\Upsilon_{B} = 33$, $\Upsilon_{B} = 100$, and $\Upsilon_{B} = 390$ are shown for three chosen spin parameter ($\rm{a}$). We consider here, the specific energy $\mathcal{E} = 0.004$ and specific angular momentum $\lambda = 3.02$, and the Kerr parameter $\rm{a} = 0.40$ (dot-dot-dash line), $\rm{a} = 0.45$ (solid line), and $\rm{a} = 0.50$ (dash line). Panel (a) shows the variation of shock strength in terms of the Mach number ratio ($M_{-} / M_{+}$), panel (b) displays the variation of the pressure ratio across the shock ($P_{+} / P_{-}$), panel (c) illustrates the post-shock to pre-shock temperature ratio ($T_{+} / T_{-}$), and panel (d) presents the compression ratio ($\rho_{+} / \rho_{-}$). We see that for all four sets of galactic mass distributions, the respective shock strengths are different, and with increasing dark matter content of the galactic configuration, the associated shock strength also increases. The spin parameter impacts are similar to earlier studies of black hole accretion studies e.g., \citet{2016NewA...43...10S}. We see shock strength increases with increasing Kerr parameter. Spin controls the location of the shock, and an increase in the Kerr parameter enhances pre-shock heating of the inflow, leading to a systematic increase in shock strength with increasing spin. The numerical results are discussed in Table \ref{tab:VE_MPTR}.
}
\label{MPTrho-para}
\end{figure*}

In Figure \ref{MPTrho-para} we plot the variation in shock strength for the vertical equilibrium (VE) disc model, with different galactic configurations $\Upsilon_{B}$ (14, 33, 100, and 390) for varying black hole spin parameter ($\rm{a}$). For the CH and CO disc geometries, we see no noticeable changes in the shock-induced variables. In this study, we keep the specific energy fixed at $\mathcal{E} = 0.004$ and the specific angular momentum at $\lambda = 3.02$, while changing the spin parameter values among $\rm{a} = 0.40$, $0.45$, and $0.50$. Panel (a) shows how the shock strength, measured by the ratio of Mach numbers before and after the shock ($M_{-} / M_{+}$), changes with spin, as this defines how much the speed of the flow changes across the shock. Panel (b) shows how the pressure ratio ($P_{+} / P_{-}$) changes across the shock, indicating how much the pressure jumps due to the shock. Panel (c) presents the ratio of temperatures after and before the shock ($T_{+} / T_{-}$) which reflects how hot the flow becomes after the shock, and panel (d) shows the density compression ratio ($\rho_{+} / \rho_{-}$), showing how much denser the flow gets after passing through the shock. All these figures show that the shock strength ratio differs for distinct galactic configurations, denoted by  $\Upsilon_{B}$. In Table~\ref{tab:VE_MPTR}, we present our numerical results for a representative case with a fixed spin parameter $\rm{a}=0.50$ for figure \ref{MPTrho-para}. As discussed earlier, the chosen set of galactic configurations exhibits distinctly different shock strengths across all four categories. In particular, for the fourth set of galactic parameters, corresponding to  $\Upsilon_{B} = 390$, the shock strength attains its maximum value. Consistent with our earlier observations (see Figure \ref{shockBHvsVE} \& Table \ref{tab:Elambda_relative}), we find that an increase in dark matter content along with hot gas mass enhances the galactic influence on the shock strength. We also observe that variations in the central black hole spin lead to corresponding changes in the shock strength. Our results show that shocks get stronger at higher black hole spin. The black hole spin influences the shock location and increases the thermal content of the pre-shock flow, which increases the strength of the shock at higher spin.

\begin{table}[H]
\centering
\renewcommand{\arraystretch}{1.1}
\setlength{\tabcolsep}{6pt}
\begin{tabular}{|l|c|c|c|c|}
\hline
Potential & $M_{-} / M_{+}$ & $P_{+} / P_{-}$ & $T_{+} / T_{-}$ & $\rho_{+} / \rho_{-}$ \\
\hline
$\Upsilon_{B=14}$  & 4.801 & 5.517  & 3.049 & 3.165 \\
$\Upsilon_{B=33}$  & 5.328 & 6.332  & 3.168 & 3.399 \\
$\Upsilon_{B=100}$ & 6.342 & 8.023  & 3.414 & 3.805 \\
$\Upsilon_{B=390}$ & 8.620 & 12.462 & 4.055 & 4.533 \\
\hline
\end{tabular}

\caption{This table numerically present the case of spin $\rm{a}=0.50$ from the figure \ref{MPTrho-para}. The galactic configurations are denoted by $\Upsilon_B$ and their respective shock strength presented by the ratio of mach number ($M_{-} / M_{+}$), pressure ($P_{+} / P_{-}$), temperature ($T$), and density ($\rho$) for the VE disc model are given in the respective column of the table, Here the $\mathcal{E}=0.004$, and $\lambda=3.02$.}
\label{tab:VE_MPTR}
\end{table}

We subsequently analyze the variation of the shock location for a fixed black hole spin parameter of $\rm{a} = 0.50$ in axisymmetric transonic accretion processes and plot it in Fig. \ ref {shocklocationCHCOVE} for three different disc geometries: CH, CO, and VE, respectively. Here, we also did the study for our four sets of galactic mass distribution given by $\Upsilon_{B} = 14, 33, 100$, and $390$. We see that the radial position of the shock location is influenced by both the disc geometry and the surrounding galactic potential. The shock front gradually changes for $\Upsilon_B=100$ and $\Upsilon_B=390$, with noticeable variations between the three models, particularly noticeable in the VE disc model. We numerically present the results from this figure in Table \ref{tab:rsh_percentage}. Here, a clear distinction among various potentials and the associated shift in shock location can be seen. We also calculate the shift in $r_{\mathrm{sh}}$  with different $\Upsilon_B$ with respect to black hole potential by estimating $\Delta r_{\mathrm{sh}} = r_{\mathrm{sh}}^{\Upsilon_B} - r_{\mathrm{sh}}^{\mathrm{BH}}$. To provide the galactic impact, we present the effect sizes in terms of the fractional change of the shock location by estimating $\Delta r_{\mathrm{sh}} / r_{\mathrm{sh}}^{\mathrm{BH}}$. From the table and the figure, we see that for $\Upsilon_B=390$, the fractional change in shock location is almost $45\%$, which is the maximum impact of the galactic potential.
\begin{figure*}
\centering
\begin{minipage}{0.48\linewidth}
\centering
\includegraphics[width=\linewidth]{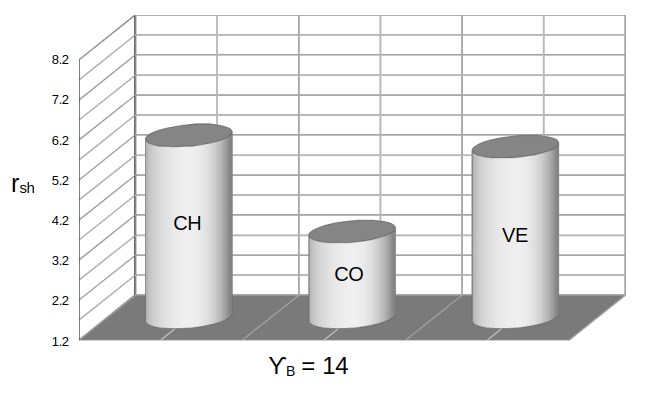}
\small (a)
\label{shokM}
\end{minipage}
\hfill
\begin{minipage}{0.48\linewidth}
\centering
\includegraphics[width=\linewidth]{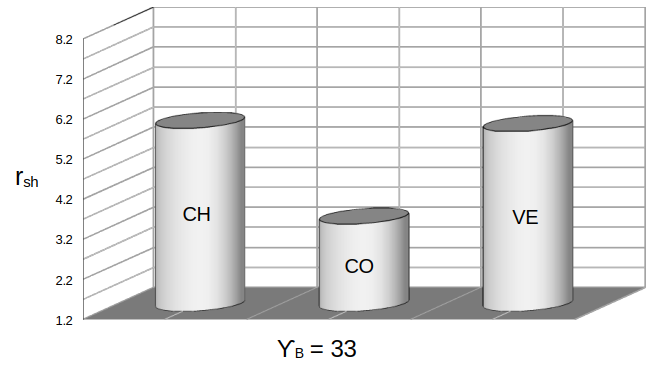}
\small (b)
\label{shokP}
\end{minipage}

\vspace{2ex}

\begin{minipage}{0.48\linewidth}
\centering
\includegraphics[width=\linewidth]{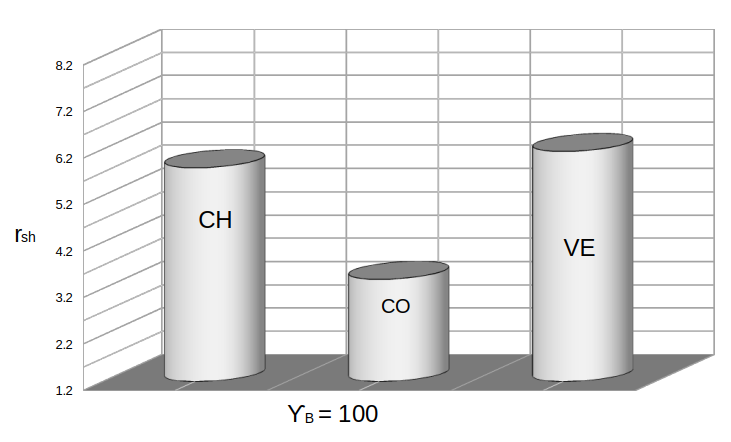}
\small (c)
\label{shokrho}
\end{minipage}
\hfill
\begin{minipage}{0.48\linewidth}
\centering
\includegraphics[width=\linewidth]{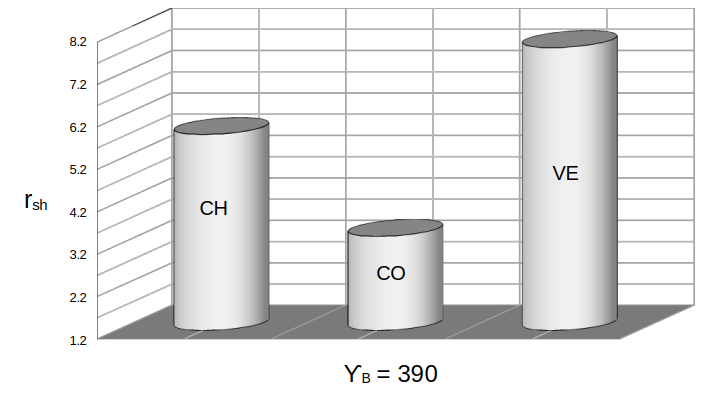}
\small (d)
\label{shokT}
\end{minipage}

\caption{We present here the variation of shock location ($r_{sh}$) for the three different accretion flow geometries: CH, CO, and VE, considering the black hole spin parameter $\rm{a} = 0.50$, energy $\mathcal{E}=0.004$, and angular momentum $\lambda=2.98$. Panels (a), (b), (c), and (d) respectively present the cases for the galactic parameters: $\Upsilon_{B} = 14$, $\Upsilon_{B} = 33$, $\Upsilon_{B} = 100$, and $\Upsilon_{B} = 390$. The differences that appear due to different galactic parameters are presented numerically in Table \ref{tab:rsh_percentage}. }
\label{shocklocationCHCOVE}
\end{figure*}

\begin{table}[H]
\centering
\renewcommand{\arraystretch}{1.0}
\setlength{\tabcolsep}{3pt}
\begin{tabular}{|l|c|c|c|c|}
\hline
Potential &
$r_{\mathrm{sh}}$ &
$\Delta r_{\mathrm{sh}} = r_{\mathrm{sh}}^{\Upsilon_B} - r_{\mathrm{sh}}^{\mathrm{BH}}$ &
$\Delta r_{\mathrm{sh}} / r_{\mathrm{sh}}^{\mathrm{BH}}$ &
$\%\ \text{Change}$ \\
\hline
BH              & 5.475 & --     & --     & --    \\
\hline
$\Upsilon_{B=14}$   & 5.475 & 0.0    & 0.0    & 0.0  \\
\hline
$\Upsilon_{B=33}$   & 5.698 & 0.226  & 0.0413 & 4.13  \\
\hline
$\Upsilon_{B=100}$  & 6.157 & 0.685  & 0.1251 & 12.51 \\
\hline
$\Upsilon_{B=390}$  & 7.877 & 2.475  & 0.4523 & 45.23 \\
\hline
\end{tabular}
\caption{This table presents the changes in shock location for the VE disc model and the adiabatic case in the presence of a galactic potential in terms of $ r _ {\ mathrm {sh}} $ with respect to the  BH case. We consider the spin parameter $\rm{a}=0.50$. The fractional changes with respect to the BH case are given by $\Delta r_{\mathrm{sh}} / r_{\mathrm{sh}}^{\mathrm{BH}}$. Variation of shock location due to the galactic component $\Upsilon_B$ relative to the black hole (BH) case figure \ref {shocklocationCHCOVE}.}
\label{tab:rsh_percentage}
\end{table}

\subsection{Acoustic surface gravity}
\label{cpe}
\begin{figure*}
 \centerline{\includegraphics[width=0.8\linewidth,clip]{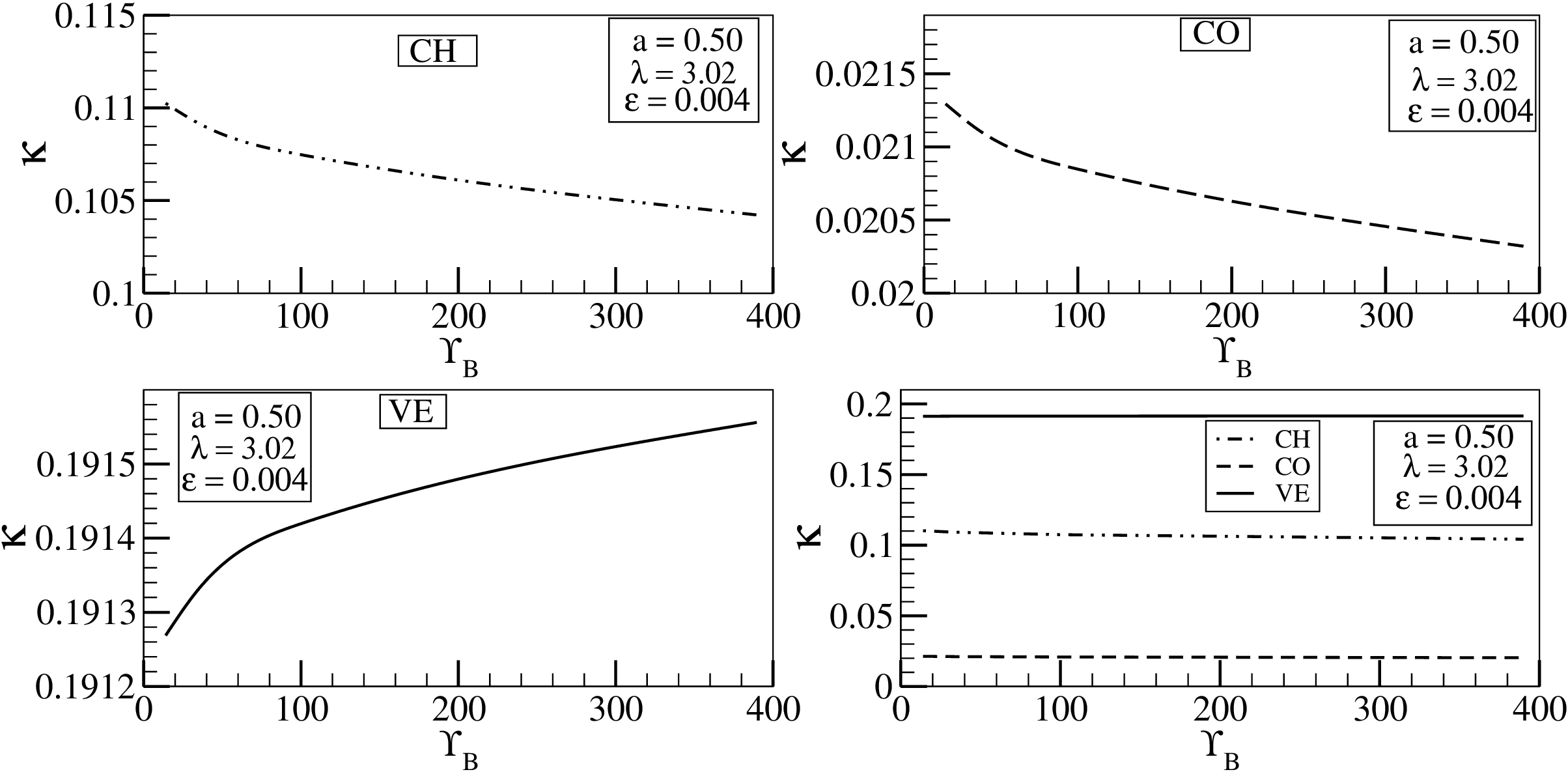}}
 \caption{ In these figures, we plot the variation of acoustic surface gravity $\kappa$ with the galactic parameter $\Upsilon_B$.  We choose adiabatic flow and the black hole spin parameter $\rm{a}=0.50$. The panel marked by CH shows the case for constant height, panel CO shows conical height, and panel VE shows the variation for the vertical equilibrium model. For all the panels, the specific energy $\mathcal{E} = 0.004$ and specific angular momentum $\lambda = 3.02$. From these figures, we see that for the CH and CO disc models, $\kappa$ decreases; however, for the VE disc model, it increases for the selected galactic mass distribution sets. The fourth panel shows the differences in values of $\kappa$ for the three disc geometries. We see acoustic surface gravity possesses a maximum value for the VE disc model.
}
\label{k-vs-para-a05}
\end{figure*}

Transonic accretion flows naturally give rise to an acoustic horizon, where the radial flow speed equals the local sound speed, closely resembling the event horizon of a black hole. The surface gravity associated with this acoustic horizon characterizes the strength of the horizon and governs the behaviour of linear perturbations in the flow. In the following, we investigate the dependence of the acoustic surface gravity on the black hole spin and the galactic background potential in the context of axisymmetric accretion. In Figure \ref{k-vs-para-a05}, we present the variation of acoustic surface gravity $\kappa$ as a function of galactic mass distribution denoted by $\Upsilon_B$. We consider the three disc configurations, i.e., CH, CO, and VE, and present them accordingly in the respective subpanels. The fourth panel presents the comparison of $\kappa$ for the three disc models and shows that $\kappa$ has a maximum value for VE disc geometries. We choose  $\mathcal{E} = 0.004$,  $\lambda = 3.02$, and spin $\rm{a}=0.50$.  From the figure, we see that for the CH and CO disc models, $\kappa$ decreases; however, for the VE disk model, it increases for the selected galactic mass distribution sets. From our selected galactic model, we see that $\Upsilon_B=100$ \& $390$ contain more dark matter contribution. The decreases in $\kappa$ values from Set-1 to Set-4 of $\Upsilon_B$ values indicate that acoustic surface gravity decreases with increasing dark matter content. Among Set-3 \& 4, both contain the same dark matter mass; however, the later posses slightly higher hot gas mass and results in more decreement in $\kappa$ values. Moreover, by comparing across different accretion geometries in the comparison subpanel, we see that acoustic surface gravity depends on the chosen disc model.

\begin{figure}
 \centerline{\includegraphics[width=1.0\linewidth,clip]{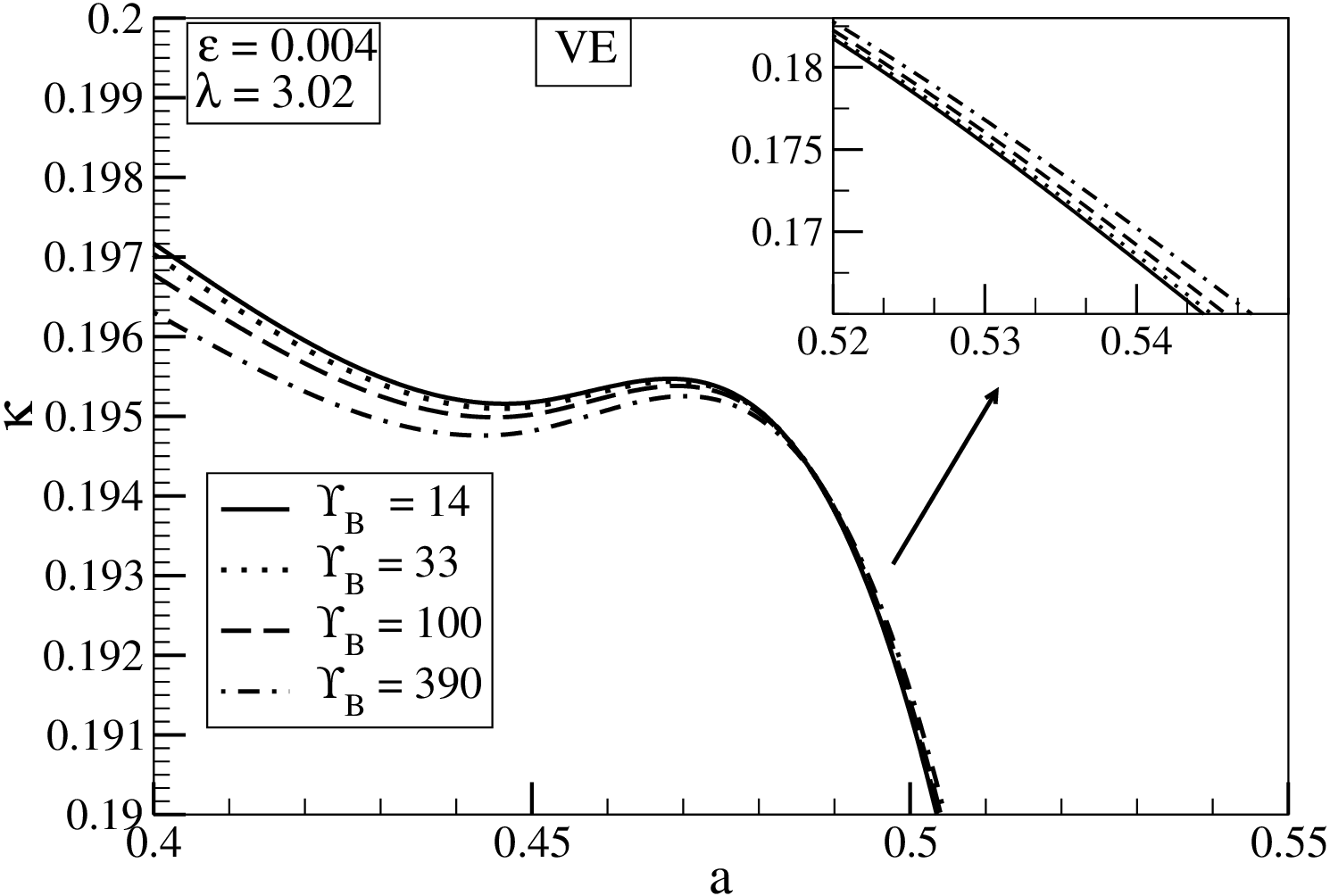}}
 \caption{This figure presents the variation of the acoustic surface gravity ($\kappa$) in the VE accretion model with the black hole spin parameter $\rm{a}$. The analysis is performed for a range of galactic parameters: $\Upsilon_{B} = 14$, $33$, $100$, and $390$. The subplot shows a zoomed-in view highlighting the variation of $\kappa$ for different values of $\Upsilon_{B}$. All four scenarios clearly have different values and appear to be the same for higher spin parameters. The zoomed-in subpanel however, clearly shows that the $\kappa$ profile differs for different galaxy parameters, even for larger black hole spin values.
}
\label{fig14}
\end{figure}

In order to investigate the impact of central black hole spins on the acoustic surface gravity in the accreting medium under $\Upsilon_B$ impact, we further investigate the variation of the acoustic surface gravity $\kappa$ with black hole spin in Figure \ref{fig14}. We consider four selected sets of galactic parameters, e.g., $\Upsilon_{B}$: $14$, $33$, $100$, and $390$. For this plot, we assume the VE accretion flow model. In the plot, $\Upsilon_{B}=14$  is shown by a solid line, $33$ by a dotted line, $100$ by a dashed line, and $390$ by a dot-dashed line. All four cases clearly possess different values and seem to be the same for higher spin parameters. However, the zoomed-in subpanel clearly visualizes that the $\kappa$ profile is distinct for different galactic parameters even for higher black hole spin values.

 We comment on the overall conclusion from the results that we obtained in the next section.

\section{Concluding Remarks}
\label{conclu}

In the present work, we have investigated low-angular-momentum, axisymmetric, inviscid accretion onto a rotating SMBH embedded in a multi-component galactic potential comprising the central BH, stellar bulge, DM halo, and hot gas, considering the CH, CO, and VE disc geometries for both adiabatic and isothermal equations of state.
The present study demonstrates that the host galaxy is not merely the environment in which accretion occurs, but an active dynamical component that regulates the transonic topology, shock properties, and analogue horizon characteristics of the flow. Our analysis leads to the following main conclusions.

\begin{enumerate}
\item The inclusion of the galactic potential modifies the balance between gravity, centrifugal support, and pressure gradients that governs transonic accretion. As a result, the sonic points and the corresponding entropy accretion rates are shifted, while the familiar X-type and O-type critical points are preserved. This leads to a systematic deformation of the multi-transonic parameter space, demonstrating that the boundaries separating mono-critical and multi-critical solutions in the $\mathcal{E}$--$\lambda$ plane depend not only on the black-hole spin and flow angular momentum but also on the large-scale mass distribution of the host galaxy.
\item The galactic environment exerts a significant influence on the shock properties of the flow by enhancing the inward gravitational acceleration and thereby weakening the effectiveness of the centrifugal barrier, especially for VE disc geometry. Consequently, the region of parameter space that permits stationary shocks is systematically reduced, with the strongest suppression occurring in dark-matter-dominated environments, where shocks migrate closer to the black hole. At the same time, the surviving shocks become systematically stronger, exhibiting larger jumps in Mach number, density, temperature, and pressure. The galactic potential therefore acts both as a regulator of shock formation and as a filter that preferentially selects stronger hydrodynamic discontinuities.
 \item Black-hole spin, multi-component galactic potential, and disc geometry together determine the transonic and shock properties of the flow. Increasing black-hole spin enhances rotational support through relativistic frame dragging, thereby promoting shock formation at lower angular momentum, whereas the galactic potential acts oppositely by increasing radial confinement and reducing the shock-permitting parameter space. The response of the flow further depends on the disc geometry through its influence on the pressure distribution and vertical force balance, with the vertical equilibrium configuration producing the strongest shocks. These findings show that neither black-hole properties nor disc geometry alone determine the flow behaviour; the surrounding galactic gravitational field is an equally important dynamical component.
 \item The stability analysis shows that the inclusion of the galactic potential modifies the stationary transonic solutions without compromising their dynamical stability. The bounded behaviour of the linear perturbations demonstrates that these embedded transonic accretion solutions remain physically realizable despite the additional gravitational complexity introduced by the host galaxy.
 \item An important implication of the analogue gravity analysis is that acoustic horizons are determined by the locations of sonic points, and sonic transitions are themselves governed by the global gravitational environment through which the fluid propagates. Consequently, the properties of acoustic horizons inherit information about the large-scale galactic mass distribution.
\end{enumerate}
Although the present study is primarily a theoretical investigation of how the host-galaxy gravitational field modifies low-angular-momentum transonic accretion flows, the underlying transonic framework is itself well established both theoretically and observationally. Transonic solutions with standing shocks have successfully explained the spectral and timing properties of black-hole accretion systems, with the post-shock region acting as the Comptonizing corona responsible for hard X-ray emission and variability, and also serving as a potential launching site for outflows and jets \citep{1995ApJ...455..623C,1996ApJ...457..805M,2008ApJ...689L..17M,2005ApJ...632..476L,Chattopadhyay:2011rmr,Chakrabarti:2016iln}. The existence and persistence of shocks have furthermore been demonstrated in numerical simulations of transonic accretion flows \citep{2013MNRAS.430.2836G,2015MNRAS.448.3221G}, while transonic models have recently been applied successfully to explain the variability and observational properties of supermassive black-hole systems, including Sgr A* and AGN sources \citep{2018MNRAS.478.3544R,2023ApJ...954...56O,2024ApJ...967....4D,2019ApJ...877...65N,2021Galax...9...21M,2021MNRAS.506.3111N,2022mgm..conf..231M,2022JApA...43...90M,2022A&A...662A..77M,2022A&A...663A.178M}. Within this established framework, our results predict that the host-galaxy potential systematically modifies the transonic structure by shifting sonic points, shrinking the multi-transonic and shock-permitting parameter space, moving surviving shocks closer to the black hole, and producing stronger shocks. These predictions can be tested in future hydrodynamic and magnetohydrodynamic simulations and may have important implications for spectral and timing models of AGN accretion embedded within realistic galactic environments. Whether these galactic-scale modifications generate distinct observational signatures remains an open question for future work.

The present results should be interpreted within the assumptions of our model. We have considered axisymmetric, inviscid accretion in a pseudo-Kerr potential, neglecting viscosity, turbulence, magnetic fields, radiative processes, and non-axisymmetric instabilities \citep{1991ApJ...376..214B,1994ApJ...428L..13N,1984MNRAS.208..721P,1996ApJ...461..565A}. The galactic environment has also been represented by static, spherically symmetric mass distributions, while the thermodynamic and stability analyses have been restricted to adiabatic and isothermal equations of state and linear radial perturbations \citep{2008gady.book.....B,1995ApJ...452..710N,2008bhad.book.....K,1988ApJ...326..277B,1994MNRAS.270..871N}. Although these assumptions may alter the quantitative properties of the solutions, the central physical conclusion—that the host galaxy modifies the effective force balance and thereby regulates the transonic and shock structure of the flow—is expected to remain robust.

Several natural extensions follow from the present work. Future studies should examine the sensitivity of our results to alternative pseudo-Kerr prescriptions and ultimately to fully general relativistic hydrodynamic and magnetohydrodynamic models \citep{1992MNRAS.256..300C,1998IJMPD...7..471L,1999A&A...343..325S,2002ApJ...581..427M,2002ApJ...577..880D,2006MNRAS.369..976C,2007ApJ...667..367G,2014MNRAS.445.4463G,2014bhns.work..121K}. It will also be important to investigate more realistic galactic mass distributions, including different dark matter halo models and cosmological backgrounds \citep{1986ApJ...301...27B,1996ApJ...462..563N,1997ApJ...490..493N,1999PhRvL..83.1719G,2012A&A...540A..70R,2013PhRvD..88f3522S,2015IJMPD..2450084G,2018MNRAS.479.3011R}, and to extend the analogue gravity analysis to nonlinear, dissipative, and magnetized flows \citep{1981PhRvL..46.1351U,1998CQGra..15.1767V,2004gr.qc....11006D}. Such investigations will provide a more complete understanding of galactic environments, black-hole accretion, and emergent spacetime phenomena.

\section*{Data availability statement}
No observational data are analysed in this work. Representative numerical datasets used to generate the results and figures are available at \url{https://doi.org/10.5281/zenodo.21400775}. The complete simulation data are available from the corresponding author upon reasonable request.

\section*{Acknowledgement}
Author RS gratefully acknowledges the Department of Physics, Dhruba Chand Halder College. Author SC acknowledges the ANRF, Government of India, under the National Post-Doctoral Fellowship (NPDF) scheme (Ref. No. PDF/2025/001982). The author, SN likes to acknowledge Dr. Shubhrangshu Ghosh, SRM University, Sikkim, for attracting attention to these components of galactic potentials available in the literature.

\appendix
\twocolumngrid

\section{Critical Values for Adiabatic flows}
\label{Vcric}
For constant height (CH) disc geometry,
\begin{equation}
H(r) \simeq C, \quad \text{with } C \text{ being a constant}.
\end{equation}
The respective energy and entropy accretion rate are ~\citep{Bilić2014},
\begin{equation}
\mathcal{E}_{\mathrm{CH}} - \frac{1}{2}\left(\frac{\gamma+1}{\gamma-1}\right)\left[r_{c}\left(\Phi'_{\mathrm{Gal}}\right)_{r_{c}}-\frac{\lambda^{2}}{r_{c}^{2}}\right]+\Phi_{\mathrm{Gal}}\left(r_{c}\right)+\frac{\lambda^{2}}{2 r_{c}^{2}} = 0,
\end{equation}

\begin{equation}
 \dot{\mathcal{M}}_{\mathrm{CH}}^{2} = 4 \pi^{2} r_{\mathrm{c}}^{2} \left( r_{\mathrm{c}} \Phi'_{\mathrm{Gal}}(r_{\mathrm{c}}) - \frac{\lambda^2}{r_{\mathrm{c}}^2} \right)^{2n+1}.
\end{equation}

The spatial gradient of the dynamical velocity, $\frac{dv}{dr}$, and the gradient of the sound speed, $\frac{dc_s}{dr}$, are,
\begin{equation}
\left. \frac{d c_{s}}{d r} \right|_{\rm CH}^{\rm adi} =
(1 - \gamma) \frac{c_{s}}{v} \left( \frac{1}{2} \frac{d v}{d r} + \frac{v}{2 r} \right),
\end{equation}

\begin{equation}
\left. \frac{d v}{d r} \right|_{\rm CH}^{\rm adi} =
\frac{v \left( \frac{c_{s}^{2}}{r} + \frac{\lambda^{2}}{r^{3}} - \Phi'_{\mathrm{Gal}}(r) \right)}{v^{2} - c_{s}^{2}}
\label{ch1}.
\end{equation}
At the critical location i.e. $r_c$,
\begin{equation}
\text{For CH:} \qquad
v_{r_c} = c_{s_{r_c}}, \qquad
c_{s_{r_c}} = \sqrt{\, r_c\,\Phi'_{\mathrm{Gal}}(r_c) - \frac{\lambda^{2}}{r_c^{2}} \,}
\label{eq:CH_condition}.
\end{equation}

Applying L'H\^opital's rule to Eq.~(\ref{ch1}) at
$r=r_c$,  yields

\begin{equation}
\left( \frac{dv}{dr} \right)^{\mathrm{CH}}_{r_c}= \left( \frac{dv}{dr} \right)^{\mathrm{CH}}_{0} \pm \sqrt{ \left( \frac{dv}{dr} \right)^{\mathrm{CH}}_{1}- \left( \frac{dv}{dr} \right)^{\mathrm{CH}}_{2}},
\end{equation}

where,

$\left(\frac{dv}{dr} \right)^{\mathrm{CH}}_{0}=\frac{1}{r_c} \left(\frac{1-\gamma}{1+\gamma}\right)\sqrt{r_c \Phi'_{\mathrm{Gal}}(r_c) - \frac{\lambda^2}{r_c^2}}$,

$ \left( \frac{dv}{dr} \right)^{\mathrm{CH}}_{1}= \frac{1}{r_c^2} \left( \frac{1 - \gamma}{1 + \gamma} \right)^2
\left( r_c \Phi'_{\mathrm{Gal}}(r_c) - \frac{\lambda^2}{r_c^2} \right)$  and

$ \left( \frac{dv}{dr} \right)^{\mathrm{CH}}_{2}= \frac{ \frac{\gamma}{r_c^2} \left( r_c \Phi'_{\mathrm{Gal}}(r_c) - \frac{\lambda^2}{r_c^2} \right)
+ \frac{3 \lambda^2}{r_c^4} + \Phi''_{\mathrm{Gal}}(r_c) }
{ \sqrt{ r_c \Phi'_{\mathrm{Gal}}(r_c) - \frac{\lambda^2}{r_c^2} } }$.

For conical flow model (CO), where the disc height varies linearly with radius as,
\begin{equation}
H(r) \simeq D r, \quad \text{where} \hspace{0.2cm} D \text{ is constant}.
\end{equation}
The respective sets of equations are,
\begin{equation}
\mathcal{E}_{\mathrm{CO}} - \frac{1}{4}\left(\frac{\gamma+1}{\gamma-1}\right)\left[r_{c}\left(\Phi'_{\mathrm{Gal}}\right)_{r_{c}}-\frac{\lambda^{2}}{r_{c}^{2}}\right]+\Phi_{\mathrm{Gal}}\left(r_{c}\right)+\frac{\lambda^{2}}{2 r_{c}^{2}} = 0,
\end{equation}

\begin{equation}
\dot{\mathcal{M}}_{\mathrm{CO}}^2 = 4 \pi^2 D^2 r_{\mathrm{c}}^4 \left( \frac{1}{2} \left[ r_{\mathrm{c}} \Phi'_{\mathrm{Gal}}(r_{\mathrm{c}}) - \frac{\lambda^2}{r_{\mathrm{c}}^2} \right] \right)^{2n+1},
\end{equation}

\begin{equation}
\left. \frac{d c_{s}}{d r} \right|_{\rm CO}^{\rm adi} =
(1 - \gamma) \frac{c_{s}}{v} \left( \frac{1}{2} \frac{d v}{d r} + \frac{v}{r} \right),
\end{equation}
\begin{equation}
\left. \frac{d v}{d r} \right|_{\rm CO}^{\rm adi} =
\frac{v \left( \frac{2 c_{s}^{2}}{r} + \frac{\lambda^{2}}{r^{3}} - \Phi'_{\mathrm{Gal}}(r) \right)}{v^{2} - c_{s}^{2}}
\label{co1}.
\end{equation}

\begin{equation}
\text{For CO:} \qquad
v_{r_c} = c_{s_{r_c}}, \qquad
c_{s_{r_c}} = \sqrt{ \frac{r_c \, \Phi'_{\mathrm{Gal}}(r_c)}{2} - \frac{\lambda^2}{2 r_c^2} }.
\end{equation}

Applying L'H\^opital's rule to Eq.~(\ref{co1}) at the critical point,

\begin{equation}
 \left( \frac{dv}{dr} \right)^{\mathrm{CO}}_{r_c}= \left( \frac{dv}{dr} \right)^{\mathrm{CO}}_{0} \pm  \sqrt{ \left( \frac{dv}{dr} \right)^{\mathrm{CO}}_{1}- \left( \frac{dv}{dr} \right)^{\mathrm{CO}}_{2}},
\end{equation}

where,

$\left( \frac{dv}{dr} \right)^{\mathrm{CO}}_{0}=\frac{2}{r_c} \left( \frac{1 - \gamma}{1 + \gamma} \right)
\sqrt{ \frac{r_c \Phi'_{\mathrm{Gal}}(r_c)}{2} - \frac{\lambda^2}{2 r_c^2} }$,

$ \left( \frac{dv}{dr} \right)^{\mathrm{CO}}_{1}=  \frac{4}{r_c^2} \left( \frac{1 - \gamma}{1 + \gamma} \right)^2
\left( \frac{r_c \Phi'_{\mathrm{Gal}}(r_c)}{2} - \frac{\lambda^2}{2 r_c^2} \right)$ and,

$ \left( \frac{dv}{dr} \right)^{\mathrm{CO}}_{2}=\frac{ \frac{2(2 \gamma - 1)}{r_c^2} \left( \frac{r_c \Phi'_{\mathrm{Gal}}(r_c)}{2} - \frac{\lambda^2}{2 r_c^2} \right)
+ \frac{3 \lambda^2}{r_c^4} + \Phi''_{\mathrm{Gal}}(r_c) }
{ (1 + \gamma) \sqrt{ \frac{r_c \Phi'_{\mathrm{Gal}}(r_c)}{2} - \frac{\lambda^2}{2 r_c^2} } }$.

For vertical equilibrium (VE) model, the disc height,
\begin{equation}
H(r) = c_{\mathrm{s}} \left( \frac{r}{\gamma \, \Phi'_{\mathrm{Gal}}} \right)^{1/2}.
\end{equation}
The respective sets of equations are,
\begin{equation}
\mathcal{E}_{\mathrm{VE}} - \frac{2 \gamma}{\gamma-1}\frac{\left[r_{c} \Phi'_{\mathrm{Gal}}\left(r_{c}\right)-\frac{\lambda^{2}}{r_{c}^{2}}\right]}{\left[3-r_{c}\left(\frac{\Phi''_{\mathrm{Gal}}}{ \Phi'_{\mathrm{Gal}}}\right)_{r_{c}}\right]} +\Phi_{\mathrm{Gal}}\left(r_{c}\right)+\frac{\lambda^{2}}{2 r_{c}^{2}} = 0 \label{EVE},
\end{equation}

\begin{multline}
\label{dotmfix}
\dot{\mathcal{M}}_{\mathrm{VE}}^2 = \frac{8 \pi^2  r_{\mathrm{c}}^3}
{\Phi^{\prime}_{\rm Gal}(r_{\mathrm{c}})\gamma(\gamma + 1)}
\left\{ \frac{1}{(\gamma + 1)}
\left[ r_{\mathrm{c}} \Phi^{\prime}_{\rm Gal}(r_{\mathrm{c}})
- \frac{\lambda^2}{r_{\mathrm{c}}^2} \right] \times \right. \\ \left. \left[ 3 - r_{\mathrm{c}}
\frac{\Phi^{\prime \prime}_{\rm Gal}(r_{\mathrm{c}})}{\Phi^{\prime}_{\rm Gal}(r_{\mathrm{c}})}
\right]^{-1} \right\}^{2(n + 1)},
\end{multline}

\begin{equation}
\left. \frac{d c_{s}}{d r} \right|_{\rm VE}^{\rm adi} =
\left( \frac{1 - \gamma}{1 + \gamma} \right) \frac{c_{s}}{v} \left[
\frac{d v}{d r} + \frac{v}{2} \left( \frac{3}{r} - \frac{\Phi''_{\mathrm{Gal}}(r)}{\Phi'_{\mathrm{Gal}}(r)} \right)
\right],
\end{equation}

\begin{equation}
\left. \frac{d v}{d r} \right|_{\rm VE}^{\rm adi} =
\frac{v \left[ \frac{c_{s}^{2}}{1 + \gamma} \left( \frac{3}{r} - \frac{\Phi''_{\mathrm{Gal}}(r)}{\Phi'_{\mathrm{Gal}}(r)} \right) + \frac{\lambda^{2}}{r^{3}} - \Phi'_{\mathrm{Gal}}(r) \right]}{v^{2} - \frac{2}{1 + \gamma} c_{s}^{2}}
\label{ve1}.
\end{equation}

\begin{multline}
\text{For VE:} \qquad
v_{r_c} = \sqrt{ \frac{2}{\gamma + 1} } \, c_{s_{r_c}}, \qquad \\
 c_{s_{r_c}} = \sqrt{ (\gamma + 1)
\frac{ \Phi'_{\mathrm{Gal}}(r_c) - {\displaystyle \frac{\lambda^2}{r_c^3}} }
     {{\displaystyle \frac{3}{r_c}} - {\displaystyle \frac{ \Phi''_{\mathrm{Gal}}(r_c) }{ \Phi'_{\mathrm{Gal}}(r_c) } }} }.
\end{multline}
Applying L'H\^opital's rule to Eq.~(\ref{ve1}) at the critical point,

\begin{equation}
 \left( \frac{dv}{dr} \right)^{\mathrm{VE}}_{r_c}= \left( \frac{dv}{dr} \right)^{\mathrm{VE}}_{0} \pm \sqrt{ \frac{\gamma + 1}{4 \gamma} } \sqrt{ \left( \frac{dv}{dr} \right)^{\mathrm{VE}}_{1}- \left( \frac{dv}{dr} \right)^{\mathrm{VE}}_{2}},
\end{equation}

where

$\left( \frac{dv}{dr} \right)^{\mathrm{VE}}_{0}=2 u_c \left( \frac{\gamma - 1}{8 \gamma} \right)
\left[ \frac{3}{r_c} + \frac{\Phi'''_{\mathrm{Gal}}(r_c)}{\Phi'_{\mathrm{Gal}}(r_c)} \right]$,

$ \left( \frac{dv}{dr} \right)^{\mathrm{VE}}_{1}= u_c^2 \frac{\gamma - 1}{\gamma + 1} \cdot \frac{\gamma - 1}{4 \gamma}
\left( \frac{3}{r_c} + \frac{\Phi''_{\mathrm{Gal}}(r_c)}{\Phi'_{\mathrm{Gal}}(r_c)} \right)^2$ and,

$\left( \frac{dv}{dr} \right)^{\mathrm{VE}}_{2}=u_c^2 \frac{1 + \gamma}{2} \left(\phi_{31}-A\phi_{31}^2+B\phi_{21}-C\right)-\Phi''_{\mathrm{Gal}}(r_c) + \frac{3 \lambda^2}{r_c^4}$,

where, $\phi_{21}=\frac{\Phi''_{\mathrm{Gal}}(r_c)}{\Phi'_{\mathrm{Gal}}(r_c)}$, $\phi_{31}=\frac{\Phi'''_{\mathrm{Gal}}(r_c)}{\Phi'_{\mathrm{Gal}}(r_c)}$, $A=\frac{2 \gamma}{(1 + \gamma)^2}$, $B=\frac{6 (\gamma - 1)}{\gamma (1 + \gamma)^2}$, and $C=\frac{6 (2 \gamma - 1)}{\gamma^2 (1 + \gamma)^2}$.

\section{Critical Values for Isothermal Flow}
\label{Vcriciso}
For isothermal accretion, the sound speed remains constant
throughout the flow. The velocity gradients for the three disc geometries are given by,
\begin{eqnarray}
\left. \frac{dv}{dr} \right|_{\rm CH}^{\rm iso} &=
\frac{v \left( \frac{c_s^2}{r} - \Phi'_{\mathrm{Gal}}(r) + \frac{\lambda^2}{r^3} \right)}{v^2 - c_s^2} \label{ch2}\\
\left. \frac{dv}{dr} \right|_{\rm CO}^{\rm iso} &=
\frac{v \left( \frac{2 c_s^2}{r} - \Phi'_{\mathrm{Gal}}(r) + \frac{\lambda^2}{r^3} \right)}{v^2 - c_s^2} \label{co2}\\
\left. \frac{dv}{dr} \right|_{\rm VE}^{\rm iso} &=
\frac{v \left[ \frac{c_s^2}{2} \left( \frac{3}{r} - \frac{\Phi''_{\mathrm{Gal}}(r)}{\Phi'_{\mathrm{Gal}}(r)} \right)
- \Phi'_{\mathrm{Gal}}(r) + \frac{\lambda^2}{r^3} \right]}{v^2 - c_s^2}
\label{ve2}
\end{eqnarray}
\begin{equation}
\left( \frac{dv}{dr} \right)^{\mathrm{CH}}_{r_c}= \left( \frac{dv}{dr} \right)^{\mathrm{CH}}_{0} \pm \sqrt{ \left( \frac{dv}{dr} \right)^{\mathrm{CH}}_{1}- \left( \frac{dv}{dr} \right)^{\mathrm{CH}}_{2}}
\end{equation}

where,

$\left(\frac{dv}{dr} \right)^{\mathrm{CH}}_{0}=\frac{1}{r_c} \left(\frac{1-\gamma}{1+\gamma}\right)\sqrt{r_c \Phi'_{\mathrm{Gal}}(r_c) - \frac{\lambda^2}{r_c^2}}$,

$ \left( \frac{dv}{dr} \right)^{\mathrm{CH}}_{1}= \frac{1}{r_c^2} \left( \frac{1 - \gamma}{1 + \gamma} \right)^2
\left( r_c \Phi'_{\mathrm{Gal}}(r_c) - \frac{\lambda^2}{r_c^2} \right)$  and

$ \left( \frac{dv}{dr} \right)^{\mathrm{CH}}_{2}= \frac{ \frac{\gamma}{r_c^2} \left( r_c \Phi'_{\mathrm{Gal}}(r_c) - \frac{\lambda^2}{r_c^2} \right)
+ \frac{3 \lambda^2}{r_c^4} + \Phi''_{\mathrm{Gal}}(r_c) }
{ \sqrt{ r_c \Phi'_{\mathrm{Gal}}(r_c) - \frac{\lambda^2}{r_c^2} } }$

which provide the following critical point conditions:
\begin{equation}
\text{For CH:} \quad
v_{r_c} = c_{s_{r_c}} = \sqrt{ \frac{k_B}{\mu m_H} } \, T^{1/2}
= \sqrt{ r_c \left[ \Phi'_{\mathrm{Gal}} \right]_{r_c} - \frac{\lambda^2}{r_c^2} }
\end{equation}

\begin{equation}
\text{For CO:} \quad
 v_{r_c} = c_{s_{r_c}} = \sqrt{ \frac{k_B}{\mu m_H} } \, T^{1/2}
= \sqrt{ \frac{1}{2} \left( r_c \left[ \Phi'_{{\mathrm{Gal}}}\right]_{r_c} - \frac{\lambda^2}{r_c^2} \right) }
\end{equation}

\begin{equation}
\text{For VE:} \quad
 v_{r_c} = c_{s_{r_c}} = \sqrt{ \frac{k_B}{\mu m_H} } \, T^{1/2} =
\sqrt{\displaystyle \frac{{2 \left( r_c \left[ \Phi'_{{\mathrm{Gal}}} \right]_{r_c} - \frac{\lambda^2}{r_c^2} \right)}}{
 \left( 3 - r_c \left[ \frac{ \Phi''_{{\mathrm{Gal}}} }{ \Phi'_{{\mathrm{Gal}}} } \right]_{r_c} \right)}}
\end{equation}

Applying L'H\^opital's rule at the critical point yields the
critical velocity gradients

\begin{equation}
\left|
\left(
\frac{dv}{dr}
\right)^{\rm iso}_{\rm CH}
\right|_{r_c}
=
\pm
\frac{1}{\sqrt{2}}
\sqrt{
-\Phi''_{\rm Gal}(r_c)
-
\left(
\frac{c_s^2}{r_c^2}
+
\frac{3\lambda^2}{r_c^4}
\right)
},
\end{equation}

\begin{equation}
\left|
\left(
\frac{dv}{dr}
\right)^{\rm iso}_{\rm CO}
\right|_{r_c}
=
\pm
\frac{1}{\sqrt{2}}
\sqrt{
-\Phi''_{\rm Gal}(r_c)
-
\left(
\frac{2c_s^2}{r_c^2}
+
\frac{3\lambda^2}{r_c^4}
\right)
},
\end{equation}

and

\begin{equation}
\left|
\left(
\frac{dv}{dr}
\right)^{\rm iso}_{\rm VE}
\right|_{r_c}
=
\pm
\frac{1}{\sqrt{2}}
\sqrt{
\frac{c_s^2}{2}
\left(
\frac{du}{dr}
\right)^{\rm VE}_{1}
-
\left(
\frac{du}{dr}
\right)^{\rm VE}_{2}
}.
\end{equation}

where
\begin{equation}
\left(
\frac{dv}{dr}
\right)^{\rm VE}_{1}
=
\left(
\frac{\Phi''_{\rm Gal}(r_c)}
{\Phi'_{\rm Gal}(r_c)}
\right)^2
-
\left(
\frac{\Phi'''_{\rm Gal}(r_c)}
{\Phi'_{\rm Gal}(r_c)}
\right),
\end{equation}

\begin{equation}
 \left(
\frac{dv}{dr}
\right)^{\rm VE}_{2}
=
\Phi''_{\rm Gal}(r_c)
+
\frac{3c_s^2}{2r_c^2}
+
\frac{3\lambda^2}{r_c^4}.
\end{equation}

\section{Shock}
\label{shok-F}
For adiabatic flow, which conserves energy across a shock, it must be of Rankine-Hugoniot type. The  Rankine-Hugoniot conditions can then be written as ~\citep{1959flme.book.....L},
\begin{eqnarray}
\left[\!\left[\rho v\right]\!\right] & = & 0, \nonumber\\
\left[\!\left[p + \rho v^{2}\right]\!\right] & = & 0, \nonumber\\
\left[\!\left[\frac{v^{2}}{2} + h\right]\!\right] & = & 0.
\label{RH3}
\end{eqnarray}

These represent the conservation of pre- and post-shock mass, energy \& momentum.  For flows in CH \& CO   ~\citep{2016NewA...43...10S}, the shock condition reduces to:
\begin{equation}
\label{shockCO}
\left[\left[ {\displaystyle \frac{(\gamma M^2 + 1)^2}{2M^2 + (\gamma - 1)M^4}} \right]\right] = 0,
\end{equation}
while for flows in VE geometry,
\begin{equation}
\left[\left[ {\displaystyle \frac{[M^2(3\gamma - 1) + 2]^2}{2M^2 + (\gamma - 1)M^4}} \right]\right] = 0.
\end{equation}
For isothermal shocks, that are dissipative in nature, as energy is not conserved in this case, the shock conditions are found to be independent of the disc flow geometry, and can be expressed as,

\begin{eqnarray}
\rho_+ v_+ = \rho_- v_-  ,\nonumber \\
P_+ + \rho_+ v_+^2 = P_- + \rho_- v_-^2  ,\nonumber \\
M_+ + \frac{1}{M_+} = M_- + \frac{1}{M_-} . \label{disshok}
\end{eqnarray}

We use the subscripts `$-$' and `$+$' to denote pre-shock and post-shock quantities, respectively.

\bibliography{agn-acc}{}
\bibliographystyle{aasjournal}

\end{document}